\documentclass[12pt]{article}
\usepackage{amssymb,amsmath}

\usepackage{amsmath}
\usepackage{graphicx}
\usepackage{subfig}
\usepackage{multirow}

\begin{document}

\begin{titlepage}

\vspace{-4cm}

\title{
   {\LARGE The Mass Spectra, Hierarchy  and Cosmology\\[.1in]
    of $B$-$L$ MSSM Heterotic Compactifications\\[.5cm]  }}
                       
\author{{\bf
   Michael Ambroso 
   and Burt A.~Ovrut}\\[5mm]
   {\it Department of Physics, University of Pennsylvania} \\
   {\it Philadelphia, PA 19104--6396}}

\date{}

\maketitle

\begin{abstract}
\noindent

The matter spectrum of the MSSM, including three right-handed neutrino supermultiplets and one pair of Higgs-Higgs conjugate superfields, can be obtained by compactifying the $E_{8} \times E_{8}$ heterotic string and M-theory on Calabi-Yau manifolds with specific $SU(4)$ vector bundles. These theories have the standard model gauge group augmented by an additional gauged $U(1)_{B-L}$. 
Their minimal content requires that the $B$-$L$ gauge symmetry be spontaneously broken by a vacuum expectation value of at least one right-handed sneutrino. In previous papers, we presented the results of a quasi-analytic renormalization group analysis showing that $B$-$L$ gauge symmetry is indeed radiatively broken with an appropriate $B$-$L$/electroweak hierarchy.  In this paper, we extend these results by 1) enlarging the initial parameter space and 2) explicitly calculating the renormalization group equations  numerically. The regions of the initial parameter space leading to realistic vacua are presented and the $B$-$L$/electroweak hierarchy computed over these regimes. At representative points, the mass spectrum for all sparticles and Higgs fields is calculated and shown to be consistent  with present experimental bounds. Some fundamental phenomenological signatures of a non-zero right-handed sneutrino expectation value are discussed, particularly the cosmology and proton lifetime arising from induced lepton and baryon number violating interactions.

\vspace{.3in}
\noindent
\end{abstract}

\thispagestyle{empty}

\end{titlepage}

\section{Introduction}

$E_{8} \times E_{8}$ heterotic strings \cite{a} and heterotic $M$-theory \cite{b}-\cite{g} offer perhaps the simplest approach for deriving realistic particle physics from superstrings. Their compactification on Calabi-Yau threefolds with slope-stable holomorphic vector bundles leads to $N=1$ supersymmetric effective theories in four-dimensions. Such vacua include complete intersection and elliptically fibered Calabi-Yau spaces admitting vector bundles constructed using monads \cite{Ca}-\cite{Ce}, spectral covers \cite{Gaaa}-\cite{Gc} and extension of lower rank bundles \cite{Hb,Hd}. The formalism for computing the low energy spectrum in each case has been developed, and presented in \cite{Ia,Ib}, \cite{Ja,Jb} and \cite{Kb,Kc} respectively. Cohomological methods have been used to calculate the texture of Yukawa couplings and other parameters in these contexts  \cite{Lb}-\cite{Ld}. Finally, non-perturbative string instanton contributions to the superpotential are computed in \cite{Maaa}-\cite{Mc} and used to discuss moduli stability, supersymmetry breaking and the cosmological constant \cite{Nb}. These methods underlie the theory of ``brane universes'' \cite{f,Ob}.

For certain choices of Calabi-Yau manifolds with non-trivial homotopy, along with vector bundles with the appropriate structure group and topology, the derived low energy theory can be phenomenologically viable \cite{Qa}-\cite{donagi}. Specifically, for Calabi-Yau manifolds with $\mathbb{Z}_{3} \times \mathbb{Z}_{3}$ homotopy and a vector bundle with $SU(4)$ structure group, it has been shown \cite{exact, lukas} that the low-energy theory can have exactly the matter and Higgs spectrum of the MSSM, including three right-handed neutrino supermultiplets, one per family, along with a relatively small number of uncharged geometric and vector bundle moduli. The gauge group of this effective theory is $SU(3)_{C} \times SU(2)_{L} \times U(1)_{Y} \times U(1)_{B-L}$, that is, the standard model gauge group times an additional $B$-$L$ Abelian gauge symmetry. We will refer to this as the $B$-$L$ MSSM theory.

The existence of the extra $U(1)_{B-L}$ gauge factor, far from being being extraneous or problematical, is precisely what is required to make a heterotic vacuum with $SU(4)$ structure group phenomenologically viable. The reason is the following. As is well-known, four-dimensional $N=1$ supersymmetric theories generically contain two lepton number violating and one baryon number violating dimension four operators in the superpotential. The former, if too large, can create serious cosmological difficulties, such as in  baryogenesis and primordial nucleosynthesis \cite{Sakharov}-\cite{Dreiner}, as well as coming into conflict with direct measurements of lepton violating decays \cite{Barbier}. The latter can produce extremely rapid proton decay, far in excess of the observed bound on its lifetime \cite{Barbier, Ub}. To avoid these problems, it is traditional in low-energy $N=1$ supersymmetric theories to impose a discrete ``matter parity'', the supersymetric version of ``R-parity'' \cite{Ub}-\cite{Uc}. This $\mathbb{Z}_{2}$ finite symmetry disallows these dimension four operators from appearing in the superpotential, thus solving all the above problems. The $B$-$L$ MSSM theory naturally contains matter parity as a $\mathbb{Z}_{2}$ subgoup of $U(1)_{B-L}$. As long as this subgroup is unbroken, or weakly broken, the theory will be phenomenologically and cosmologically viable. Importantly, however, since a gauged $B$-$L$ Abelian symmetry is not observed at low energy, it is essential that $U(1)_{B-L}$ be spontaneously broken above the electroweak scale.

In pre-superstring supersymmetric grand unified theories (GUTs), this phenomenon was discussed within the context of the unification group $SO(10)$ \cite{Va}-\cite{Vf}. It was noted that certain $SU(3)_{C} \times SU(2)_{L} \times U(1)_{Y}$ singlets carrying non-zero $3(B-L)$ even charge would, if they obtained a non-vanishing vacuum expectation value (VEV), spontaneously break $U(1)_{B-L}$ down to a residual $\mathbb{Z}_{2}$ symmetry, that is, matter parity. The superpotential is then constructed so that their VEVs are very large, usually near the unification scale. Thus, these multiplets are heavy and disappear from low-energy physics, leaving behind unbroken matter parity. However, as first pointed out in \cite{BLLetter}, this compelling mechanism {\it cannot occur in smooth heterotic compactifications}. The reason is straightforward. The requisite $3(B-L)$ even charged singlets  can only arise in the decomposition of 
$SO(10)$ representations with dimension 126 and higher. Although unconstrained GUTs can always add such multiplets, they can {\it never appear as zero-modes of the Dirac operator in smooth $SU(4)$ heterotic compactifications}. All such zero-modes must arise from the decomposition of the $\bf 2 48$ representation of $E_{8}$ under $SU(4)$, whose largest dimensional $SO(10)$ representation is the $\bf 45$. It follows that all $SU(3)_{C} \times SU(2)_{L} \times U(1)_{Y} \times U(1)_{B-L}$ multiplets that appear after Wilson line breaking will have charges $3(B-L)=\pm1, \pm3, 0$. Hence, the above mechanism to obtain matter parity never occurs.

What, then, can one do in smooth heterotic $SU(4)$ compactifications? 
Note that the only 
$SU(3)_{C} \times SU(2)_{L} \times U(1)_{Y}$ singlets carrying non-zero $U(1)_{B-L}$ charge 
are the right-handed sneutrinos, each of which has $3(B-L)=3$. Were at least one sneutrino to develop a non-zero VEV, it would spontaneously break the $U(1)_{B-L}$ symmetry. However, since its $B$-$L$ charge is odd, the $\mathbb{Z}_{2}$ matter parity will also be broken. It follows that a viable theory can only emerge if at least one right-handed sneutrino develops a non-vanishing VEV that is 1) larger than the electroweak scale but 2) small enough that the broken matter parity can sufficiently prevent large lepton and baryon number violation. That is, one should have a reasonable $B$-$L$/electroweak hierarchy. It was shown in \cite{BLLetter, BLPaper} that this radiative hierarchy can, indeed, occur. Specifically, using a quasi-analytic solution to the renormalization group equations (RGEs) it was found that a non-zero VEV can develop in any right-handed sneutrino followed, at a sufficiently lower scale,  by the standard radiative breaking of electroweak symmetry via Higgs VEVs. That is, the $B$-$L$ MSSM theory can be a viable theory of nature with interesting cosmological consequences \cite{tamaz}. However, the analysis in \cite{BLLetter, BLPaper} was restricted in two important ways. First, in order to attain a quasi-analytic solution it was necessary to choose realistic, but constrained, initial data for the soft supersymmetry breaking parameters. Second, the quasi-analytic solution required that relatively ``small''  terms were dropped in the calculation. 
Hence, for example, the superparticle mass spectra presented in \cite{BLLetter, BLPaper} were only approximate, as were the calculations of the $B$-$L$/electroweak hierarchy.

In this paper, we present a completely numerical calculation of the RGEs in the $B$-$L$ MSSM theory.
Hypercharge and $B$-$L$ kinetic mixing does occur in this theory and, generically, should be included in the analysis. However, we show in Appendix C that for the regions of parameter space of emphasis here, this mixing is small and, to leading order, can be ignored. This allows a great simplification in a theory which already has a large number of input parameters. With this caveat, all RGEs are solved without approximation.
Furthermore, one can now explore the complete initial soft supersymmetry breaking parameter space and find the sub-regions that  give realistic particle physics and cosmology.
Specifically, we do the following. In Section 2, the exact MSSM spectrum and the associated 
$SU(3)_{C} \times SU(2)_{L} \times U(1)_{Y} \times U(1)_{B-L}$ quantum numbers are presented, along with the superpotential, $D$-terms and soft supersymmetry breaking quadratic and cubic terms. Section 3 is devoted to extending the ideas in \cite{BLLetter, BLPaper} relevant to ensuring the spontaneous breaking of $U(1)_{B-L}$ through right-handed sneutrino VEVs, as well as specifying some physically less interesting initial parameters. The number of initial parameters is reduced to four, related to the squarks, right-handed sneutrinos, the $\mu$ parameter and $\tan\beta$. Three phenomenological constraints are then presented. The first is two inequalities that ensure radiative breaking of electroweak symmetry through up- and down-Higgs VEVs. Second, we give the constraints required to make the $B$-$L$/electroweak vacua local minima of the potential energy. Third, the lower bounds on the masses of all superpartners, as well as the Higgs fields, are presented. The calculations in this paper will satisfy all three constraints. 
Our main numerical results are presented in Section 4. This section is broken into three subsections and a brief summary. The three subsections reflect the fact that all $B$-$L$ MSSM vacua can be catagorized by the sign of the left-handed squark and right-handed slepton squared masses; that is, 1) all $m^{2}>0$, 2) all masses positive except $m_{Q_{i}}^{2}<0$ and 3) all masses positive except $m_{e_{i}}^{2}<0$.
In each case, we find the complete region of parameter space for which one obtains a realistic theory, compute the $B$-$L$/electroweak hierarchy over each acceptable region and, at some representative points, explicitly compute the sparticle and Higgs mass spectrum. We verify that phenomenological mass constraints are indeed satisfied at these points. A discussion of the formalism used in this paper to calculate fermion and scalar masses is presented in Appendix A. The present experimental constraints on the Higgs masses are reviewed in Appendix B.

Our results are {\it predictive}, since many low-energy phenomena arise from the radiative breaking of a right-handed sneutrino. Perhaps the most striking aspect of this is that the non-vanishing sneutrino VEV
``grows back'' the previously disallowed lepton number violating dimension four terms in the superpotential, each with an explicitly calculable coefficient. Following \cite{Buchmuller, Takayama}, we confront our results with various cosmological constraints, such as baryon asymmetry and primordial nucleosynthesis. We find that they are all easily satisfied in the $B$-$L$ MSSM theory. Furthermore, we show that our theory is consistent with gravitino dark matter and rapidly decaying standard model sparticles. Another important aspect of breaking $B$-$L$ symmetry with a right-handed sneutrino is that the previously disallowed baryon violating dimension four operator does {\it not} grow back from the dimension four superpotential. It can only reappear from higher dimensional operators with calculable, and naturally suppressed, coefficients. Putting in our calculated results, we find that proton decay through dimension four operators can be sufficiently suppressed to satisfy all bounds on the proton lifetime. These lepton and baryon number violating results are presented in Section 5.

Finally, we want to point out that the $B$-$L$ MSSM theory, with spontaneous breaking of $U(1)_{B-L}$ through right-handed sneutrinos, was presented from a ``bottom up'' point of view in \cite{Spina}-\cite{Spinc}. These authors discussed various phenomenological predictions and applied similar ideas to other low-energy theories \cite{Spind}.

\section{The $N=1$ Supersymmetric Theory}

We will consider an $N=1$ supersymmetric theory with gauge group
\begin{equation}
G=SU(3)_{C} \times SU(2)_{L} \times U(1)_{Y} \times U(1)_{B-L}
\label{1}
\end{equation}
and the associated vector superfields. The gauge parameters are denoted by $g_{3}$, $g_{2}$, $g_{Y}$ and $g_{B-L}$ respectively. The matter spectrum consists of three families of quark and lepton chiral superfields, each family with a {\it right-handed neutrino}. They transform under the gauge group in the standard manner as
\begin{equation}
Q_{i}=({\bf 3},{\bf 2},1/3,1/3), \quad u_{i}=({\bf \bar{3}}, {\bf 1}, -4/3, -1/3), \quad d_{i}=({\bf \bar{3}}, {\bf 1}, 2/3, -1/3)
\label{2}
\end{equation}
for the left and right-handed quarks and
\begin{equation}
L_{i}=({\bf 1},{\bf 2},-1,-1), \quad \nu_{i}=({\bf 1}, {\bf 1}, 0, 1), \quad e_{i}=({\bf 1}, {\bf 1}, 2, 1)
\label{3}
\end{equation}
for the left  and right-handed leptons, where $i=1,2,3$. In addition, the spectrum has one pair of Higgs-Higgs conjugate chiral superfields transforming as
\begin{equation}
H=({\bf 1},{\bf 2},1,0), \qquad \bar{H}=({\bf 1},{\bf 2}, -1,0).
\label{4}
\end{equation}
When necessary, the left-handed $SU(2)_{L}$ doublets will be written as 
\begin{equation}
Q_{i}=(U_{i}, D_{i}), \quad L_{i}=(N_{i}, E_{i}), \quad H=(H^{+},H^{0}), \quad \bar{H}=({\bar{H}}^{0}, {\bar{H}}^{-}).
\label{5}
\end{equation}
There are {\it no other fields in the spectrum}.

The supersymmetric potential energy is given by the usual sum over the modulus squared of the $F$
and $D$-terms. In principle, the $F$-terms are determined from the most general superpotential  invariant under the gauge group,
\begin{equation}
W=\mu H\bar{H} +{\sum_{i,j=1}^{3}}\left(\lambda_{u, ij} Q_{i}Hu_{j}+\lambda_{d, ij} Q_{i}\bar{H}d_{j}+\lambda_{\nu, ij} L_{i}H\nu_{j}+\lambda_{e, ij} L_{i}\bar{H}e_{j}\right)
\label{6}
\end{equation}
Note that the quadradic mixing term of the form $L_{i}H$,
as well as the dangerous lepton and baryon number violating interactions
\begin{equation}
L_{i}L_{j}e_{k}, \quad L_{i}Q_{j}d_{k}, \quad u_{i}d_{j}d_{k}
\label{7}
\end{equation}
which generically would lead, for example, to rapid nucleon decay, are disallowed by the $U(1)_{B-L}$ gauge symmetry. To simplify the upcoming calculations, we will assume that we are in a mass-diagonal basis where
\begin{equation}
\lambda_{u, ij}=\lambda_{d, ij}=\lambda_{\nu, ij}=\lambda_{e, ij}=0, \quad i \neq j.
\label{8}
\end{equation}
Note that once these off-diagonal couplings vanish just below the compactification scale, they will do so at all lower energy-momenta. We will denote the diagonal Yukawa couplings by $\lambda_{ii}=\lambda_{i}$, $i=1,2,3$. Next, observe that a constant, field-independent $\mu$ parameter cannot arise in a supersymmetric string vacuum  since the Higgs fields are zero modes. However, the $H{\bar{H}}$ bilinear can have higher-dimensional couplings to moduli through both holomorphic and non-holomorphic interactions in the superpotential and Kahler potential respectively. When moduli acquire VEVs due to non-perturbative effects, these can induce non-vanishing supersymmetric contributions to $\mu$. A non-zero $\mu$ can also be generated by gaugino condensation in the hidden sector. Why this induced $\mu$-term should be small enough to be consistent with electroweak symmetry breaking is a difficult, model dependent problem. In this paper, we will not discuss this ``$\mu$-problem''.  Instead, we will consider the $\mu$ parameter as an input to our analysis and consider a range of possible values.

The $SU(3)_{C}$ and $SU(2)_{L}$ $D$-terms are of the standard form. We present the $U(1)_{Y}$ 
and $U(1)_{B-L}$ $D$-terms,
\begin{equation}
D_{Y}= \xi_{Y} +g_{Y}{\phi}_{A}^{\dagger}\left({\bf{Y}\rm}/2\right)_{AB}{\phi}_{B}
\label{9}
\end{equation}
and 
\begin{equation}
D_{B-L}= \xi_{B-L} +g_{B-L}{\phi}_{A}^{\dagger}\left({\bf{Y_{B-L}}\rm}\right)_{AB}{\phi}_{B}
\label{10}
\end{equation}
where the index $A$ runs over all scalar fields ${\phi}_{A}$, to set the notation for the hypercharge and $B$-$L$ charge generators and to remind the reader that each of these $D$-terms potentially has a Fayet-Iliopoulos (FI) additive constant.  However, as with the $\mu$ parameter, constant field-independent FI terms cannot occur in string vacua since the low energy fields are zero modes. Field-dependent FI terms can occur in some contexts, see for example \cite{Lara}. However, since both the hypercharge and $B$-$L$ gauge symmetries are anomaly free, such field-dependent FI  terms are not generated in the supersymmetric effective theory. We include them in (\ref{9}),(\ref{10}) since they can, in principle, arise at a lower scale from radiative corrections once supersymmetry is softly broken \cite{Yd}. Be that as it may, if calculations are done {\it in the $D$-eliminated formalism, which we use in this paper, these FI parameters can be consistently absorbed into the definition of the soft scalar masses} and their beta functions. Hence, we will no longer consider them.

In addition to the supersymmetric potential, the Lagrangian density also contains explicit ``soft'' supersymmetry violating terms. These arise from the spontaneous breaking of supersymmetry in  
a hidden sector that has been integrated out of the theory. This breaking can occur in either $F$-terms, $D$-terms or both in the hidden sector. In this paper, for simplicity, we will restrict our discussion to soft supersymmetry breaking terms arising exclusively from  $F$-terms. The form of these terms is well-known and, in the present context, given by \cite{W}-\cite{Zc}
\begin{equation}
V_{\rm soft}=V_{2s}+V_{3s}+V_{2f},
\label{11}
\end{equation}
where $V_{2s}$ are scalar mass terms
\begin{eqnarray}
V_{2s} & =& { \sum_{i=1}^{3}} (m^{2}_{Q_{i}}|{Q}_{i}|^{2}+m^{2}_{u_{i}}|{u}_{i}|^{2}+
             m^{2}_{d_{i}}|{d}_{i}|^{2}  +m^{2}_{L_{i}}|{L}_{i}|^{2}+m^{2}_{\nu_{i}}|{\nu}_{i}|^{2}     \nonumber \\ 
             &    &+m^{2}_{e_{i}}|{e}_{i}|^{2})+m_{H}^{2}|H|^{2} +m_{\bar{H}}^{2}|\bar{H}|^{2}  -(BH\bar{H}+hc), \label{12}
\end{eqnarray}
 $V_{3s}$ are scalar cubic couplings
\begin{equation}
V_{3s}=\sum_{i=1}^{3} (A_{u_{i}} {Q}_{i}H{u}_{i} +A_{d_{i}} {Q}_{i}{\bar{H}}{d}_{i} +A_{\nu_{i}} {L}_{i}H{\nu}_{i}+A_{e_{i}} {L}_{i}{\bar{H}}{e}_{i} +{\rm hc})
\label{13}
\end{equation}
and $V_{2f}$ contains the gaugino mass terms
\begin{equation}
V_{2f}= \frac{1}{2} M_{3} \lambda_{3} \lambda_{3}+ \frac{1}{2} M_{2} \lambda_{2} \lambda_{2}+ \frac{1}{2} M_{Y} \lambda_{Y} \lambda_{Y}+ \frac{1}{2} M_{B-L} \lambda_{B-L} \lambda_{B-L}+
{\rm hc}.
\label{14}
\end{equation}
As above, to simplify the calculation we assume the parameters in (\ref{12}) and (\ref{13}) are flavor-diagonal. This is consistent since once the off-diagonal parameters vanish just below the compactification scale, they will do so at all lower energy-momenta. Finally, note that lepton and baryon violating scalar cubic terms of the form \eqref{7} are disallowed in $V_{3s}$ by the $U(1)_{B-L}$ gauge symmetry.

\section{Initial Parameter Space}
 
The four-dimensional effective theory described in the previous section arises at an initial energy-momentum just below the compactification scale given by the inverse Calabi-Yau radius. In order to carry out a detailed renormalization group analysis, we must specify this initial energy-momentum precisely.
We will do this as follows.

\subsection{Gauge Coupling Parameters}

It is well known that precision measurements \cite{Zd}-\cite{Ze}  carried out at the electroweak scale indicate that the $SU(3)_{C} \times SU(2)_{L} \times U(1)_{Y}$ gauge couplings, $g_{3}$,$g_{2}$ and $g_{1}=\sqrt{\frac{5}{3}}g_{Y}$ respectively, unify to 
\begin{equation}
g(0) \simeq .726  
\label{15}
\end{equation}
at scale
\begin{equation}
M_{u} \simeq 3 \times 10^{16} GeV  \ .
\label{16}
\end{equation}
For simplicty, so that we can ignore a discussion of threshold effects, we will assume that the initial energy momentum for our effective theory is precisely the unification scale $M_{u}$. In addition, since the $SU(4)$ vector bundle breaks $E_{8}$ to $SO(10)$, we will take the $U(1)_{B-L}$ gauge coupling $g_{4}=\sqrt{\frac{4}{3}}g_{B-L}$ to unify with the three other couplings at $M_{u}$.\\

\noindent Having fixed the initial energy-momentum as $M_{u}$, one must now specify the initial values of all parameters in the effective theory at this scale. In principle, string theory would predict these parameters as functions of the moduli VEVs. In this paper, however, we will be content with simply choosing the initial parameters subject to the dictates of simplicity, the ``universality'' of some parameters observed in minimal supergravity and simple string compactifiacations \cite{Zc, Yg} and the necessity to break $U(1)_{B-L}$ through a VEV of at least one right-handed sneutrino. Having chosen all the initial parameters, their values at any lower scale, specified by
\begin{equation}
t=ln(\frac{\mu}{M_{u}}) \ ,
\label{17}
\end{equation}
are determined by the associated renormalization group equations (RGEs). These are discussed in detail in several reviews, see, for example \cite{Yd}-\cite{Yf}, and were generalized to include the $U(1)_{B-L}$ symmetry in our previous papers \cite{BLLetter, BLPaper}. In this paper, all calculations will be carried out at the one-loop level.

The initial unified gauge coupling is given in \eqref{15}. We now turn to specifying the initial values for all other parameters in our effective low-energy theory. We begin with the dimensionful parameters.

\subsection{Gaugino Mass Parameters }

Consider the soft supersymmetry breaking  gaugino mass parameters that appear in $V_{2f}$ in \eqref{14}. Following standard notation, we henceforth denote $M_{Y}$ = $M_{1}$ and $M_{B-L}$ = $M_{4}$.  We now make the assumption that at the compactification scale the gaugino masses unify, that is,
\begin{equation}
|M_{1}(0)| = |M_{2}(0)|  =  |M_{3}(0)| =  |M_{4}(0)| \ .
\label{18}
\end{equation}
Such universal gaugino masses naturally occur in minimal supergravity \cite{Xa}-\cite{Xe} and simple string theories \cite{Xb, Xd}. Here, we choose \eqref{18} for reasons of simplicity.

\subsection{Higgs, Squark and Slepton Masses}

The RGEs for the soft supersymmetry breaking Higgs, squark and slepton masses all contain a term proportional to $g_{1}^{2}{\cal{S}}$ where
\begin{eqnarray}
{\cal{S}} & =& Tr({\frac{\bf Y}{2} {\bf m}^{2}})   \label{19} \\
&=& m_{H}^{2}-m_{\bar{H}}^{2}+\sum_{i=1}^{3}(m_{Q_{i}}^{2}-2m_{u_{i}}^{2}+m_{d_{i}}^{2}-m_{L_{i}}^{2}+m_{e_{i}}^{2}) \ .\nonumber 
\end{eqnarray}
It greatly simplifies the boundary condtions of these RGEs to choose the initial soft breaking masses so that ${\cal{S}}(0)=0$. A natural way to achieve this is to impose a separate unification of the Higgs masses, squark masses and the left doublet/down right singlet slepton masses. That is, we henceforth choose
\begin{equation}
 m_{H}(0)^{2}=m_{\bar{H}}(0)^{2}, \quad m_{Q_{i}}(0)^{2}=m_{u_{j}}(0)^{2}=m_{d_{k}}(0)^{2}
 \label{20}
\end{equation}
and
\begin{equation}
 m_{L_{i}}(0)^{2}=m_{e_{j}}(0)^{2} 
 \label{21}
\end{equation} 
for all $i,j,k=1,2,3$. In addition to the hypercharge induced $g_{1}^{2}{\cal{S}}$ term, the gauged $U(1)_{B-L}$ symmetry of our effective theory introduces a new term into the RGEs for the squarks and slepton soft supersymmetry breaking masses. This term is of the form $g_{4}^{2}{\cal{S}}'$ where
\begin{equation}
{\cal{S}'} = Tr( { {\bf Y}_{B-L} {\bf m}^{2}}) = {\cal{S}}_{0}'+  {\cal{S}}_{1}'  
 \label{22} 
\end{equation}
and
\begin{equation}
{\cal{S}}_{0}'=\sum_{i=1}^{3}(2m_{Q_{i}}^{2}-m_{u_{i}}^{2}-m_{d_{i}}^{2}-m_{L_{i}}^{2}+m_{e_{i}}^{2}) , \quad {\cal{S}}_{1}' = \sum_{i=1}^{3}(-m_{L_{i}}^{2}+m_{\nu_{i}}) .
\label{23}
\end{equation}
It follows from \eqref{20} and \eqref{21} that ${\cal{S}}_{0}'(0)=0$. Note, however, that unlike ${\cal{S}}$ and ${\cal{S}}'_{0}$, the ${\cal{S}}_{1}' $ term depends on the soft supersymmetry breaking right-handed sneutrino masses. We choose the initial values of these parameters {\it not} to be degenerate with the other slepton masses, that is,
\begin{equation}
 m_{L_{i}}(0)^{2}=m_{e_{j}}(0)^{2} \neq m_{\nu_{k}}(0)^{2}
 \label{24}
\end{equation} 
for all $i,j,k=1,2,3$. It follows that
\begin{equation}
{\cal{S}}_{1}'(0)=\sum_{i=1}^{3}(-m_{L_{i}}(0)^{2}+m_{\nu_{i}}(0)^{2}) \neq 0 \ .
 \label{25}
\end{equation}
{\it This asymmetry is an important ingredient in generating radiative breaking of the $U(1)_{B-L}$ symmetry}.  We point out that soft scalar masses need not be ``universal'' in string theories, since they are not generically ``minimal''.

\subsection{The A and B Parameters}

Now consider the soft supersymmetry breaking up/down $A_{i}$ and B parameters in equations (\ref{13}) and (\ref{12}) respectively.  As already stated, we take the $A_{i}$  coefficients to be flavor diagonal.  In addition, it is conventional \cite{Zc} to let
\begin{equation}
A_{u_{i}}=\lambda_{u_{i}} {\tilde{A}}_{u_{i}}, \ A_{d_{i}}=\lambda_{d_{i}} {\tilde{A}}_{d_{i}},  \ 
A_{\nu_{i}}=\lambda_{\nu_{i}} {\tilde{A}}_{\nu_{i}}, \  A_{e_{i}}=\lambda_{e_{i}} {\tilde{A}}_{e_{i}} 
\label{26}
\end{equation}
for $i=1,2,3$, where $\lambda_{i}$ are the Yukawa couplings and the dimensionful $\tilde{A}_{i}$ parameters are chosen to be of order the supersymmetry breaking scale.  
This is {\it not} a requirement in the ``non-minimal'' string vacua that we are discussing. Be that as it may,
for simplicity of presentation we will assume (\ref{26}) for the remainder of this paper. The input Yukawa parameters will be discussed below. In this paper, we will, for simplicity, assume the ${\tilde{A_{i}}}$ parameters unify at the scale $M_{u}$. That is,
\begin{equation}
 {\tilde{A}}_{u_{i}}(0)={\tilde{A}}_{d_{j}}(0)= {\tilde{A}}_{\nu_{k}}(0)={\tilde{A}}_{e_{l}}(0) 
\label{27}
\end{equation}
for all $i,j,k,l=1,2,3$.

The initial value of the soft breaking B parameter, $B(0)$,  is taken to be arbitrary. However, in our analysis B will be treated differently than the other dimensionful parameters.  As will be shown below,
{\it rather than choosing the value of the B parameter, we will instead input tan$\beta$ and the supersymmetry breaking scale}.  This will dynamically fix the value of B for any given set of initial conditions.  

\subsection{The $\mu$ Parameter}

The supersymetric $\mu$ parameter has a fundamentally different origin than the soft supersymmetry breaking dimensionful couplings discussed above. In this paper, we will simply allow its initial value $\mu(0)$ to be arbitrary.  As in conventional radiative breaking scenarios, to be compatible with electroweak symmetry breaking we expect it to be of ${\cal{O}}(100)GeV$. However, we make no attempt to solve this ``$\mu$-problem''.  Having discussed the initial values for the dimensionful parameters, we now consider the dimensionless parameters in our effective theory.

\subsection{Tan$\beta$ and the Yukawa Couplings}

As with any MSSM-like model, our low energy theory requires two Higgs chiral supermultiplets, $H$ and $\bar{H}$, whose VEVs $\langle H \rangle$ and $\langle {\bar{H}} \rangle$ break electroweak symmetry and give mass to the $W^{\pm}$ and $Z$ vector bosons. The experimentally measured vector boson masses put a constraint on these VEVs. In terms of the $Z$ mass, this is 
\begin{equation}
 \langle H \rangle^{2} + \langle \bar{H} \rangle^{2} = \frac{2 M_{Z}^{2}}{g_{Y}^{2} + g_{2}^{2}} \simeq (\frac{246}{\sqrt{2}} GeV)^{2} \ .
 \label{28}
\end{equation}
Hence, giving one Higgs VEV completely determines the other. It is conventional to re-express the remaining Higgs VEV in terms of the ratio
\begin{equation}
\tan\beta=\frac{\langle H \rangle}{\langle {\bar{H}} \rangle} \ .
\label{29}
\end{equation}
If the value of $\tan\beta$ is given, one can easily find both Higgs VEVs using \eqref{28} and \eqref{29}.
The result is
\begin{equation}
\langle H \rangle=(\frac{246}{\sqrt{2}} GeV)\frac{\tan\beta}{\sqrt{1+\tan^{2}\beta}} \ , \quad \langle {\bar{H}} \rangle= (\frac{246}{\sqrt{2}} GeV)\frac{1}{\sqrt{1+\tan^{2}\beta}}  \ .
\label{30}
\end{equation}
In this paper, we will take $\tan\beta$ as an input parameter

In addition to the vector bosons, the Higgs VEVs $\langle H \rangle$ and $\langle {\bar{H}} \rangle$ give mass to the up and down quarks/leptons respectively. As with the $SU(3)_{C} \times SU(2)_{L} \times U(1)_{Y}$ gauge couplings, the Yukawa couplings are highly constrained by experiment. Given a value of $\tan\beta$ and, hence, $\langle H \rangle$ and $\langle {\bar{H}} \rangle$, the known masses of the quarks/leptons completely determine the Yukawa couplings at the electroweak scale. However, unlike the gauge couplings, the Yukawa coupling do not unify at $M_{u}$. Rather, when run up to the unification scale using their RGEs, the initial values of the Yukawa couplings are a set of $\tan\beta$ dependent numbers with no particular relationship. Therefore, in this paper, rather than specifying the initial Yukawa couplings at scale $M_{u},$ we will instead input a value of $\tan\beta$ and use the associated Higgs VEVs and the measured quark/lepton masses to calculate all Yukawa parameters at the electroweak scale. These will then be run back to the unification scale and stored in our program. When required, the initial Yukawa parameters can then be input into any other RGE and scaled down along with the other relevant  parameters.

It is important to note from \eqref{30} that as $\tan\beta$ is decreased, the up Higgs VEV $\langle H \rangle$ must get smaller. This then necessitates taking larger values for the up Yukawa couplings to be consistent with the measured masses. For 
 $\langle H \rangle$ sufficiently small, the top quark Yukawa coupling will become much larger than unity and the theory becomes non-perturbative. This puts a bound on how small 
 $\langle H \rangle$ can be and, hence, a lower bound on $\tan\beta$. Similarly, increasing $\tan\beta$ requires the down Higgs VEV $\langle {\bar{H}} \rangle$ to decrease. For $\langle {\bar{H}} \rangle$ 
 sufficiently small, the bottom quark Yukawa coupling will become much larger than unity and the theory non-perturbative. This puts a bound on how small $\langle {\bar{H}} \rangle$ can be and, hence, an upper bound on $\tan\beta$. These bounds on $\tan\beta$ are typically estimated \cite{Zb, Sola} to be
 \begin{equation}
   4  \lesssim  \tan\beta  \lesssim 50  \ .
\label{31}   
\end{equation} 
When inputting $\tan\beta$ in this paper, we will always restrict it to be within these bounds.

\subsection{Parameterizing the Initial Conditions}
\label{parameterization}

Recall that, with the exception of the $\mu$ parameter, all of the dimensional coefficients discussed above occur in soft supersymmetry breaking interactions. If we denote by ${\cal{M}}$ a mass characterizing the scale of supersymmetry breaking, then each of the above coefficients can be written in the form
\begin{equation}
c_{i}(t) {\cal M } \ ,  
 \label{32}
\end{equation} 
where $c_{i}(t)$ is dimensionless. This parameterization emphasizes that the soft dimensionful  coefficients share a common supersymmetry breaking scale. The initial coefficients, $c_{i}(0)$, are arbitrary. However, naturalness would dictate that they not to be too much larger, or smaller, than unity.
The exception to this is the parameter $\mu$. This arises in the supersymmetric quadratic Higgs term
and is, a priori, unrelated to the scale ${\cal{M}}$. However, it can always be written in the form \eqref{32}. In this case, however, one does not expect the associated coefficient to be of order unity. Be that as it may, the ``$\mu$-problem'' specifies that appropriate radiative electroweak breaking will require $\mu$ to be close to the scale ${\cal{M}}$.

Specifically, this parameterization of the dimensionful parameters allows us to write
the initial value for the gaugino masses as
\begin{equation}
|M_{1}(0)| = |M_{2}(0)|  =  |M_{3}(0)| =  |M_{4}(0)| =c_{M}(0){\cal{M}} \ ,
\label{33}
\end{equation}
as well as 
\begin{equation}
 m_{H}(0)=m_{\bar{H}}(0)=c_{H}(0){\cal{M}} 
 \label{34}
\end{equation}
for the initial Higgs parameters. Similarly, the initial squark and doublet/ down singlet slepton  masses are
\begin{equation}
m_{Q_{i}}(0)=m_{u_{j}}(0)=m_{d_{k}}(0)=c_{q}(0){\cal{M}}
 \label{35}
\end{equation} 
and
\begin{equation}
m_{L_{i}}(0)=m_{e_{j}}(0) =c_{e}(0){\cal{M}} 
 \label{36}
\end{equation} 
respectively. However, for the reasons discussed below, we will allow the initial right-handed sneutrino masses to have the texture
\begin{equation}
 m_{\nu_{1}}(0) = m_{\nu_{2}}(0) = c_{\nu_{1,2}}(0) {\cal M}  \ ,
 \quad   m_{\nu_{3}}(0) = c_{\nu_{3}}(0) {\cal M} \ .
 \label{37}
\end{equation}
Finally, we write
\begin{equation}
 {\tilde{A}}_{u_{i}}(0)={\tilde{A}}_{d_{j}}(0)= {\tilde{A}}_{\nu_{k}}(0)={\tilde{A}}_{e_{l}}(0) =c_{\tilde{A}}(0){\cal{M}}
\label{38}
\end{equation}
and 
\begin{equation}
\mu (0) = c_{\mu}(0) {\cal M} \ ,   \quad  \quad  B(0) = c_{B}^{2}(0) {\cal M}^{2}  
\label{39}
\end{equation}
for the initial dimension-one $\tilde{A}$, $\mu$ parameters and the dimension-two $B$ parameter respectively. That is, there is a total of nine 
dimensionless $c_{i}(0)$ parameters arising from the dimensionful parameters in our effective theory. However, this number can be reduced as follows.

First, note that all mass parameters scale with the same factor ${\cal{M}}$. Hence, one can always redefine ${\cal{M}}$ so as to absorb one of these coefficients. Without loss of generality, we can choose this to be the Higgs parameter. That is, set
\begin{equation}
c_{H}(0)=1 \ .
\label{40}
\end{equation}
Second, in minimal supergravity and simple superstring vacua, the unified initial $\tilde{A}$ and gaugino mass parameters are numbers of order unity times the supersymmetry breaking scale 
${\cal{M}}$. We will assume this in our calculation as well. For simplicity, choose
\begin{equation}
c_{\tilde{A}}(0)=1 \ .
\label{41}
\end{equation}
The initial value for $c_{M}$ is more subtle to determine. We have done an extensive numerical analysis of phenomenologically acceptable initial conditions allowing $c_{M}(0)$ to vary freely. The result is a bound given by $0.1 < c_{M}(0) < 1.2$. In this paper, for simplicity of presentation, we fix this initial parameter to a value in the middle of this range given by
\begin{equation}
c_{M}(0)=0.6 \ .
\label{42}
\end{equation}
Finally, we will also specify the coefficients $c_{e}(0)$ and $ c_{\nu_{1,2}}(0) $ as follows.

In a previous paper \cite{BLPaper}, we presented a quasi-analytic solution to the RGEs in the $B$-$L$ MSSM theory subject to certain initial conditions on the parameters. To obtain an analytic solution, the initial parameters chosen were considerably more constrained than they are in this paper. Be that as it may, the generalized parameter space discussed here contains these initial conditions as a small subset. 
Specifically, we showed that at the $B$-$L$ scale $M_{B-L}\simeq 10^{4} GeV$ the right-handed down slepton and right-handed sneutrino soft mass parameters are given by
\begin{equation}
m_{e_{i}}(t_{B-L})^{2} =  m_{e_{i}}(0)^{2}-(3.35 \times 10^{-2}){\cal{S}}_{1}'(0) \ ,
\label{43} 
\end{equation}
\begin{equation}
m_{\nu_{i}}(t_{B-L})^{2} =  m_{\nu_{i}}(0)^{2}-(3.35 \times 10^{-2}){\cal{S}}_{1}'(0) 
\label{44}
\end{equation}
for $i=1,2,3$ where, using \eqref{25}, \eqref{36} and \eqref{37}, one can write
\begin{equation}
{\cal{S}}_{1}'(0)=(1+2C^{2}-3A^{2})m_{\nu_{3}}(0)^{2}
\label{45}
\end{equation}
with
\begin{equation}
C=\frac{c_{\nu_{1,2}}(0)}{c_{\nu_{3}}(0)} \ , \quad A=\frac{c_{e}(0)}{c_{\nu_{3}}(0)} \ .
\label{46}
\end{equation}
For specificity, let us choose
\begin{equation}
(3.35 \times 10^{-2})(1+2C^{2}-3A^{2})=5 \ .
\label{47}
\end{equation}
Then one obtains the simple result that
\begin{equation}
m_{\nu_{3}\rm }(t_{B-L})^{2}=-4 m_{\nu_{3}}(0)^{2} \ ,
\label{48}
\end{equation}
leading to a non-zero VEV in the $\nu_{3}$ direction. In this way, we guarantee radiative $U(1)_{B-L}$ breaking in the theory. Similarly, using \eqref{47} we find that
\begin{equation}
m_{\nu_{1,2}\rm }(t_{B-L})^{2}=(C^{2}-5)m_{\nu_{3}}(0)^{2} 
\label{49}
\end{equation}
and
\begin{equation}
m_{e_{i}\rm }(t_{B-L})^{2}=(A^{2}-5)m_{\nu_{3}}(0)^{2} 
\label{50}
\end{equation}
for $i=1,2,3$. The simplest vacuum structure occurs when all $m_{e_{i}\rm}(t_{B-L})^{2}$ are positive. For this to be the case, the coefficient $A$ must satisfy $A^{2}-5 > 0$. Again, for specificity we will choose
\begin{equation}
A=\sqrt{6} \ ,
\label{51}
\end{equation}
which yields the simple result that
\begin{equation}
m_{e_{i}\rm }(t_{B-L})^{2}=m_{\nu_{3}}(0)^{2} 
\label{52}
\end{equation}
for $i=1,2,3$. Putting $A=\sqrt{6}$ into expression \eqref{47} gives 
\begin{equation}
C=9.12 \ .
\label{53}
\end{equation}
It then follows from (\ref{49}) that both $m_{\nu_{1,2}\rm }(t_{B-L})^{2}$ are positive and given by
\begin{equation}
m_{\nu_{1,2}\rm }(t_{B-L})^{2}=78.2\ m_{\nu_{3}}(0)^{2} \ .
\label{54}
\end{equation}
 We conclude that the choice of the $A$ and $C$ parameters given in \eqref{51} and \eqref{53} respectively leads to a vacuum that has positive soft squared masses and, hence, vanishing VEVs for all sleptons with the exception of the third family right-handed sneutrino. This acquires a non-zero VEV which radiatively breaks $U(1)_{B-L}$ symmetry. It is clear that these choices for $A$ and $C$ are far from unique, and that a wide range of values would still lead to a vacuum with appropriate $U(1)_{B-L}$ symmetry breaking. Be that as it may, we find it convenient to continue to use \eqref{51} and \eqref{53} in the present paper. It then follows from \eqref{46} that we will choose
 \begin{equation}
 c_{e}(0) = \sqrt{6} \ c_{\nu_{3}}(0) \ ,  \quad  c_{\nu_{1,2}}(0) = 9.12 \ c_{\nu_{3}}(0) \ .
  \label{55}
 \end{equation}

The constraints given in \eqref{40}, \eqref{41}, \eqref{42} and \eqref{55} reduce the number of free parameters down to six-- four $c_{i}$ parameters as well as ${\cal{M}}$ and $\tan\beta$. There are, however, important  phenomenological constraints on these parameters, to which we now turn.

\subsection{Phenomenological Constraints}
\label{phenom const}

It is well-known \cite{Zc} that for an MSSM-like theory with two Higgs doublets, $H$ and $\bar{H}$, to have a stable vacuum solution that breaks electroweak symmetry, the parameters of the theory have to satisfy two constraints {\it at the electroweak scale} $M_{EW}\simeq 10^{2} GeV$. These are
\begin{equation}
B^{2}  >  (\vert \mu \vert^{2} + m_{H}^{2}) (\vert \mu \vert^{2} + m_{\bar{H}}^{2}) \ ,
\label{56}
\end{equation}
which ensures that one linear combination of $H$ and $\bar{H}$ has a negative squared mass, thus enabling a non-zero Higgs VEV to form, and
\begin{equation}
2 B  <  2 \vert \mu \vert^{2} + m_{H}^{2} + m_{\bar{H}}^{2} \ ,
\label{57}
\end{equation}
which guarantees that the quadratic part of the potential energy is positive along the $D$-flat directions and, hence, that the potential energy is bounded from below. Once these conditions are satisfied, the theory has a stable Higgs vacuum specified by the two minimization equations. Their solutions can be put in the form
\begin{equation}
  \sin(2 \beta) = \frac{2 B}{m_{H}^{2} + m_{\bar{H}}^{2}  + 2 \vert \mu \vert^{2}} 
  \label{58}
\end{equation}
and
\begin{equation}
 M_{Z}^{2} = \frac{\vert m_{\bar{H}}^{2} - m_{H}^{2} \vert }{\sqrt{1 - \sin^{2}(2\beta)}} - m_{\bar{H}}^{2} - m_{H}^{2} - 2 \vert \mu \vert^{2} 
  \label{59}
\end{equation}
with the parameters evaluated {\it at the electroweak scale}.

In many analyses of electroweak breaking, all the soft masses and the $\mu$ parameter are given as input with $\tan\beta$ and $M_{Z}$ generated as solutions of \eqref{58} and \eqref{59}.
However, as discussed above, it is convenient in this paper to take $\tan\beta$ as an input parameter. It follows that equation \eqref{58} should be viewed as yet another constraint on the soft breaking parameters. Specifically, we will use \eqref{58} to solve for $B$ as a function of $\tan\beta$, $m_{H}^{2}$, $m_{\bar{H}}^{2}$ and $\mu$ at the electroweak scale. This is possible 
since the RGEs for $m_{H}^{2}$, $m_{\bar{H}}^{2}$ and $\mu$ \cite{Yb} and, hence, the value of these parameters at the electroweak scale do not depend implicitly on $B$. 
Written in terms of the notation introduced in the previous section, it follows that
\begin{equation}
c_{B}^{2}= \frac{sin(2\beta)}{2}(c_{H}^{2}+c_{\bar{H}}^{2}+2|c_{\mu}|^{2}) \ .
\label{60}
\end{equation}
We can then scale this parameter back up to the $M_{u}$ to determine the initial value $B(0)$.

Similarly, we can input the experimental value of $M_{Z}$ into \eqref{59} and use this to put a further constraint on the initial parameters. In terms of the above parameterization, \eqref{59} can be re-written as
\begin{equation}
  M_{Z}^{2} = \Big( \frac{\vert c_{\bar{H}}^{2} - c_{H}^{2} \vert }{\sqrt{1 - \sin^{2}(2\beta)}} - c_{\bar{H}}^{2} - c_{H}^{2} - 2 \vert c_{\mu} \vert^{2} \Big){\cal{M}}^{2}   .
  \label{61}
\end{equation}
From this equation we see that, given the initial values of $c_{i}$ and $\tan\beta$, one can use the experimentally derived value for $M_{Z}$ to fix ${\cal{M}}$ and, thus, the soft breaking scale.  
Note that for fixed values of $c_{H}$, $c_{\bar{H}}$ and $\tan\beta$, mass ${\cal{M}}$ is a minimum as $c_{\mu}\rightarrow 0$ and becomes arbitrarily large as 
\begin{equation}
 \vert c_{\mu} \vert^{2} \longrightarrow \frac{1}{2} \Big( \frac{\vert c_{\bar{H}}^{2} - c_{H}^{2} \vert }{\sqrt{1 - \sin^{2}(2\beta)}} - c_{\bar{H}}^{2} - c_{H}^{2}\Big) \ .
\label{62}
\end{equation}
It follows that the value of the supersymmetry breaking parameter is not particularly restricted by constraint \eqref{61}. Be that as it may, obtaining its minimum value and, in particular, a large value requires fine-tuning $c_{\mu}$ to zero and \eqref{62} respectively. Without fine-tuning, the typical value for ${\cal{M}}$ is set by the $Z$-mass and, for the initial parameters in this paper, found to be of order a few hundred GeV up to order 10 TeV.

Applying constraints \eqref{60} and \eqref{61}  to fix the values of $c_{B}(0)$ and ${\cal M}$ respectively, we are now left with four free parameters.  They are
\begin{equation}
 c_{q}(0),~c_{\nu_{3}}(0), ~c_{\mu}(0),~\tan\beta \ .
  \label{63}
\end{equation}
In the remainder of this paper, we analyze the vacuum state and mass spectrum of the $B$-$L$ MSSM theory over this four-dimensional initial parameter space. This is accomplished by numerically solving all RGEs for a given choice of initial conditions, scaling down from $M_{u}$ to $M_{EW}$. In doing this, however, we will impose several important phenomenological constraints-- rejecting the initial parameters if the results fail to satisfy these constraints and accepting them if they are satisfied. In this way, one can map out the allowed region of the four-dimensional initial parameter space.\\

\noindent The phenomenological constraints we impose are the following.
\begin{itemize}

\item  To ensure that a stable electroweak breaking vacuum can develop at low energy-momenta,  we impose the constraint that inequalities \eqref{56} and \eqref{57} be satisfied. This should be understood as a consistency check on our assumption, implicit in using $\tan\beta$ and the experimental value of $M_{Z}$ as input parameters,  that a stable electroweak breaking vacuum described by \eqref{58} and \eqref{59} exists. 
In terms of the parameterization introduced in Subsection 3.7, these constraints are 
\begin{equation}
c_{B}^{4} > (|c_{\mu}|^{2}+c_{H}^{2})(|c_{\mu}|^{2}+c_{\bar{H}}^{2})
\label{64}
\end{equation}
and
\begin{equation}
2c_{B}^{2} < c_{H}^{2}+c_{\bar{H}}^{2}+2|c_{\mu}|^{2}
\label{65}
\end{equation}
respectively.

\item As discussed above, condition \eqref{55} ensures that a vacuum expectation value develops in the third right-handed sneutrino. To guarantee that this is a stable local minimum, we impose the constraint that the effective squared masses of all squarks and sleptons evaluated at the $B$-$L$ breaking 
VEV $\langle \nu_{3} \rangle$, for example,
\begin{equation}
\langle m_{Q_{i}}^{2} \rangle = m_{Q_{i}}^{2}+\frac{1}{4} g_{4}^{2} \langle \nu_{3} \rangle^{2} \ , \quad\langle m_{L_{i}}^{2} \rangle  =  m_{L_{i}\rm}^{2}-\frac{3}{4}g_{4}^{2}\langle \nu_{3} \rangle^{2}
\label{66}
\end{equation}
are positive {\it over the entire scaling range}. It follows that color and charge symmetry are never
spontaneously broken. Note that imposing the positivity of the effective masses does {\it not} necessarily 
restrict the soft squared masses to be positive. For example, the positivity of $\langle m_{Q_{i}}^{2} \rangle$ does not require that $m_{Q_{i}}^{2}$ be positive. On the other hand, $m_{L_{i}\rm}^{2}$ must be positive to ensure that $\langle m_{L_{i}}^{2} \rangle$ is. This allows us to classify the $B$-$L$ MSSM vacua in terms of the signs of the soft squared masses at the electroweak scale. This will be discussed in detail later in the paper.

\begin{table}
\begin{center}
  \begin{tabular}{ | c | c | c | c | c | c | }
    \hline
    Particle & Symbol  & Mass [GeV] & Particle & Symbol  & Mass [GeV] \\  \hline
    \multirow{3}{*}{Squarks} & $ \tilde{q}_{1,2}$ & 379 & \multirow{3}{*}{Higgs}  & $(h,H)^{0}$ & $114.4$ \\
    & $\tilde{b}$ &  89 & & $A^{0}$ & $93$ \\
    & $\tilde{t}$ & 96  & & $H^{\pm}$ & $80$ \\
    \hline
    \multirow{3}{*}{Sleptons} & $\tilde{e}$ & 73 & \multirow{4}{*}{Neutralinos} & $\tilde{N}^{0}_{1}$ & $46$  \\
    & $\tilde{\mu}$ & 94 &  & $\tilde{N}^{0}_{2}$ & $62$\\
    & $ \tilde{\tau} $ & 82 &  & $\tilde{N}^{0}_{3}$ & $100$\\ \cline{1-3}
    Charginos & $\tilde{\chi}^{\pm}$, $\tilde{ \chi}^{\prime \pm}$ & $94$ & & $\tilde{N}^{0}_{4}$ & $116$\\
    \hline
    Gluinos & $\tilde{g}$ & $~300$ & $Z^{\prime}$ Boson & $A_{B-L}$ & $800$ \\     
    \hline    
  \end{tabular}
\end{center}
    \caption{{\scriptsize Experimental lower bounds on the Higgs fields and sparticles in the MSSM.  The $Z^{'}$ mass is for an additional $U(1)$ gauge boson arising from spontaneously broken $SO(10)$.  See Appendix B for a discussion of the bounds on the Higgs scalars.}}
\label{table2}
\end{table}

\item An important phenomenological constraint is that our results be consistent with the observed bounds on the masses of the Higgs fields, Higgsinos and all squarks, sleptons and gauginos.  Note that such bounds are partially dependent on the theory in which they are analyzed. The determination of the exact experimental bounds
within the context of the $B$-$L$ MSSM  theory is left to future work.  In this paper, we use the fact that this theory is a minimal extension of the MSSM and, hence, expect most bounds to be similar to those found for the MSSM.  The most conservative bounds for the MSSM are given in the Summary Tables contained in the Particle Data Group review \cite{Zb} and reproduced in Table \ref{table2}. We emphasize that these serve as guidelines rather than strict bounds, since we are working with a model that is somewhat different than the MSSM.  

The theoretical calculation of the masses in the $B$-$L$ MSSM involves considerable mixing of the fields induced by the $\nu_{3}$ and $H$,$\bar{H}$ VEVs. This presents a challenge in our analysis.  Details of the mass matrices, the diagonalization process, as well as a discussion of the role of the spontaneously broken $B$-$L$ gauge symmetry, are presented in Appendix A. In this paper, we compute the mass eigenvalues for the Higgs fields and all sparticles and compare the results to the values in Table \ref{table2}. We disallow all initial conditions that violate these bounds. This requires particular care for the Higgs fields, and is discussed in detail in Appendix B.

\end{itemize}

\section{Numerical Analysis}

We now turn to the numerical analysis of the low-energy vacua associated with the four initial parameters given in \eqref{63}. Even though the number of these parameters has been reduced to four, a systematic study of this space is still labor intensive. Happily, there is a natural splitting  into two two-dimensional spaces. To see this, 
note that one of the physical properties we are most concerned with is the hierarchy between the 
$B$-$L$ and electroweak breaking.  This hierarchy can be described in several ways \cite{BLPaper}. Here, we will define the hierarchy as the ratio of the mass of the $U(1)_{B-L}$ gauge boson, given by
\begin{equation}
   M_{A_{B-L}}=\sqrt{2} \ g_{B-L} \langle \nu_{3} \rangle  \ , \quad   \langle \nu_{3} \rangle= 
   \frac{|m_{\nu_{3}}|}{g_{B-L}} 
\label{67}   
\end{equation}
evaluated at the electroweak scale, and the $Z$-boson mass given in \eqref{59}. Written in terms of the parameterization introduced in Subsection \ref{parameterization}, the hierarchy becomes
\begin{equation}
  \frac{M_{A_{B-L}}}{M_{Z}} = \frac{   \sqrt{2} |c_{\nu_{3}}| }{ \Big( \frac{\vert c_{\bar{H}}^{2} - c_{H}^{2} \vert }{\sqrt{1 - \sin^{2}(2\beta)}} - c_{\bar{H}}^{2} - c_{H}^{2} - 2 \vert c_{\mu} \vert^{2} \Big)^{1/2} }  \ .
\label{68}
\end{equation}
The factor of ${\cal M}$ occurs in both the numerator and the denominator and, hence, cancels out of this expression. Of the five parameters in \eqref{68}, only  $c_{\nu_{3}}$, $c_{\mu}$ and 
$\tan\beta$ have arbitrary initial conditions. Noting that all $c_{i}$ coefficients, even when evaluated at the electroweak scale,  are essentially of order unity, we see that the most influential factors in the size of the hierarchy are $c_{\mu}$ and $\tan\beta$.
This is because for fixed $\tan\beta$ one can drive the denominator in \eqref{68} to zero, and, hence, the hierarchy to be arbitrarily large, by fine-tuning $c_{\mu}$.
For this reason, we will examine the two-dimensional $c_{\mu}(0)$-$\tan\beta$ plane for different values of $c_{q}(0)$ and $c_{\nu_{3}}(0)$.  This naturally splits the four-dimensional space of initial values into two two-dimensional surfaces, greatly simplifying the analysis.

\subsection{All $m^{2}>0$}
\label{sec:msquared}

\subsubsection*{Phenomenologically Allowed Regions and the Mass Spectrum:}

We first present our analysis subject to the following additional condition.

\begin{itemize}
\item With the exception of $m_{\nu_{3}}^{2}$, {\it all squark and slepton soft squared masses 
are constrained to be  positive} over the entire scaling range.
\end{itemize}

\noindent To illustrate the procedure, pick an arbitrary point
\begin{equation}
c_{q}(0)= 0.75 \ , \quad c_{\nu_{3}}(0)=0.75
\label{69}
\end{equation}
in the $c_{q}(0)$-$c_{\nu_{3}}(0)$ plane. For these initial values, we scan over the $c_{\mu}(0)$-$\tan\beta$ plane, first imposing the positive squark/slepton squared mass condition and then
analyzing each point relative to the constraints discussed in the previous section. The results are shown in Figure 1.\footnote{Here, and throughout the remainder of this paper, the irregularity of the plotted lines in the $c_{\mu}(0)$-$\tan\beta$ and $c_{q}(0)$-$c_{\nu_{3}}(0)$ planes reflects the granularity of the grid structure used in the calculation.  Each $c_{\mu}(0)$-$\tan\beta$ figure roughly represents $120^{2}$ data points, whereas there are roughly $200^{2}$ points in each $c_{q}(0)$-$c_{\nu_{3}}(0)$ plot.  } The positive squared mass condition is satisfied everywhere in the depicted region.

\begin{figure}
 \centering
  \subfloat[]{\includegraphics[scale=0.2]{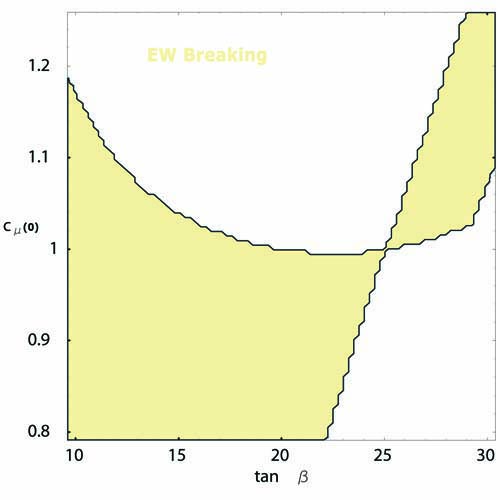}}
  \subfloat[]{\includegraphics[scale=0.2]{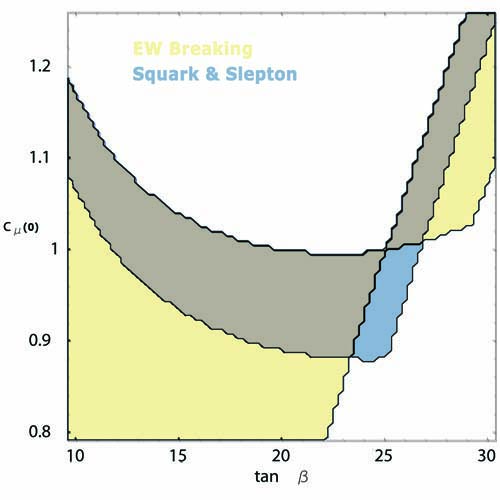}}
  \\
  \subfloat[]{\includegraphics[scale=0.2]{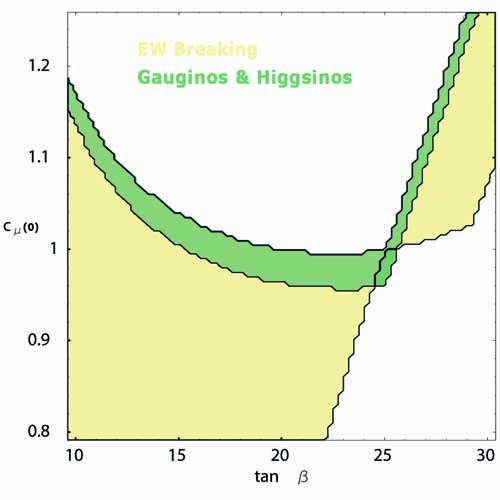}}
  \subfloat[]{\includegraphics[scale=0.2]{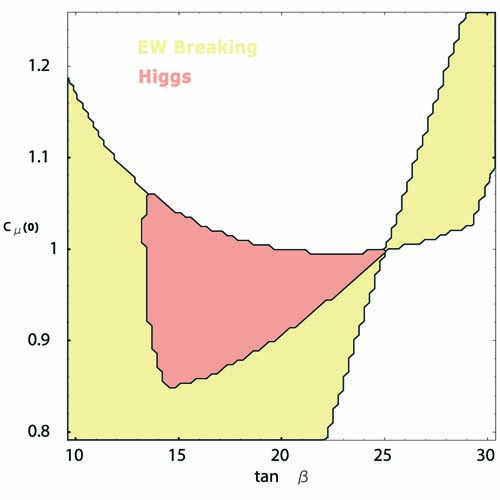}}
  \\
  \subfloat[]{\includegraphics[scale=0.3]{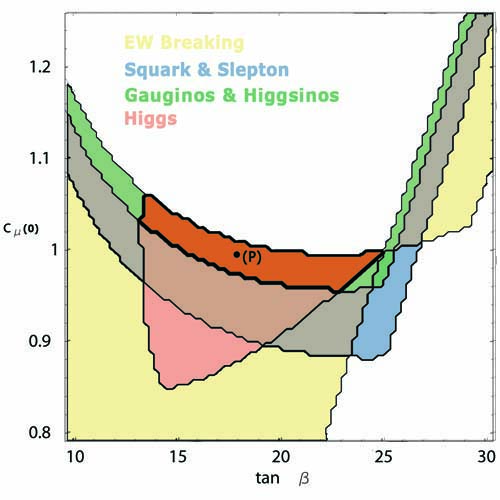}}
 \caption{{\scriptsize The $c_{\mu}(0)$-$\tan\beta$ plane corresponding to the point  $c_{q}(0)= 0.75, c_{\nu_{3}}(0)=0.75$. The yellow and white regions of (a) indicate where electroweak symmetry is and is not broken respectively. The individual regions satisfying the present experimental bounds for squarks and sleptons, gauginos and Higgs fields are shown in (b),(c) and (d), while their intersection is presented in (e). The dark brown area of (e) is the phenomenologically allowed region where electroweak symmetry is broken and all experimental mass bounds are satisfied. We present our predictions for the sparticle and Higgs masses at point (P).}}
 \label{fig:hier Reg2}
\end{figure}

Figure 1(a) shows the regions where electroweak symmetry is and is not radiatively broken, indicated in yellow and white respectively. The yellow region is defined as the locus of points where {\it both} inequalities \eqref{64} and \eqref{65} are {\it satisfied}, whereas in any white region 
{\it either one or both} of these inequalities is {\it violated}. Before analyzing the individual areas, let us recall the consequences of each inequality. As discussed in Subsection 3.8, \eqref{64} guarantees that one linear combination of Higgs fields has a negative squared mass. In this case, satisfying  inequality \eqref{65} implies a stable electroweak breaking vacuum. If, however, \eqref{65} is violated, the potential energy is not bounded from below and no stable vacuum state exits. On the other hand, violating inequality \eqref{64} indicates that the origin of Higgs space is either a local minimum or a local maximum of the potential energy, depending on whether or not \eqref{65} is satisfied.

 Let us now discuss the individual regions. Anywhere in the yellow region both \eqref{64} and \eqref{65} are satisfied, leading to a stable electroweak breaking vacuum. 
Note that there are {\it two} separated areas where electroweak breaking does not occur. Our analysis shows that at any point in the {\it upper} white region it is the first inequality \eqref{64} that is violated, while \eqref{65} continues to be satisfied. This indicates a stable vacuum, but with {\it vanishing} Higgs VEVs.
The transition between the yellow and {\it upper} white regions is defined by saturating inequality \eqref{64}, that is,
\begin{equation}
c_{B}^{4} = (|c_{\mu}|^{2}+c_{H}^{2})(|c_{\mu}|^{2}+c_{\bar{H}}^{2}) \ .
\label{70}
\end{equation}
It follows from this and expression \eqref{60} that the boundary between these regions corresponds to the vanishing of $M_{Z}^{2}$ in \eqref{61}, that is,
\begin{equation}
 \frac{\vert c_{\bar{H}}^{2} - c_{H}^{2} \vert }{\sqrt{1 - \sin^{2}(2\beta)}} - c_{\bar{H}}^{2} - c_{H}^{2} - 2 \vert c_{\mu} \vert^{2} =0 \ ,
\label{71}
\end{equation}
plotted as a function of $\tan\beta$ and $c_{\mu}(0)$.
Below this boundary  $M_{Z}^{2}$ is positive, indicating electroweak symmetry breaking vacua. At and above this line, however, $M_{Z}^{2}$ vanishes, implying that electroweak symmetry is unbroken. Similarly, the {\it lower right} white region shown in Figure 1(a) also violates constraint \eqref{64} while satisfying \eqref{65}. Hence, the above analysis applies here as well. For completeness, we point out that, beyond the boundaries shown in Figure 1(a), there is a transition of this lower right region to an area where {\it both} inequalities \eqref{64} and \eqref{65} are violated. In this regime, there are no stable vacua.

Figures 1(b),(c) and (d) indicate where our calculated masses of the squarks, sleptons, Higgs and gauginos respectively exceed the experimental lower bounds presented in Table 1. Finally, Figure 1(e) superimposes all of these with the area of electroweak symmetry breaking, the dark brown region representing their intersection. Any point in this region has broken electroweak symmetry and a mass spectrum satisfying all experimental bounds. As an example, consider the point (P) indicated in this region. Our calculated values for the squark, slepton, Higgs and gaugino masses are presented in Table 2. Note that, as stated, their values all exceed the experimental bounds.

\begin{table}
\begin{center}
\hspace*{-0.12 in}
  \begin{tabular}{ | c | c | c | c | c | c | }
    \hline
    Particle & Symbol  & Mass [GeV] & Particle & Symbol  & Mass [GeV] \\  \hline
    \multirow{4}{*}{Squarks} & $ \tilde{Q}_{1,2} $ & 1080  & \multirow{4}{*}{Higgs}  & $h^{0}$ & $132$ \\
    & $\tilde{t}_{1,2}, \tilde{b}_{1,2}$ & 1012, 1140 & & $H^{0}$ & $473$ \\
    & $\tilde{b}_{3}^{(1)}, \tilde{b}_{3}^{(2)}$ &  884, 1055 & & $A^{0}$ & $472$ \\
    & $\tilde{t}_{3}^{(1)}, \tilde{t}_{3}^{(2)}$ & 665, 929  & & $H^{\pm}$ & $479$ \\
    \hline
    \multirow{3}{*}{Sleptons} & $\tilde{L}_{1,2} $ & 1216 & \multirow{4}{*}{Neutralinos} & $\tilde{N}^{0}_{1}$ & $147$  \\
    & $ \tilde{\tau}_{1,2}$ & 1185  &  & $\tilde{N}^{0}_{2}$ & $286$\\
    & $ \tilde{\tau}_{3}^{(1)}, \tilde{\tau}_{3}^{(2)} $ & 1141, 1197 &  & $\tilde{N}^{0}_{3}$ & $523$\\ \cline{1-3}
    Charginos & $\tilde{\chi}^{\pm}$, $\tilde{\chi}^{\prime\pm}$ & $286, 537$ & & $\tilde{N}^{0}_{4}$ & $536$\\
    \hline
    Gluinos & $\tilde{g}$ & $ 1074 $ & $Z^{\prime}$ & $A_{B-L}, \tilde{A}_{B-L}$ & $1252, 1302$ \\     
    \hline    
  \end{tabular}
\end{center}
    \caption{  {\scriptsize The predicted spectrum at point (P) in Figure 1(e).  The tilde denotes the superpartner of the respective particle.  The superpartners of left-handed fields are depicted by an upper case label whereas the lower case is used for right-handed fields.  The considerable mixing between the third family left- and right-handed scalar fields is incorporated into these results.    } }
\label{table3}
\end{table}

\begin{figure}
 \centering
 \includegraphics[scale=0.25]{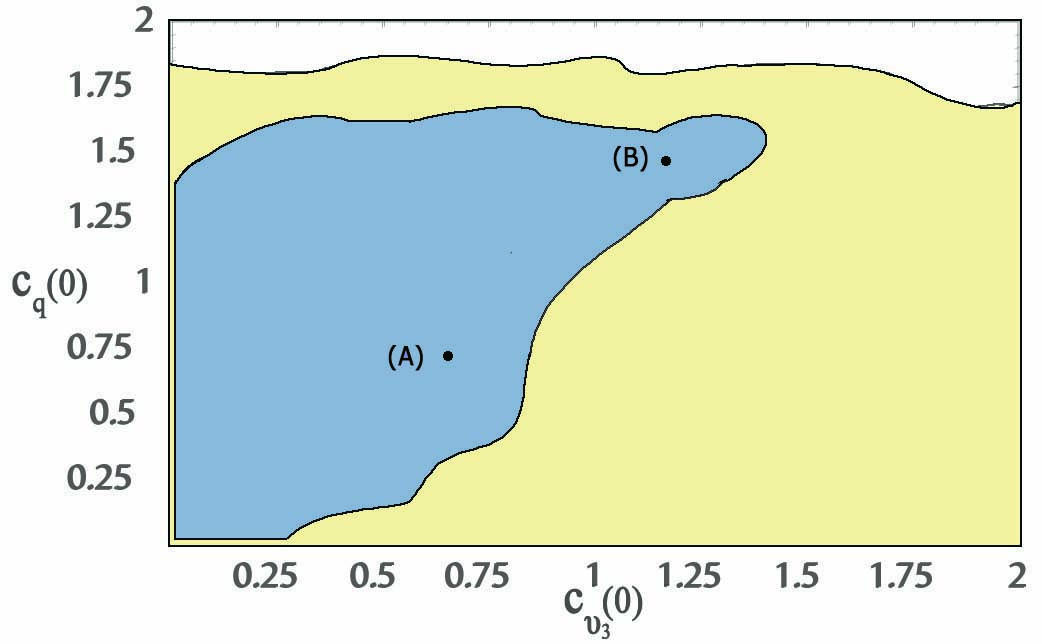}
 \caption{{\scriptsize  A plot of the $c_{q}(0)$-$c_{\nu_{3}}(0)$ plane showing physically relevant areas.  The yellow and white indicate points whose corresponding $c_{\mu}(0)$-$\tan\beta$ plane does and does not contain a region of electroweak symmetry breaking respectively. Within the yellow area, the blue shading contains all points whose $c_{\mu}(0)$-$\tan\beta$ plane has a non-vanishing region satisfying all experimental sparticle and Higgs bounds and for which {\it all soft susy breaking masses remain positive over the entire scaling range}. (A) and (B) indicate the two points analyzed in detail in the text.}}
 \label{fig:CqCnu12burt}
\end{figure}

\begin{figure}
 \centering
 \includegraphics[scale=.415]{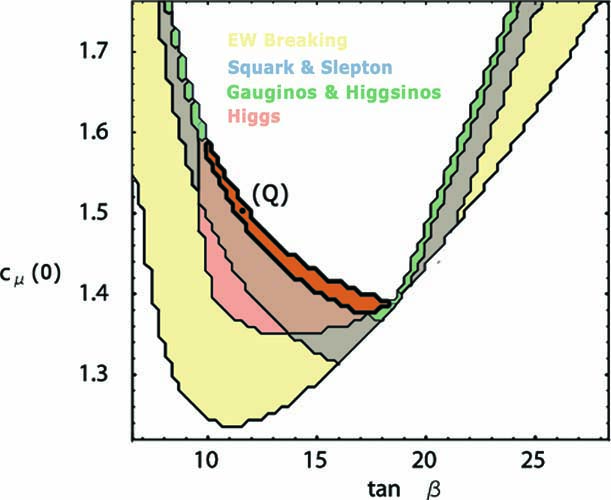}
 \caption{{\scriptsize The $c_{\mu}(0)$-$\tan\beta$ plane corresponding to the point  $c_{q}(0)= 1.4, c_{\nu_{3}}(0)=1.2$. The yellow and white regions indicate where electroweak symmetry is and is not broken respectively. The individual regions satisfying the present experimental bounds for squarks and sleptons, gauginos and Higgs fields are shown in the indicated colors. The dark brown area is their mutual intersection where electroweak symmetry is broken and all experimental mass bounds are satisfied. We present our predictions for the sparticle and Higgs masses at point (Q). }}
 \label{fig:Overlap}
\end{figure}

The above analysis was carried out for the arbitrarily chosen point \eqref{69} in the $c_{q}(0)$-$c_{\nu_{3}}(0)$ plane. 
We emphasize that although this point has a non-vanishing region in the $c_{\mu}(0)$-$\tan\beta$ plane satisfying all phenomenological bounds, this need not be the case for other points. To explore this, we now scan over  the entire  $c_{q}(0)$-$c_{\nu_{3}}(0)$ plane. At each point, we analyze the associated $c_{\mu}(0)$-$\tan\beta$ plane and see if an allowed region exists.  
The results are shown in Figure \ref{fig:CqCnu12burt}.\footnote{ Note that the blue shaded allowed region can get infinitesmally close to, but not touch, both the horizontal and vertical axes.  The somewhat irregular boundary lines reflect both the complexity of solving many RGEs, as well as the numerical limitations of our calculation.  These comments apply to all figures of the $c_{q}(0)$-$c_{\nu_{3}}(0)$ plane in this paper.  }
The white region indicates points whose corresponding  $c_{\mu}(0)$-$\tan\beta$ plane contains {\it no} locus of electroweak symmetry breaking.
The yellow area represents points whose $c_{\mu}(0)$-$\tan\beta$ plane {\it has} a region where electroweak symmetry is broken. Finally, each point in the blue area has a phenomenologically allowed region in its corresponding  $c_{\mu}(0)$-$\tan\beta$ plane satisfying the squark/slepton positive squared mass condition.
Point \eqref{69} analyzed above is indicated by (A) in the diagram. It is of interest to see how the results change as we move to different phenomenologically allowed points in the $c_{q}(0)$-$c_{\nu_{3}}(0)$ plane. For example, consider point (B) shown in Figure \ref{fig:CqCnu12burt}. This has the values
\begin{equation}
c_{q}(0) = 1.4 \ , \quad c_{\nu}(0) = 1.2 \ . 
\label{72}
\end{equation}
For this point, the regions of the $c_{\mu}(0)$-$\tan\beta$ plane corresponding to the different constraints, as well as their intersection, are shown in Figure \ref{fig:Overlap}. The positive squared mass condition is satisfied everywhere in the depicted regime.

In the yellow region both \eqref{64} and \eqref{65} are satisfied, leading to stable electroweak breaking vacua. There are two separated areas where electroweak breaking does not occur. As occurred for point (A), anywhere in the {\it upper} white region the first inequality \eqref{64} is violated, while \eqref{65} continues to be satisfied. This indicates stable vacua, but with {\it vanishing} Higgs VEVs. As discussed above, the boundary between the yellow and {\it upper} white regions corresponds to the vanishing of $M_{Z}^{2}$ in \eqref{61}.
Unlike the analysis of point (A), however, the {\it lower right} white region shown in Figure 3 violates 
{\it both} constraints \eqref{64} and \eqref{65}. Hence, the origin of Higgs space is a local maximum and the potential energy is unbounded from below. There are no stable vacua in this regime.

The regions where the squarks/sleptons, gauginos and Higgs exceed their experimental lower bounds are depicted in the indicated colors.
Any point in the intersection area, shown in dark brown, has broken electroweak symmetry and a mass spectrum satisfying all experimental bounds. As an example, consider the point (Q) indicated in this region. Our calculated values for the squark, slepton, Higgs and gaugino masses are presented in Table 3. Note that, as stated, their values all exceed the experimental bounds.

\begin{table}
\begin{center}
\hspace*{-0.12 in}
  \begin{tabular}{ | c | c | c | c | c | c | }
    \hline
    Particle & Symbol  & Mass [GeV] & Particle & Symbol  & Mass [GeV] \\  \hline
    \multirow{4}{*}{Squarks} & $ \tilde{Q}_{1,2} $ & 850  & \multirow{4}{*}{Higgs}  & $h^{0}$ & $127$ \\
    & $\tilde{t}_{1,2}, \tilde{b}_{1,2}$ & 775, 953 & & $H^{0}$ & $382$ \\
    & $\tilde{b}_{3}^{(1)}, \tilde{b}_{3}^{(2)}$ &  670, 915 & & $A^{0}$ & $381$ \\
    & $\tilde{t}_{3}^{(1)}, \tilde{t}_{3}^{(2)}$ & 456, 737  & & $H^{\pm}$ & $390$ \\
    \hline
    \multirow{3}{*}{Sleptons} & $\tilde{L}_{1,2} $ & 1255 & \multirow{4}{*}{Neutralinos} & $\tilde{N}^{0}_{1}$ & $97$  \\
    & $ \tilde{\tau}_{1,2}$ & 1237  &  & $\tilde{N}^{0}_{2}$ & $189$\\
    & $ \tilde{\tau}_{3}^{(1)}, \tilde{\tau}_{3}^{(2)} $ & 1217, 1246 &  & $\tilde{N}^{0}_{3}$ & $499$\\ \cline{1-3}
    Charginos & $\tilde{\chi}^{\pm}$, $\tilde{\chi}^{\prime\pm}$ & $190, 510$ & & $\tilde{N}^{0}_{4}$ & $509$\\
    \hline
    Gluinos & $\tilde{g}$ & $ 712 $ & $Z^{\prime}$ & $A_{B-L}, \tilde{A}_{B-L}$ & $1314, 1348$ \\     
    \hline    
  \end{tabular}
\end{center}
    \caption{  {\scriptsize The predicted spectrum at point (Q) in Figure 3.  The tilde denotes the superpartner of the respective particle.  The superpartners of left-handed fields are depicted by an upper case label whereas the lower case is used for right-handed fields.  The mixing between the third family left- and right-handed scalar fields is incorporated.   } }
\label{table4}
\end{table}

\subsubsection*{The $B$-$L$/Electroweak Hierarchy:}

We have determined the subspace of the $c_{q}(0)$-$c_{\nu_{3}}(0)$ plane for which each point has a region in the corresponding $c_{\mu}(0)$-$\tan\beta$ plane satisfying 1) the positive squark/slepton squared mass condition with 2) broken electroweak symmetry and 3) phenomenologically acceptable squark, slepton, Higgs and gaugino masses. Given such a point in the 
 $c_{q}(0)$-$c_{\nu_{3}}(0)$ plane and choosing a point in the acceptable region in the 
$c_{\mu}(0)$-$\tan\beta$ plane, we now analyze the following question: What is the $B$-$L$/electroweak hierarchy for these initial values?

An expression for the $B$-$L$/electroweak hierarchy in terms of the $c_{i}$ coefficients and $\tan\beta$ was given in \eqref{69}. We repeat it here for convenience.
\begin{equation}
  \frac{M_{A_{B-L}}}{M_{Z}} = \frac{   \sqrt{2} |c_{\nu_{3}}| }{ \Big( \frac{\vert c_{\bar{H}}^{2} - c_{H}^{2} \vert }{\sqrt{1 - \sin^{2}(2\beta)}} - c_{\bar{H}}^{2} - c_{H}^{2} - 2 \vert c_{\mu} \vert^{2} \Big)^{1/2} }  \ .
\label{73}
\end{equation}
For the specific point chosen in the initial $c_{q}(0),~c_{\nu_{3}}(0), ~c_{\mu}(0),~\tan\beta$ parameter space, one can scale all quantities down to the electroweak scale and evaluate the hierarchy using \eqref{73}. As a concrete example, consider point (A) in the $c_{q}(0)$-$c_{\nu_{3}}(0)$ plane of 
Figure 2. The corresponding regions of the $c_{\mu}(0)$-$\tan\beta$ plane were superimposed in Figure 1(e) and are presented again in Figure 4(a). The allowed region is the dark brown area. For (A) given in \eqref{70}, the $B$-$L$/electroweak hierarchy is evaluated for each point in this allowed region and plotted in Figure 4(b). We find that the hierarchy takes values of 6.30-6.36 along the lower boundary of the allowed region. 
\begin{figure}
 \centering
  \hspace*{-0.4 in}
  \subfloat[]{\includegraphics[scale=0.25]{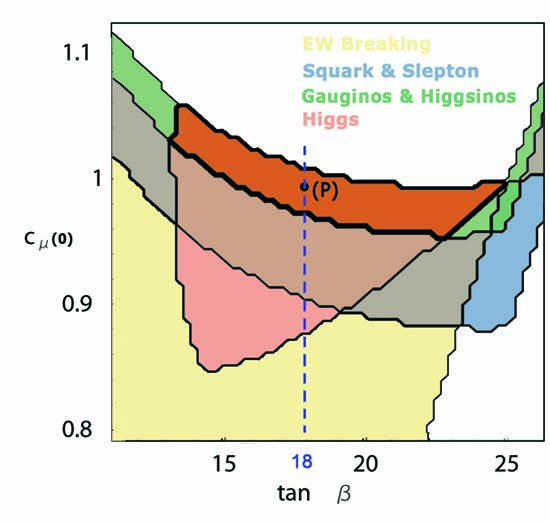}}
  \\
  \subfloat[]{\includegraphics[scale=0.4]{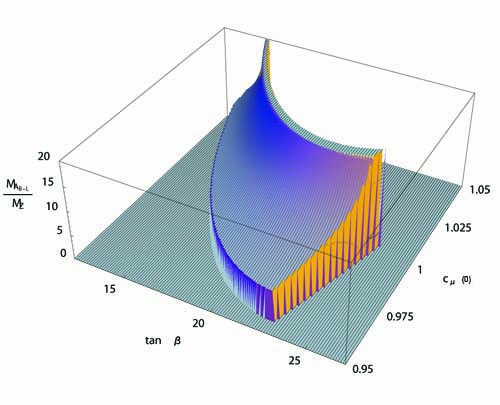}}
  \\
  \subfloat[]{\includegraphics[scale=0.35]{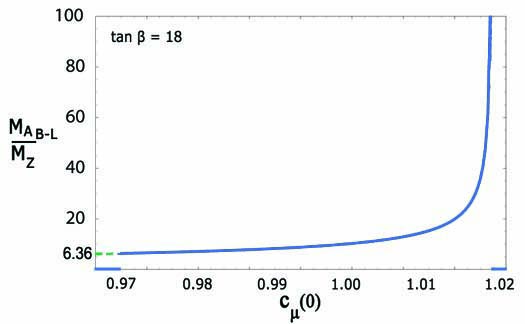}}
 \caption{{\scriptsize Plot (a) shows the $c_{\mu}(0)$-$\tan\beta$ plane corresponding to point (A) in Figure 2 with the phenomenologically allowed region indicated in dark brown. The mass spectrum at (P) was presented in Table 2. A plot of the hierarchy $M_{B-L}/M_{Z}$ over the allowed region is given in (b). Graph (c) shows the hierarchy as a function of $c_{\mu}(0)$ along the $\tan\beta=18$ line passing through (P).  }}
 \label{fig:hier Reg2}
\end{figure}
Note that below this boundary at least one of the gaugino or Higgs masses violates their experimental bound.  Hence, the lower values of the hierarchy are determined from the experimental data. On the other hand, as one approaches the boundary with the upper white region, the hierarchy becomes infinitely large. To understand this, recall from \eqref{71}
that this boundary is determined by the
vanishing of $M_{Z}^{2}$ in \eqref{62}, that is,
\begin{equation}
 \frac{\vert c_{\bar{H}}^{2} - c_{H}^{2} \vert }{\sqrt{1 - \sin^{2}(2\beta)}} - c_{\bar{H}}^{2} - c_{H}^{2} - 2 \vert c_{\mu} \vert^{2} =0 \ .
\label{74}
\end{equation}
Hence, at any point on this boundary the denominator in \eqref{73} vanishes and 
\begin{equation}
 \frac{M_{A_{B-L}}}{M_{Z}} \longrightarrow \infty \ .
\label{75}
\end{equation}
It follows that within the phenomenologically acceptable region, {\it any value of the $B$-$L$ hierarchy in the range $6.30 \lesssim M_{A_{B-L}}/M_{Z} < \infty$
can be attained}.

Another way to analyze this data is to pick a specific point in the allowed region and to compute 
\eqref{73} as a function of $c_{\mu}(0)$ along the fixed $\tan\beta$ line passing through it. For concreteness, choose the point (P) for which we calculated the mass spectrum in Table 2. This is 
shown in Figure 4(a) along with the dotted line $\tan\beta=18$ intersecting it. The $B$-$L$/electroweak hierarchy along this line is plotted in Figure 4(c). Note that this begins at 
$ M_{A_{B-L}}/M_{Z}=6.35$ at the experimentally determined lower boundary, rises slowly to 
$ M_{A_{B-L}}/M_{Z}\sim 20$ across most of the region, and then rapidly diverges to infinity as one approaches the upper boundary. Approaching both the lower and, especially, the upper boundary requires fine-tuning of $c_{\mu}(0)$. For ``typical'' values of $c_{\mu}(0)$, the hierarchy is {\it naturally} in the range
\begin{equation}
10 \lesssim  \frac{M_{A_{B-L}}}{M_{Z}} \lesssim 20 \ .
\label{76}
\end{equation}

As a second example, consider point (B) in the $c_{q}(0)$-$c_{\nu_{3}}(0)$ plane of 
Figure 2. The corresponding regions of the $c_{\mu}(0)$-$\tan\beta$ plane were superimposed in Figure 3 and presented again in Figure 5(a). The allowed region is the dark brown area. For (B) given in \eqref{73}, the $B$-$L$/electroweak hierarchy is evaluated for each point in this allowed region and plotted in Figure 5(b). We find that the hierarchy takes values of 10.00-10.21 along the lower boundary of the allowed region, below which at least one of the gaugino or Higgs masses 
\begin{figure}
 \centering
  \hspace*{-0.4 in}
  \subfloat[]{\includegraphics[scale=0.4]{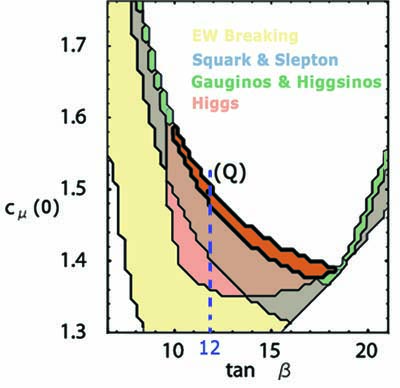}}
  \\
  \subfloat[]{\includegraphics[scale=0.4]{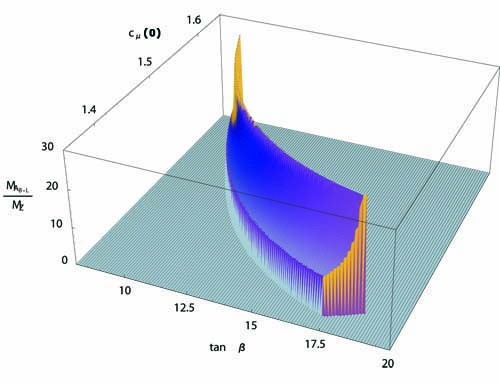}}
  \\
  \subfloat[]{\includegraphics[scale=0.35]{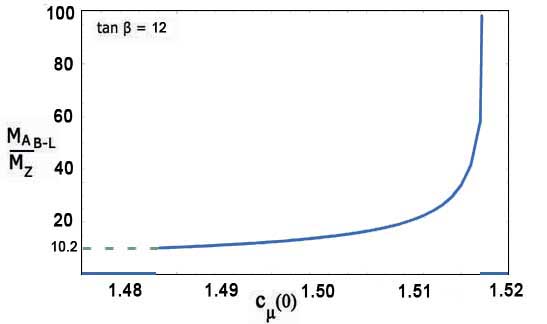}}
 \caption{{\scriptsize Plot (a) shows the $c_{\mu}(0)$-$\tan\beta$ plane corresponding to point (B) in Figure 2 with the phenomenologically allowed region indicated in dark brown. The mass spectrum at (Q) was presented in Table 3. A plot of the hierarchy $M_{B-L}/M_{Z}$ over the allowed region is given in (b). Graph (c) shows the hierarchy as a function of $c_{\mu}(0)$ along the $\tan\beta=12$ line passing through (Q). }}
 \label{fig:hier Reg2}
\end{figure}
violates their experimental bound.  Again, as one approaches the boundary with the upper white region, the hierarchy becomes infinitely large. It follows that within the phenomenologically acceptable region {\it any value of the $B$-$L$ hierarchy in the range $10 \lesssim M_{A_{B-L}}/M_{Z} < \infty$ can be attained}.

Another way to analyze this data is to pick a specific point in the allowed region and to compute 
\eqref{73} as a function of $c_{\mu}(0)$ along the fixed $\tan\beta$ line passing through it. For concreteness, choose the point (Q) for which we calculated the mass spectrum in Table 3. This is 
shown in Figure 5(a) along with the dotted line $\tan\beta=12$ intersecting it. The $B$-$L$/electroweak hierarchy along this line is plotted in Figure 5(c). Note that this begins at 
$ M_{A_{B-L}}/M_{Z}=10.15$ at the experimentally determined lower boundary, rises slowly to 
$ M_{A_{B-L}}/M_{Z}\sim 30$ across most of the region, and then rapidly diverges to infinity as one approaches the upper boundary. For ``typical'' values of $c_{\mu}(0)$ not fine-tuned near either boundary, the hierarchy is {\it naturally} in the range
\begin{equation}
15 \lesssim  \frac{M_{A_{B-L}}}{M_{Z}} \lesssim 30 \ .
\label{77}
\end{equation}

\subsection{$m_{Q_{3}}^{2}<0$}
\label{sec:l3squark}

\subsubsection*{The Potential Energy for $m_{\nu_{3}}^{2}<0$ and $m_{Q_{3}}^{2}<0$:}

For the choice of parameters in \eqref{55}, all sleptons have positive soft squared masses with the exception of the third family right-handed sneutrino, for which $m_{\nu_{3}}^{2}<0$. As noted in Subsection 3.8, imposing positivity on the effective masses of the left-handed squarks at the $B$-$L$ breaking VEV 
$\langle \nu_{3} \rangle$,  that is,
\begin{equation}
\langle m_{Q_{i}}^{2}\rangle =m_{Q_{i}}^{2}+\frac{1}{4}g_{4}^{2}\langle \nu_{3} \rangle^{2} >0 \ ,
\label{78}
\end{equation}
does {\it not} require that $m_{Q_{i}}^{2}$ be positive. In general, one or more of these soft squared masses can be negative. Despite our assumption in \eqref{20},\eqref{35} that the initial squark masses are universal, the effect of the large third family up-Yukawa coupling in the RGEs is to break this degeneracy, driving  $m_{Q_{3}}^{2}$ negative more quickly than the first and second family squark masses. Therefore, for simplicity, we explore the possibility that only the third family left-handed squark soft mass becomes negative, $m_{Q_{3}}^{2}<0$,  as it is scaled down to electroweak energy-momenta. 

The electroweak phase transition breaks the left-handed $SU(2)_{L}$ doublet $Q_{3}$ into its up- and down- quark components $U_{3}$ and $D_{3}$ respectively. The leading order contribution of the Higgs VEVs to their mass 
splits the degeneracy between these two fields, destabilizing the potential most strongly in the $D_{3}$ direction. For this reason, the relevant Lagrangian for analyzing this vacuum can be restricted to
\begin{eqnarray}
&& {\cal{L}}=|{\cal{D}}_{\nu_{3}\mu} \nu_{3}|^{2} -\frac{1}{4}F_{B-L \mu\nu}F_{B-L}^{\mu\nu}
+|{\cal{D}}_{D_{3}\mu}D_{3}|^{2} -\frac{1}{4}F_{Y \mu\nu}F_{Y}^{\mu\nu} \nonumber \\
&& \quad \  \  \ -\frac{1}{4}F_{SU(2) \mu\nu}F_{SU(2)}^{\mu\nu}-\frac{1}{4}F_{SU(3) \mu\nu}F_{SU(3)}^{\mu\nu} 
-V(\nu_{3},D_{3})
\label{79}
\end{eqnarray}
where
\begin{eqnarray}
&& {\cal{D}}_{\nu_{3}\mu}= \partial_{\mu}-i g_{B-L}A_{B-L \mu} \ , \label{80}  \\
&& {\cal{D}}_{D_{3}\mu}=\partial_{\mu}-i \frac{g_{B-L}}{3} A_{B-L \mu}-i \frac{g_{Y}}{6}A_{Y \mu}-i g_{2}A_{SU(2) \mu}-i g_{3}A_{SU(3) \mu} \nonumber
\end{eqnarray}
and 
\begin{eqnarray} 
&& V(\nu_{3},D_{3})=m_{\nu_{3}}^{2}|\nu_{3}|^{2}+m_{D_{3}}^{2}|D_{3}|^{2}+\frac{g_{B-L}^{2}}{2}(|\nu_{3}|^{2}+\frac{1}{3}|D_{3}|^{2})^{2} \label{81} \\  
&& \qquad \qquad \quad \ +\frac{1}{2}( \frac{g_{Y}^{2}}{36} +\frac{g_{2}^{2}}{4} +\frac{g_{3}^{2}}{3})|D_{3}|^{4} \ . \nonumber
\end{eqnarray}
The first two terms in the potential are the soft supersymmetry breaking masses in \eqref{12}, while the remaining terms are supersymmetric and arise from $D_{B-L}$,  $D_{Y}$ in \eqref{10}, \eqref{9} and $D_{SU(2)_{L}}$, $D_{SU(3)_{C}}$ respectively. 
Using $\lambda_{d_{3}} \simeq 5 \times 10^{-2}$, 
a hierarchy with $\langle H^{0} \rangle \ll \langle \nu_{3} \rangle$
and assuming $|m_{D_{3}}|$ is of order $|m_{\nu_{3}}|$, terms proportional to the Higgs VEVs are small and are ignored in \eqref{81}. For simplicity, we henceforth drop the small $g_{B-L}^{2}/9+g_{Y}^{2}/36$ piece of the $D$-term contribution.

If both $m_{\nu_{3}}^{2}<0, m_{D_{3}}^{2}<0$ at the electroweak scale, then the potential is unstable at the origin of field space and has two other local extrema at
\begin{equation}
\langle \nu_{3} \rangle^{2}=-\frac{m_{\nu_{3}}^{2}}{g_{B-L}^{2}} , \quad \langle D_{3} \rangle=0 \ ,
\label{82}
\end{equation}
and
\begin{equation}
\langle \nu_{3}\rangle= 0, \quad \langle D_{3}\rangle^{2} = -\frac{m_{D_{3}}^{2}}{g_{2}^{2}/4+g_{3}^{2}/3}
\label{83}
\end{equation}
respectively. Using these, potential \eqref{81} can be rewritten as
\begin{eqnarray} 
&& V(\nu_{3},D_{3})= \frac {g_{B-L}^{2}}{2}(|\nu_{3}|^{2}-\langle \nu_{3}\rangle^{2})^{2} +\frac{g_{B-L}^{2}}{3} |\nu_{3}|^{2} |D_{3}|^{2} \nonumber \\
&& \qquad \qquad \quad +\frac {g_{2}^{2}/4+g_{3}^{2}/3}{2}(|D_{3}|^{2}-\langle D_{3}\rangle^{2})^{2} \ .
\label{84}
\end{eqnarray}
Let us analyze these two extrema. Both have positive masses in their radial directions. At the sneutrino vacuum \eqref{82}, the mass squared in the $D_{3}$ direction is given by
\begin{equation}
m_{D_{3}}^{2}|_{\langle \nu_{3}\rangle}= \frac{g_{B-L}^{2}}{3}\langle \nu_{3}\rangle^{2}-(\frac{g_{2}^{2}}{4}+\frac{g_{3}^{2}}{3}) \langle D_{3}\rangle^{2}= \frac{|m_{\nu_{3}}|^{2}}{3}-|m_{D_{3}}|^{2} \ ,
\label{85}
\end{equation}
whereas at the $D_{3}$ vacuum \eqref{83}, the mass squared in the $\nu_{3}$ direction is 
\begin{equation}
m_{\nu_{3}}^{2}|_{\langle D_{3}\rangle}= \frac{g_{B-L}^{2}}{3}\langle D_{3}\rangle^{2}-g_{B-L}^{2} \langle \nu_{3}\rangle^{2}=|m_{D_{3}}|^{2} (\frac{g_{B-L}^{2}}{3g_{2}^{2}/4+g_{3}^{2}})-|m_{\nu_{3}}|^{2} \ .
\label{86}
\end{equation}
Note that either \eqref{85} or \eqref{86} can be positive, but not both. To be consistent with the hierarchy solution, we want \eqref{82} to be a stable minimum. Hence, we demand $m_{D_{3}}^{2}|_{\langle \nu_{3}\rangle}>0$ or, equivalently, that
\begin{equation}
|m_{\nu_{3}}|^{2}>3|m_{D_{3}}|^{2} \ .
\label{87}
\end{equation}
We will impose \eqref{87} as an additional condition for the remainder of this subsection. It then follows from \eqref{86} that $m_{\nu_{3}}^{2}|_{\langle D_{3}\rangle}<0$ and, hence, the $D_{3}$ extremum \eqref{83} is a saddle point. As a consistency check, note that $V|_{\langle\nu_{3}\rangle}<V|_{\langle D_{3}\rangle}$ if and only if
\begin{equation}
g_{B-L}^{2}\langle \nu_{3}\rangle^{4}>(\frac{g_{2}^{2}}{4}+\frac{g_{3}^{2}}{3}) \langle D_{3}\rangle^{4} 
\label{88}
\end{equation}
or, equivalently, 
\begin{equation}
|m_{\nu_{3}}|^{2}>|m_{D_{3}}|^{2} (\frac{g_{B-L}^{2}}{3g_{2}^{2}/4+g_{3}^{2}})^{1/2} \ .
\label{89}
\end{equation}
This follows immediately from constraint \eqref{87}. 

Finally, note that the potential descends monotonically along a path  ${\cal{C}}$ from the saddle point at \eqref{83} to the absolute minimum at \eqref{84}. Solving the $\frac{\partial V}{\partial {D_{3}}}=0$ equation, this curve is found to be
\begin{equation}
|D_{3}|_{{\cal{C}}}=\big( \langle D_{3}\rangle^{2}-|\nu_{3}|^{2}(\frac{g_{B-L}^{2}}{3g_{2}^{2}/4+g_{3}^{2}})\big)^{1/2} \ .
\label{90}
\end{equation}
Note that it begins at $ \langle D_{3}\rangle$ for $\nu_{3}=0$ and continues until it tangentially intersects the $D_{3}=0$ axis at $|\nu_{30}|=\sqrt{3}\frac{|m_{D_{3}}|}{|m_{\nu_{3}}|}\langle \nu_{3}\rangle$. From here, the path continues down this axis to the stable minimum at \eqref{82}.
We conclude that at the electroweak scale the absolute minimum of potential \eqref{81} occurs at the sneutrino vacuum given in \eqref{82}. 

\subsubsection*{Phenomenologically Allowed Regions and the Mass Spectrum:}

In this subsection, we analyze our results subject to the following additional conditions.

\begin{itemize}

\item The {\it third family left-handed down-squark soft mass squared will be constrained to be negative}, that is, 
$m_{D_{3}}^{2}<0$.  All other squark and slepton soft squared masses are positive over the entire scaling range, with the exception of $m_{\nu_{3}}^{2}$.

\item To ensure that the $B$-$L$ breaking VEV is the absolute minimum, we impose condition \eqref{87},
\begin{equation}
|m_{\nu_{3}}|^{2}>3|m_{D_{3}}|^{2} \ ,
\label{91}
\end{equation}
at the electroweak scale.

\end{itemize}
We will refer to these two conditions collectively as the $m_{D_{3}}^{2}<0$ mass condition.

As discussed in the previous subsection, we proceed by scanning over  the entire  $c_{q}(0)$-$c_{\nu_{3}}(0)$ plane, at each point analyzing the associated $c_{\mu}(0)$-$\tan\beta$ plane to see if an allowed region exists.  
The results are shown in Figure 6. As in Figure 2, the white region indicates points whose corresponding  $c_{\mu}(0)$-$\tan\beta$ plane contains {\it no} locus of electroweak symmetry breaking,
whereas the yellow area represents points whose $c_{\mu}(0)$-$\tan\beta$ plane {\it has} a region where electroweak symmetry is broken. Finally, each point in the red area has a phenomenologically allowed region in its corresponding  $c_{\mu}(0)$-$\tan\beta$ plane satisfying the $m_{D_{3}}^{2}<0$ mass condition. Note that this is distinct from the blue region in Figure 2, where all squark/slepton mass squares are positive. Let us analyze the properties of an arbitrary point in the red area. 
For example, consider point (C) shown in Figure 6. This has the values
\begin{equation}
c_{q}(0) = 1.0 \ , \quad c_{\nu}(0) = 1.1 \ . 
\label{92}
\end{equation}
For this point, the regions of the $c_{\mu}(0)$-$\tan\beta$ plane corresponding to the different constraints, as well as their intersection, are shown in Figure 7. 
\begin{figure}
 \centering
 \includegraphics[scale=0.4]{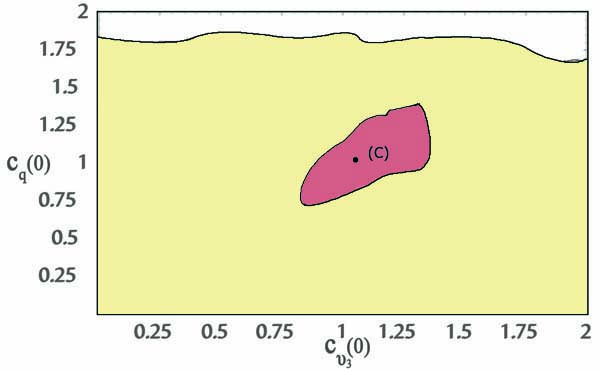}
 \caption{{\scriptsize  A plot of the $c_{q}(0)$-$c_{\nu_{3}}(0)$ plane showing physically relevant areas.  The yellow and white indicate points whose corresponding $c_{\mu}(0)$-$\tan\beta$ plane does and does not contain a region of electroweak symmetry breaking respectively. Within the yellow area, the red shading contains all points whose $c_{\mu}(0)$-$\tan\beta$ plane has a non-vanishing region satisfying all experimental sparticle and Higgs bounds and for which $m_{D_{3}}^{2}<0$. (C) indicates the point analyzed in detail in the text.}}
 \label{fig:CqCnu12}
\end{figure}
The $m_{D_{3}}^{2}<0$ mass condition is satisfied everywhere in the depicted regime.

In the yellow region both \eqref{64} and \eqref{65} are satisfied, leading to stable electroweak breaking vacua. There are two separated areas where electroweak breaking does not occur. As for point (A) in Figure 2, anywhere in the upper 
and lower right white regions the first inequality \eqref{64} is violated, while \eqref{65} continues to be satisfied. This indicates stable vacua, but with {\it vanishing} Higgs VEVs. It follows that the boundary between the yellow and white regions corresponds to $M_{Z}^{2}$ in \eqref{61} becoming zero.
The regions where the squarks/sleptons, gauginos and Higgs exceed their experimental lower bounds are depicted in the indicated colors.
Any point in the intersection area, shown in dark brown, has broken electroweak symmetry and an acceptable mass spectrum. As an example, consider the point (R) indicated in this region. Our calculated values for the squark, slepton, Higgs and gaugino masses are presented in Table 4. 
\begin{figure}
 \centering
 \includegraphics[scale=0.35]{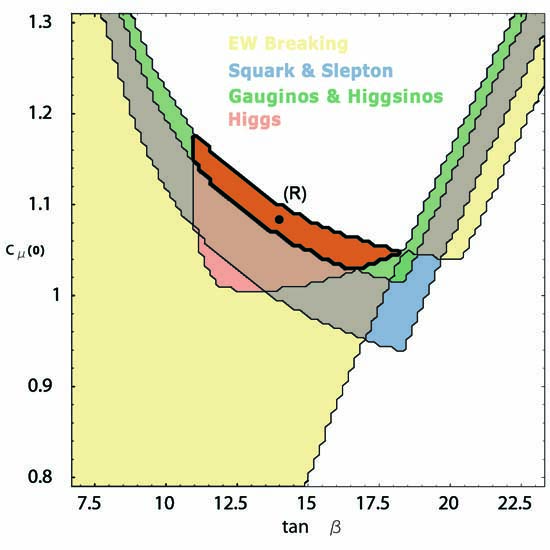}
 \caption{{\scriptsize The $c_{\mu}(0)$-$\tan\beta$ plane corresponding to the point  $c_{q}(0)= 1.0, c_{\nu_{3}}(0)=1.1$. The yellow and white regions indicate where electroweak symmetry is and is not broken respectively. The individual regions satisfying the present experimental bounds for squarks and sleptons, gauginos and Higgs fields are shown in the indicated colors. The dark brown area is their mutual intersection where electroweak symmetry is broken and all experimental mass bounds are satisfied. We present our predictions for the sparticle and Higgs masses at point (R). }}
 \label{}
\end{figure}
\\
\begin{table}
\begin{center}
\hspace*{-0.12 in}
  \begin{tabular}{ | c | c | c | c | c | c | }
    \hline
    Particle & Symbol  & Mass [GeV] & Particle & Symbol  & Mass [GeV] \\  \hline
    \multirow{4}{*}{Squarks} & $ \tilde{Q}_{1,2} $ & 778  & \multirow{4}{*}{Higgs}  & $h^{0}$ & $126$ \\
    & $\tilde{t}_{1,2}, \tilde{b}_{1,2}$ & 708, 869 & & $H^{0}$ & $271$ \\
    & $\tilde{b}_{3}^{(1)}, \tilde{b}_{3}^{(2)}$ &  640, 828 & & $A^{0}$ & $270$ \\
    & $\tilde{t}_{3}^{(1)}, \tilde{t}_{3}^{(2)}$ & 428, 687  & & $H^{\pm}$ & $282$ \\
    \hline
    \multirow{3}{*}{Sleptons} & $\tilde{L}_{1,2} $ & 1148 & \multirow{4}{*}{Neutralinos} & $\tilde{N}^{0}_{1}$ & $98$  \\
    & $ \tilde{\tau}_{1,2}$ & 1129  &  & $\tilde{N}^{0}_{2}$ & $188$\\
    & $ \tilde{\tau}_{3}^{(1)}, \tilde{\tau}_{3}^{(2)} $ & 1105, 1137 &  & $\tilde{N}^{0}_{3}$ & $382$\\ \cline{1-3}
    Charginos & $\tilde{\chi}^{\pm}$, $\tilde{\chi}^{\prime\pm}$ & $ 187, 400 $ & & $\tilde{N}^{0}_{4}$ & $398$\\
    \hline
    Gluinos & $\tilde{g}$ & $ 727 $ & $Z^{\prime}$ & $A_{B-L}, \tilde{A}_{B-L}$ & $1199, 1233$ \\     
    \hline    
  \end{tabular}
\end{center}
    \caption{  {\scriptsize The predicted spectrum at point (R) in Figure 7.  The tilde denotes the superpartner of the respective particle.  The superpartners of left-handed fields are depicted by an upper case label whereas the lower case is used for right-handed fields.  The mixing between the third family left- and right-handed scalar fields is incorporated.   } }
\label{tablereg2}
\end{table}

\subsubsection*{The $B$-$L$/Electroweak Hierarchy:}

We have determined the subspace of the $c_{q}(0)$-$c_{\nu_{3}}(0)$ plane for which each point has a region in the corresponding $c_{\mu}(0)$-$\tan\beta$ plane satisfying 1) the $m_{D_{3}}^{2}<0$ mass condition with 2) broken electroweak symmetry and 3) phenomenologically acceptable squark, slepton, Higgs and gaugino masses. Given such a point in the 
 $c_{q}(0)$-$c_{\nu_{3}}(0)$ plane and choosing a point in the acceptable region in the 
$c_{\mu}(0)$-$\tan\beta$ plane, we now analyze the $B$-$L$/electroweak hierarchy for these initial values.

An expression for this hierarchy in terms of the $c_{i}$ coefficients and $\tan\beta$ was given in \eqref{73}. 
For the specific point chosen in the initial $c_{q}(0), c_{\nu_{3}}(0), c_{\mu}(0)$, $\tan\beta$ parameter space, one can scale all quantities down to the electroweak scale and use this expression to evaluate the hierarchy. As a concrete example, consider point (C) in the $c_{q}(0)$-$c_{\nu_{3}}(0)$ plane of 
Figure 6. The corresponding regions of the $c_{\mu}(0)$-$\tan\beta$ plane were superimposed in Figure 7 and are presented again in Figure 8(a). The allowed region is the dark brown area. For (C) given in \eqref{92}, the $B$-$L$/electroweak hierarchy is evaluated for each point in this allowed region and plotted in Figure 8(b). We find that the hierarchy takes values of 8.99-9.06 along the lower boundary of the allowed region. Note that below this boundary at least one of the gaugino or Higgs masses violates their experimental bound.  Hence, the lower values of the hierarchy are determined from the experimental data. On the other hand, as one approaches the boundary with the upper white region, the hierarchy becomes infinitely large. As discussed in the previous subsection, this is explained by the
vanishing of $M_{Z}^{2}$ in \eqref{61}. 
Hence, at any point on this boundary the denominator in \eqref{73} vanishes and $M_{A_{B-L}}/M_{Z}\longrightarrow \infty$.
It follows that within the phenomenologically acceptable region, {\it any value of the $B$-$L$ hierarchy in the range $8.99 \lesssim M_{A_{B-L}}/M_{Z} < \infty$
can be attained}.

Another way to analyze this data is to pick a specific point in the allowed region and to compute 
\eqref{73} as a function of $c_{\mu}(0)$ along the fixed $\tan\beta$ line passing through it. For concreteness, choose the point (R) for which we calculated the mass spectrum in Table 4. This is 
shown in Figure 8(a) along with the dotted line $\tan\beta=14$ intersecting it. The $B$-$L$/electroweak hierarchy along this line is plotted in Figure 8(c). Note that this begins at 
$ M_{A_{B-L}}/M_{Z}=9.0$ at the experimentally determined lower boundary, rises slowly to 
$ M_{A_{B-L}}/M_{Z}\sim 40$ across most of the region, and then rapidly diverges to infinity as one approaches the upper boundary. Approaching both the lower and, especially, the upper boundary requires fine-tuning of $c_{\mu}(0)$. For ``typical'' values of $c_{\mu}(0)$, the hierarchy is {\it naturally} in the range
\begin{equation}
15 \lesssim  \frac{M_{A_{B-L}}}{M_{Z}} \lesssim 40 \ .
\label{93}
\end{equation}

\begin{figure}
 \centering
  \hspace*{-0.4 in}
  \subfloat[]{\includegraphics[scale=0.3]{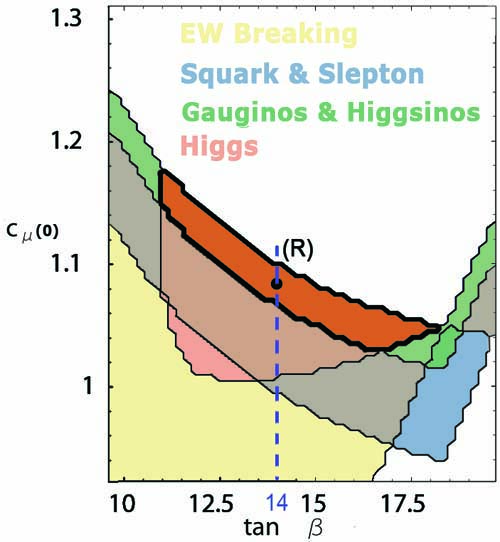}}
  \\
  \subfloat[]{\includegraphics[scale=0.3]{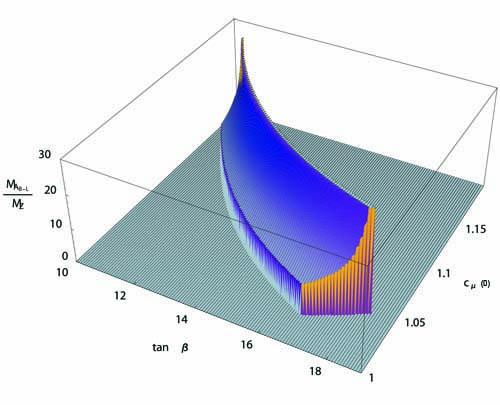}}
  \\
  \subfloat[]{\includegraphics[scale=0.35]{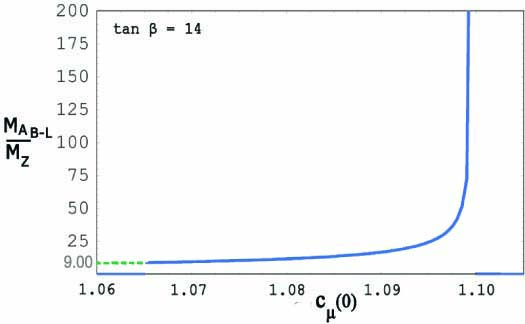}}
 \caption{{\scriptsize Plot (a) shows the $c_{\mu}(0)$-$\tan\beta$ plane corresponding to point (C) in Figure 6 with the phenomenologically allowed region indicated in dark brown. The mass spectrum at (R) was presented in Table 4. A plot of the hierarchy $M_{B-L}/M_{Z}$ over the allowed region is given in (b). Graph (c) shows the hierarchy as a function of $c_{\mu}(0)$ along the $\tan\beta=14$ line passing through (R).  }}
 \label{fig:hier}
\end{figure}

\subsubsection*{``Mixed'' $m^{2}>0$ and $m_{D_{3}}^{2}<0$ Mass Conditions:}

It is of interest to superimpose the blue region in Figure 2, satisfying the $m^{2}>0$ mass condition, with the red region of Figure 6, defined by the $m_{D_{3}}^{2}<0$ constraint. This is shown in Figure 9. 
\begin{figure}
 \centering
 \includegraphics[scale=0.4]{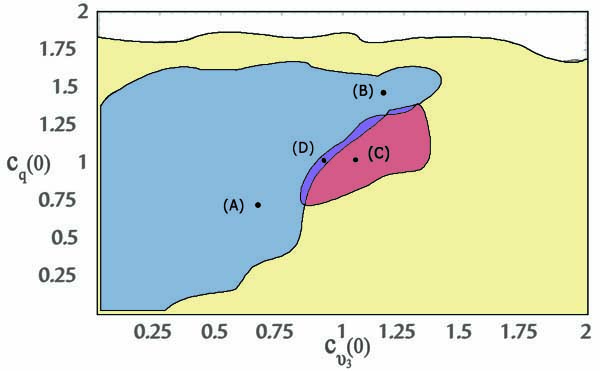}
 \caption{{\scriptsize  A plot of the 
 $c_{q}(0)$-$c_{\nu_{3}}(0)$ 
 plane showing both the blue and red regions presented in Figures 2 and 6 respectively. They have a non-vanishing intersection, indicated in purple. Any point in this overlap 
has an allowed region in the 
$c_{\mu}(0)$-$\tan\beta$ 
plane that is divided into two areas--one  with 
$m^{2}>0$ and the second with $m_{D_{3}}^{2} <0$. 
(D) indicates a point in this overlap region analyzed in 
detail in the text. }}
 \label{fig:CqCnu124}
\end{figure}
Note that there is a non-vanishing intersection between these two areas. This is comprised of points in the $c_{q}(0)$-$c_{\nu_{3}}(0)$ plane whose phenomenologically allowed regions in the corresponding
$c_{\mu}(0)$-$\tan\beta$ plane are each divided into two regimes--one satisfying the $m^{2}>0$ mass condition and the other the $m_{D_{3}}^{2}<0$ constraint. As a specific example, consider the point (D) shown in Figure 9. This has the values
\begin{equation}
c_{q}(0) = 1.0 \ , \quad c_{\nu}(0) = 0.9 \ . 
\label{94}
\end{equation}
For this point, the areas of the $c_{\mu}(0)$-$\tan\beta$ plane 
corresponding to the different constraints, as well as their intersection, are shown in Figure 10. The regions where the squarks/sleptons, gauginos and Higgs exceed their 
\begin{figure}
 \centering
 \includegraphics[scale=0.35]{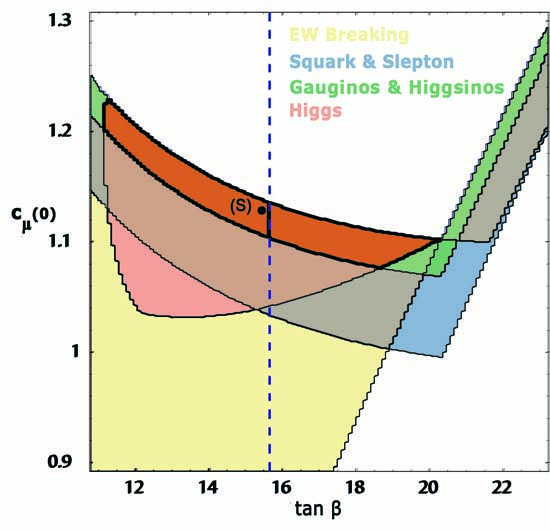}
 \caption{{\scriptsize The $c_{\mu}(0)$-$\tan\beta$ plane corresponding to the point  $c_{q}(0)= 1.0, c_{\nu_{3}}(0)=0.9$. The yellow and white regions indicate where electroweak symmetry is and is not broken respectively. The individual regions satisfying the present experimental bounds for squarks and sleptons, gauginos and Higgs fields are shown in the indicated colors. The dark brown area is their mutual intersection where electroweak symmetry is broken and all experimental mass bounds are satisfied. We present our predictions for the sparticle and Higgs masses at point (S). The dotted line passing to the right of (S) separates the $m^{2}>0$ region, to the left of this line, from the area where $m_{D_{3}}^{2}<0$, to the right.   }}
 \label{}
\end{figure}
\begin{table}
\begin{center}
\hspace*{-0.12 in}
  \begin{tabular}{ | c | c | c | c | c | c | }
    \hline
    Particle & Symbol  & Mass [GeV] & Particle & Symbol  & Mass [GeV] \\  \hline
    \multirow{4}{*}{Squarks} & $ \tilde{Q}_{1,2} $ & 2228  & \multirow{4}{*}{Higgs}  & $h^{0}$ & $142$ \\
    & $\tilde{t}_{1,2}, \tilde{b}_{1,2}$ & 2062, 2417 & & $H^{0}$ & $971$ \\
    & $\tilde{b}_{3}^{(1)}, \tilde{b}_{3}^{(2)}$ &  1831, 2270 & & $A^{0}$ & $970$ \\
    & $\tilde{t}_{3}^{(1)}, \tilde{t}_{3}^{(2)}$ & 1310, 1850  & & $H^{\pm}$ & $974$ \\
    \hline
    \multirow{3}{*}{Sleptons} & $\tilde{L}_{1,2} $ & 2899 & \multirow{4}{*}{Neutralinos} & $\tilde{N}^{0}_{1}$ & $289$  \\
    & $ \tilde{\tau}_{1,2}$ & 2837  &  & $\tilde{N}^{0}_{2}$ & $577$\\
    & $ \tilde{\tau}_{3}^{(1)}, \tilde{\tau}_{3}^{(2)} $ & 2768, 2865 &  & $\tilde{N}^{0}_{3}$ & $1150$\\ \cline{1-3}
    Charginos & $\tilde{\chi}^{\pm}$, $\tilde{\chi}^{\prime\pm}$ & $ 577, 1155 $ & & $\tilde{N}^{0}_{4}$ & $1155$\\
    \hline
    Gluinos & $\tilde{g}$ & $ 2102 $ & $Z^{\prime}$ & $A_{B-L}, \tilde{A}_{B-L}$ & $3005, 3096$ \\     
    \hline    
  \end{tabular}
\end{center}
    \caption{  {\scriptsize The predicted spectrum at point (S) in Figure 10.  The tilde denotes the superpartner of the respective particle.  The superpartners of left-handed fields are depicted by an upper case label whereas the lower case is used for right-handed fields.  The mixing between the third family left- and right-handed scalar fields is incorporated.   } }
\label{tableoverlap}
\end{table}
experimental lower bounds are depicted in the indicated colors. Any point in the intersection area, shown in dark brown, has broken electroweak symmetry and an acceptable mass spectrum. 
Importantly, however, note the dotted line dividing this plane. We find that the $m^{2}>0$ mass condition is satisfied everywhere to the left of this line, whereas the $m_{D_{3}}^{2}<0$ constraint holds at all points to the right--consistent with (D) being a point in the intersection of the blue and red regions.   The dotted line is vertical since, to leading order the $D_{3}$ mass squared, although a function of $\tan\beta$, is independent of $c_{\mu}(0)$.  The sparticle and Higgs mass spectrum for point (S) in the allowed region is presented in Table 5.

\subsection{$m_{e_{3}}^{2}<0$}

\subsubsection*{The Initial Conditions and Potential Energy for $m_{\nu_{3}}^{2}<0$ and $m_{e_{3}}^{2}<0$:}

As discussed in Subsection 3.8, to guarantee that the $B$-$L$ vacuum is a stable local minimum, we impose the constraint that the effective squared masses of all squarks and sleptons evaluated at $\langle \nu_{3} \rangle$ are positive over the entire scaling range. Similarly to the left-handed squark mass condition \eqref{77}, imposing positivity on the effective right-handed down slepton masses at the $B$-$L$ breaking VEV $\langle \nu_{3} \rangle$, that is,
\begin{equation}
\langle m_{e_{i}}^{2} \rangle = m_{e_{i}}^{2} +\frac{3}{4}g_{4}^{2} \langle \nu_{3} \rangle^{2} >0 \ ,
\label{95}
\end{equation}
does {\it not} require that $m_{e_{i}}^{2}$ be positive. In general, one or more of these soft squared masses can be negative. Recall from \eqref{21},\eqref{36} that we have assumed that all left-handed and right-handed down sleptons have a universal initial mass. This is similar to the initial condition on squark  masses. Unlike the squarks, however, the down-Yukawa couplings of sleptons are all too small to greatly effect the RGE running of their soft masses. It follows that, at a low scale, the three families of right-handed down sleptons mass squares tend to be all positive or all negative. Splitting this degeneracy, for example, to drive only $m_{e_{3}}^{2}<0$,  requires considerable fine-tuning. Therefore, if one wishes to consider the case where only the third family squared mass turns negative, it is necessary to alter the initial slepton mass conditions given in Section 3. This is easily accomplished as follows.

As discussed in Subsection 3.3, the boundary conditions for the RGEs of the Higgs, squarks and sleptons squared masses are greatly simplified if one chooses the initial soft masses so that both ${\cal{S}}(0)=0$ and ${\cal{S}}'_{0}(0)=0$, with ${\cal{S}}$ and ${\cal{S}}'_{0}$ given in \eqref{19} and \eqref{23} respectively. Hence, in this paper we always choose the initial parameters to satisfy these two conditions. However, the specific choices made in Subsection 3.3 were overly constraining, since they imposed unification of all three families of squarks and sleptons, whereas the unification of each family separately is sufficient. In particular, condition \eqref{21} sets
\begin{equation}
m_{L_{i}}(0)^{2}=m_{e_{j}}(0)^{2}
\label{96}
\end{equation}
for all $i,j=1,2,3$. This leads to the difficulty discussed above. However, this constraint can easily be weakened. The simplest example is to take
\begin{equation}
m_{L_{1,2}}(0)^{2}=m_{e_{1,2}}(0)^{2} \ , \quad m_{L_{3}}(0)^{2}=m_{e_{3}}(0)^{2}
\label{97}
\end{equation}
which clearly continues to solve both ${\cal{S}}(0)=0$ and ${\cal{S}}'_{0}(0)=0$. 
Expression \eqref{36} then generalizes to 
\begin{equation}
m_{L_{1,2}}(0) = m_{e_{1,2}}(0) = c_{e_{1,2}}(0) {\cal M}  \ ,
~~m_{L_{3}}=m_{e_{3}}(0) = c_{e_{3}}(0) {\cal M} \ .
 \label{98}
\end{equation}
In terms of these parameters, \eqref{45} becomes
\begin{equation}
{\cal{S}}_{1}'(0)=(1+2C^{2}-2A^{2}-A_{3}^{2})m_{\nu_{3}}(0)^{2} \ ,
\label{99}
\end{equation}
where
\begin{equation}
C=\frac{c_{\nu_{1,2}}(0)}{c_{\nu_{3}}(0)} \ , \quad A=\frac{c_{e_{1,2}}(0)}{c_{\nu_{3}}(0)} \ , \quad A_{3}=\frac{c_{e_{3}}(0)}{c_{\nu_{3}}(0)} \ .
\label{100}
\end{equation}
To stay as close as possible to our previous analysis, we continue to use the values
\begin{equation}
A=\sqrt{6} \ , \quad C=9.12
\label{101}
\end{equation}
introduced in \eqref{51} and \eqref{53} respectively. In addition, let us choose
\begin{equation}
A_{3}=\sqrt{3} \ ,
\label{102}
\end{equation}
thus minimally changing the value of  \eqref{47} from 5 to 5.1. It follows that 
equations \eqref{48}, \eqref{54}, and the conclusions thereof for $U(1)_{B-L}$ breaking, do not change substantially. Similarly, equation \eqref{52} for $i=1,2$ is minimally altered to
\begin{equation}
m_{e_{1,2}}(t_{B-L})^{2}=((\sqrt{6})^{2}-5.1)~m_{\nu_{3}}(0)^{2}= 0.9~m_{\nu_{3}}(0)^{2} \ .
\label{103}
\end{equation}
However, we now find that
\begin{equation}
m_{e_{3}}(t_{B-L})^{2}=((\sqrt{3})^{2}-5.1)~m_{\nu_{3}}(0)^{2}=-2.1~m_{\nu_{3}}(0)^{2} \ .
\label{104}
\end{equation}
That is, splitting the slepton coefficient into $A=\sqrt{6}$ and $A_{3}=\sqrt{3}$ allows the mass squares of the first two families to remain positive while constraining $m_{e_{3}}^{2}<0$, as desired. 
Henceforth, \eqref{55} is replaced by
 \begin{equation}
 c_{e_{1,2}}(0) = \sqrt{6} \ c_{\nu_{3}}(0) \ , \quad  c_{e_{3}}(0) = \sqrt{3} \ c_{\nu_{3}}(0) \ , \quad  c_{\nu_{1,2}}(0) = 9.12 \ c_{\nu_{3}}(0) \ . 
  \label{105}
 \end{equation}
Despite these changes in the initial conditions, $c_{q}(0)$, $c_{\nu_{3}}(0)$, $c_{\mu}(0)$ and 
$\tan\beta$   in \eqref{63} remain the four independent parameters of our analysis.

The new set of initial parameters just discussed allows for the possibility that, at the electroweak scale, all soft squared masses are positive with the exception of $m_{\nu_{3}}^{2}<0$ and $m_{e_{3}}^{2}<0$. The relevant potential for discussing the vacuum of $\nu_{3}$ and $e_{3}$ is given by
\begin{equation} 
V(\nu_{3},e_{3})=m_{\nu_{3}}^{2}|\nu_{3}|^{2}+m_{e_{3}}^{2}|e_{3}|^{2}+\frac{g_{B-L}^{2}}{2}(|\nu_{3}|^{2}+|e_{3}|^{2})^{2} +\frac{g_{Y}^{2}}{2}|e_{3}|^{4} \ .
\label{106}
\end{equation}
The first two terms in the potential are the soft supersymmetry breaking mass terms in \eqref{12}, while the third and fourth terms are supersymmetric and arise from the $D_{B-L}$ and $D_{Y}$ in \eqref{10} and \eqref{9} respectively. Contributions to \eqref{106} from the relevant Yukawa couplings in \eqref{6} are suppressed, since $\lambda_{\nu_{3}}$ and $\lambda_{e_{3}}$ are of order $10^{-10}$ and $10^{-2}$ respectively. Hence, we ignore them. 

If both $m_{\nu_{3}}^{2}<0, m_{e_{3}}^{2}<0$ at the electroweak scale, then the potential is unstable at the origin of field space and has two other local extrema at
\begin{equation}
\langle \nu_{3} \rangle^{2}=-\frac{m_{\nu_{3}}^{2}}{g_{B-L}^{2}} , \quad \langle e_{3} \rangle=0
\label{107}
\end{equation}
and
\begin{equation}
\langle \nu_{3}\rangle= 0, \quad \langle e_{3}\rangle^{2}=-\frac{m_{e_{3}}^{2}}{g_{B-L}^{2}+g_{Y}^{2}} 
\label{108}
\end{equation}
respectively. Using these, potential \eqref{106} can be rewritten as
\begin{eqnarray} 
&& V(\nu_{3},e_{3})= \frac {g_{B-L}^{2}}{2}(|\nu_{3}|^{2}-\langle \nu_{3}\rangle^{2})^{2} +g_{B-L}^{2} |\nu_{3}|^{2} |e_{3}|^{2} \nonumber \\
&& \qquad \qquad +\frac {g_{B-L}^{2} 
+g_{Y}^{2}}{2}(|e_{3}|^{2}-\langle e_{3}\rangle^{2})^{2} \ .
\label{109}
\end{eqnarray}

Let us analyze these two extrema. Both have positive masses in their radial directions. At the sneutrino vacuum \eqref{107}, the mass squared in the $e_{3}$ direction is given by
\begin{equation}
m_{e_{e}}^{2}|_{\langle \nu_{3}\rangle}= g_{B-L}^{2}\langle \nu_{3}\rangle^{2}-(g_{B-L}^{2}+g_{Y}^{2})\langle e_{3}\rangle^{2}= |m_{\nu_{3}}|^{2}-|m_{e_{3}}|^{2} \ ,
\label{110}
\end{equation}
whereas at the stau vacuum \eqref{108}, the mass squared in the $\nu_{3}$ direction is 
\begin{equation}
m_{\nu_{3}}^{2}|_{\langle e_{3}\rangle}= g_{B-L}^{2}\langle e_{3}\rangle^{2}-g_{B-L}^{2} \langle \nu_{3}\rangle^{2}= |m_{e_{3}}|^{2}(1+\frac{g_{Y}^{2}}{g_{B-L}^{2}})^{-1}-|m_{\nu_{3}}|^{2} \ .
\label{111}
\end{equation}
Note that either \eqref{110} or \eqref{111} can be positive, but not both. To be consistent with the hierarchy solution, we want \eqref{107} to be a stable minimum. Hence, we demand $m_{e_{3}}^{2}|_{\langle \nu_{3}\rangle}>0$ or, equivalently, that
\begin{equation}
|m_{\nu_{3}}|^{2}>|m_{e_{3}}|^{2} \ .
\label{112}
\end{equation}
We will impose  \eqref{112} as an additional condition for the remainder of this subsection. It then follows from \eqref{111} that $m_{\nu_{3}}^{2}|_{\langle e_{3}\rangle}<0$ and, hence, the stau extremum \eqref{108} is a saddle point. As a consistency check, note that $V|_{\langle\nu_{3}\rangle}<V|_{\langle e_{3}\rangle}$ if and only if 
\begin{equation}
g_{B-L}^{2}\langle \nu_{3}\rangle^{4}>(g_{B-L}^{2}+g_{Y}^{2})\langle e_{3}\rangle^{4} 
\label{113}
\end{equation}
or, equivalently, 
\begin{equation}
|m_{\nu_{3}}|^{2}>|m_{e_{3}}|^{2}(1+\frac{g_{Y}^{2}}{g_{B-L}^{2}})^{-1/2} \ .
\label{114}
\end{equation}
This follows immediately from constraint \eqref{112}. Finally, note that the potential descends monotonically along a path  ${\cal{C}}$ from the saddle point at \eqref{108} to the absolute minimum at \eqref{107}. Solving the $\frac{\partial V}{\partial {e_{3}}}=0$ equation, this curve is found to be
\begin{equation}
|e_{3}|_{{\cal{C}}}=( \langle e_{3}\rangle^{2}-|\nu_{3}|^{2}(1+\frac{g_{Y}^{2}}{g_{B-L}^{2}})^{-1})^{1/2} \ .
\label{115}
\end{equation}
Note that it begins at $ \langle e_{3}\rangle$ for $\nu_{3}=0$ and continues until it tangentially intersects the $e_{3}=0$ axis at $|\nu_{30}|=\frac{|m_{e_{3}}|}{|m_{\nu_{3}}|}\langle \nu_{3}\rangle$. From here, the path continues down this axis to the stable minimum at \eqref{95}.
We conclude that at the electroweak scale the absolute minimum of potential \eqref{106} occurs at the sneutrino vacuum given in \eqref{107}. 

\subsubsection*{Phenomenologically Allowed Regions and the Mass Spectrum:}

In this subsection, we analyze our results subject to the following additional conditions.

\begin{itemize}

\item The {\it third family right-handed slepton soft mass squared will be constrained to be negative}, that is, 
$m_{e_{3}}^{2}<0$.  All other squark and slepton soft squared masses are positive over the entire scaling range, with the exception of $m_{\nu_{3}}^{2}$.

\item To ensure that the $B$-$L$ breaking VEV is the absolute minimum, we impose condition \eqref{112},
\begin{equation}
|m_{\nu_{3}}|^{2}>|m_{e_{3}}|^{2} \ ,
\label{116}
\end{equation}
at the electroweak scale.

\end{itemize}
We will refer to these two conditions collectively as the $m_{e_{3}}^{2}<0$ mass condition.

As discussed in previous subsections, we proceed by scanning over  the entire  $c_{q}(0)$-$c_{\nu_{3}}(0)$ plane, at each point analyzing the associated $c_{\mu}(0)$-$\tan\beta$ plane to see if an allowed region exists.  
The results are shown in Figure 11. As in Figures 2 and 6, the white region indicates points whose corresponding  $c_{\mu}(0)$-$\tan\beta$ plane contains {\it no} locus of electroweak symmetry breaking,
whereas the yellow area represents points whose $c_{\mu}(0)$-$\tan\beta$ plane {\it has} a region where electroweak symmetry is broken. Finally, each point in the green area has a phenomenologically allowed region in its corresponding  $c_{\mu}(0)$-$\tan\beta$ plane satisfying the $m_{e_{3}}^{2}<0$ mass condition. 
Since some of the initial parameters are now different to allow for a negative stau squared mass, this green region cannot be superimposed with the blue and red regions discussed previously. Let us analyze the properties of an arbitrary point in the green area. 
For example, consider point (E) shown in Figure 11. This has the values
\begin{equation}
c_{q}(0) = 1.1 \ , \quad c_{\nu}(0) = 0.5 \ . 
\label{117}
\end{equation}
For this point, the regions of the $c_{\mu}(0)$-$\tan\beta$ plane corresponding to the different constraints, as well as their intersection, are shown in Figure 12. 
The $m_{e_{3}}^{2}<0$ mass condition is satisfied everywhere in the depicted regime.

In the yellow region both inequalities \eqref{64} and \eqref{65} are satisfied, leading to stable electroweak breaking vacua. There are two separated areas where 
\begin{figure}
 \centering
 \includegraphics[scale=0.4]{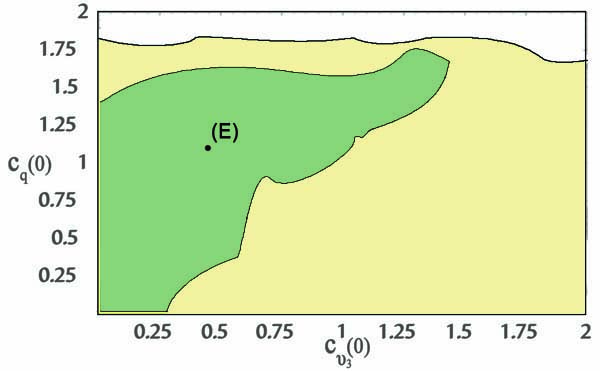}
 \caption{{\scriptsize  A plot of the $c_{q}(0)$-$c_{\nu_{3}}(0)$ plane showing physically relevant areas.  The yellow and white indicate points whose corresponding $c_{\mu}(0)$-$\tan\beta$ plane does and does not contain a region of electroweak symmetry breaking respectively. Within the yellow area, the green shading contains all points whose $c_{\mu}(0)$-$\tan\beta$ plane has a non-vanishing region satisfying all experimental sparticle and Higgs bounds and for which $m_{e_{3}}^{2}<0$. (E) indicate the point analyzed in detail in the text.}}
 \label{fig:CqCnu34}
\end{figure}
\begin{figure}
 \centering
 \includegraphics[scale=0.3535]{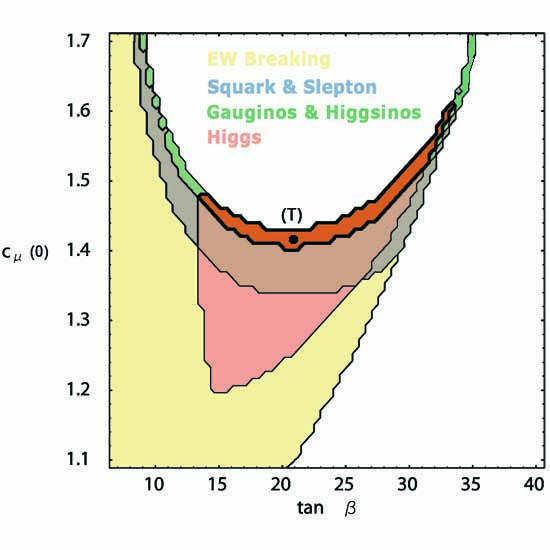}
 \caption{{\scriptsize The $c_{\mu}(0)$-$\tan\beta$ plane corresponding to the point  $c_{q}(0)= 1.1, c_{\nu_{3}}(0)=0.5$. The yellow and white regions indicate where electroweak symmetry is and is not broken respectively. The individual regions satisfying the present experimental bounds for squarks and sleptons, gauginos and Higgs fields are shown in the indicated colors. The dark brown area is their mutual intersection where electroweak symmetry is broken and all experimental mass bounds are satisfied. We present our predictions for the sparticle and Higgs masses at point (T). }}
 \label{}
\end{figure}
electroweak breaking does not occur. As before, anywhere in the {\it upper} white region the first inequality \eqref{64} is violated, while \eqref{65} continues to be satisfied. This indicates stable vacua, but with {\it vanishing} Higgs VEVs. It follows that the boundary between the yellow and {\it upper} white regions corresponds to the vanishing of $M_{Z}^{2}$ in \eqref{61}.
However, as at point (B), for example, the {\it lower right} white region shown in Figure 12 violates 
{\it both} constraints \eqref{64} and \eqref{65}. Hence, the origin of Higgs space is a local maximum and the potential energy is unbounded from below. There are no stable vacua in this regime.

The regions where the squarks/sleptons, gauginos and Higgs exceed their experimental lower bounds are depicted in the indicated colors.
Any point in the intersection area, shown in dark brown, has broken electroweak symmetry and an acceptable mass spectrum. As an example, consider the point (T) indicated in this region. Our calculated values for the squark, slepton, Higgs and gaugino masses are presented in Table 6. 
\\
\begin{table}
\begin{center}
\hspace*{-0.12 in}
  \begin{tabular}{ | c | c | c | c | c | c | }
    \hline
    Particle & Symbol  & Mass [GeV] & Particle & Symbol  & Mass [GeV] \\  \hline
    \multirow{4}{*}{Squarks} & $ \tilde{Q}_{1,2} $ & 1176  & \multirow{4}{*}{Higgs}  & $h^{0}$ & 132 \\
    & $\tilde{t}_{1,2}, \tilde{b}_{1,2}$ & 1136, 1184 & & $H^{0}$ & 640 \\
    & $\tilde{b}_{3}^{(1)}, \tilde{b}_{3}^{(2)}$ &  932, 1057 & & $A^{0}$ & 640 \\
    & $\tilde{t}_{3}^{(1)}, \tilde{t}_{3}^{(2)}$ & 770, 986  & & $H^{\pm}$ & 645 \\
    \hline
    \multirow{3}{*}{Sleptons} & $\tilde{L}_{1,2} $ & 806 & \multirow{4}{*}{Neutralinos} & $\tilde{N}^{0}_{1}$ & 146  \\
    & $ \tilde{\tau}_{1,2}$ & 768  &  & $\tilde{N}^{0}_{2}$ & 290 \\
    & $  \tilde{\tau}_{3}^{(1)}, \tilde{\tau}_{3}^{(2)} $ & 519, 606 &  & $\tilde{N}^{0}_{3}$ & 726 \\ \cline{1-3}
    Charginos & $\tilde{\chi}^{\pm}_{1}, \tilde{\chi}^{\pm}_{2} $ & 290, 734 & & $\tilde{N}^{0}_{4}$ & 734 \\
    \hline
    Gluinos & $\tilde{g}$ & $ 1066 $ & $Z^{'}$ Boson & $A_{B-L}, \tilde{A}_{B-L}$ & 1776, 1860 \\     
    \hline    
  \end{tabular}
\end{center}
    \caption{  {\scriptsize The spectrum at point (T) in Figure 12.  The tilde denotes the superpartner of the respective particle.  The superpartners of left-handed fields are depicted by an upper case label whereas the lower case is used for the right-handed fields.  The mixing between the third family left- and right-handed scalar fields is incorporated.    } }
\label{  }
\end{table}

\subsubsection*{The $B$-$L$/Electroweak Hierarchy:}

We have determined the subspace of the $c_{q}(0)$-$c_{\nu_{3}}(0)$ plane for which each point has a region in the corresponding $c_{\mu}(0)$-$\tan\beta$ plane satisfying 1) the $m_{e_{3}}^{2}<0$ mass condition with 2) broken electroweak symmetry and 3) phenomenologically acceptable squark, slepton, Higgs and gaugino masses. Given such a point in the 
 $c_{q}(0)$-$c_{\nu_{3}}(0)$ plane and choosing a point in the acceptable region in the 
$c_{\mu}(0)$-$\tan\beta$ plane, one can analyze the $B$-$L$/electroweak hierarchy for these initial values. The analysis proceeds exactly as in previous subsections, so we simply present the results.

For point (E) in Figure 11, we have computed the hierarchy everywhere in the dark brown area of Figure 12. We find that this takes values of 7.60-7.74 along the lower boundary of the allowed region. Note that below this boundary at least one of the gaugino or Higgs masses violates their experimental bound.  Hence, the lower values of the hierarchy are determined from the experimental data. On the other hand, as one approaches the boundary with the upper white region, the hierarchy becomes infinitely large for the reasons previously discussed. 
It follows that within the phenomenologically acceptable region, {\it any value of the $B$-$L$ hierarchy in the range $7.60 \lesssim M_{A_{B-L}}/M_{Z} < \infty$
can be attained}.

Another way to analyze this data is to pick a specific point in the allowed region and to compute 
\eqref{73} as a function of $c_{\mu}(0)$ along the fixed $\tan\beta$ line passing through it. For concreteness, choose the point (T) with $\tan\beta=22$ for which we calculated the mass spectrum in Table 6.
We find that the hierarchy begins at $ M_{A_{B-L}}/M_{Z}=7.65$ at the experimentally determined lower boundary, rises slowly to 
$ M_{A_{B-L}}/M_{Z}\sim 30$ across most of the region, and then rapidly diverges to infinity as one approaches the upper boundary. Approaching both the lower and, especially, the upper boundary requires fine-tuning of $c_{\mu}(0)$. For ``typical'' values of $c_{\mu}(0)$, the hierarchy is {\it naturally} in the range
\begin{equation}
10 \lesssim  \frac{M_{A_{B-L}}}{M_{Z}} \lesssim 30 \ .
\label{118}
\end{equation}

\subsection{Summary}

We first note that the above classification of vacua using the sign of $m_{Q_{i}}^{2}$ and 
$m_{e_{i}}^{2}$ is complete. The only other squared masses are for right-handed squarks and left-handed sleptons, which enter the effective masses at the $B$-$L$ breaking VEV $\langle \nu_{3} \rangle$ as
\begin{equation}
\langle m_{u_{i}}^{2} \rangle= m_{u_{i}}^{2}-\frac{1}{4}g_{4}^{2}\langle \nu_{3} \rangle^{2} \ , \quad
\langle m_{d_{i}}^{2} \rangle= m_{d_{i}}^{2}-\frac{1}{4}g_{4}^{2}\langle \nu_{3} \rangle^{2} 
\label{118A}
\end{equation}
and
\begin{equation}
\langle m_{L_{i}}^{2} \rangle= m_{L_{i}}^{2}-\frac{3}{4}g_{4}^{2}\langle \nu_{3} \rangle^{2}
\label{118B}
\end{equation}
respectively. Since all of these effective masses must be positive to ensure that the vacuum is a stable minimum, it follows from the minus signs in each expression that $m_{u_{i}}^{2}$,$m_{d_{i}}^{2}$, and 
$m_{L_{i}}^{2}$ must all be positive. Therefore, all $m^{2}>0$, $m_{Q_{i}}^{2}<0$, and $m_{e_{i}}^{2}<0$ in subsections 4.1, 4.2 and 4.3 respectively are the only possibilities.

From the above analysis, several broad conclusions can be made. For the reasons discussed above, we limited our search to the four-dimensional space of parameters listed in \eqref{63}. By combining the results in the $m^{2}>0$, $m_{Q_{i}}^{2}<0$, and $m_{e_{i}}^{2}<0$ regimes, we can find the generic region of this parameter space for which one obtains a phenomenologically acceptable vacuum. 
The full range of allowed values for the $c_{q}(0)$ and $c_{\nu_{3}}(0)$ parameters were presented in Figures \ref{fig:CqCnu124} and \ref{fig:CqCnu34}.  From these, we observe a maximum range of 
\begin{equation}
0 < c_{q}(0) < 1.8 \ , \quad 0 < c_{\nu_{3}}(0) < 1.5 \ .
\label{118C}
\end{equation}
Similarly, by examining the $c_{\mu}(0)$-$\tan\beta$ plane over the allowed values of $c_{q}(0)$ and $c_{\nu_{3}}(0)$, the range of phenomenologically allowed values is found to be 
\begin{equation}
0.8 < c_{\mu} (0)< 1.75 \ , \quad 8 < \tan\beta < 33.   
\label{118D}
\end{equation}
To obtain this result, we computed the allowed regions for numerous points in the $c_{q}(0)$-$c_{\nu_{3}}(0)$ plane including, but not limited to,  (A)-(E) presented in the text.
Thus, even with our restrictive premises in Section 3, a phenomenologically viable $B$-$L$ MSSM vacuum exhibiting an acceptable hierarchy occurs for a reasonably wide space of initial parameters. Lifting some of the above constraints, such as allowing {\it all} $m_{Q_{i}}^{2}<0$ instead of just $m_{Q_{3}}^{2}<0$, will clearly allow a significant enlargement of the acceptable initial parameter space.

\section{Some $\langle \nu_{3} \rangle \neq 0$ Phenomenology}

The results presented in this paper allow one to compute any quantity in our $B$-$L$ MSSM theory at any energy scale. In particular, we have shown that for a wide range of initial conditions there is a stable vacuum which breaks both $B$-$L$ and electroweak symmetry with an acceptable sparticle and Higgs mass spectrum and $B$-$L$/electroweak hierarchy. These are important necessary conditions on the theory, but are not sufficient to guarantee that it is phenomenologically viable. In this section, we explore two more important constraints arising from lepton number and baryon number violation respectively.

\subsection{Lepton Number Violation}

The most general superpotential invariant under gauge group $SU(3)_{C} \times SU(2)_{L} \times U(1)_{Y} \times U(1)_{B-L}$ is presented in \eqref{6}. Assuming a flavor diagonal basis, the superpotential becomes
\begin{equation}
W=\mu H\bar{H} +{\sum_{i=1}^{3}}\left(\lambda_{u, i} Q_{i}Hu_{i}+\lambda_{d, i} Q_{i}\bar{H}d_{}+\lambda_{\nu, i} L_{i}H\nu_{i}+\lambda_{e, i} L_{i}\bar{H}e_{i}\right).
\label{119}
\end{equation}
Recall that since $U(1)_{B-L}$ contains matter parity, the dangerous lepton and baryon number violating terms in \eqref{7} are forbidden. Note, however, that these results are only valid at high scales where the gauge symmetry, in particular $U(1)_{B-L}$, is exact. At low energy-momentum the gauged $B$-$L$ symmetry is spontaneously broken, potentially allowing these operators to ``grow back''.
This can be analyzed by expanding the third family right-handed sneutrino around its VEV, that is, 
let $\nu_{3}= \langle \nu_{3} \rangle+\nu_{3}'$. Note that
\begin{equation}
\mu H\bar{H} +\lambda_{\nu_{3}} L_{3}H\nu_{3}=\mu H(\bar{H}+\epsilon_{3}L_{3}) +\dots \ ,
\label{120}
\end{equation}
where
\begin{equation}
\epsilon_{3}=\lambda_{\nu_{3}}\frac{\langle \nu_{3} \rangle}{\mu} \ .
\label{121}
\end{equation}
This motivates performing a rotation of the down Higgs and third family lepton doublet superfields given, to leading order, by
\begin{equation}
\bar{H}'=\bar{H}+\epsilon_{3}L_{3} \ , \quad L_{3}'=L_{3}-\epsilon_{3}\bar{H} \ .
\label{122}
\end{equation}
Written in terms of these new superfields, and then dropping the $'$ for simplicity, 
the superpotential becomes
\begin{equation}
{\cal{W}}=W+\epsilon_{3}{\sum_{i=1}^{3}}\lambda_{e,i}L_{3}L_{i}e_{i}+\epsilon_{3} {\sum_{i=1}^{3}}\lambda_{d,i}L_{3}Q_{i}d_{i} \ ,
\label{123}
\end{equation}
where $W$ is given in \eqref{119}. As expected, the lepton number violating terms of the form 
\begin{equation}
L_{3}L_{i}e_{i} \ , \quad L_{3}Q_{i}d_{i}
\label{124}
\end{equation}
have grown back. Note, however, that the baryon violating terms $u_{i}d_{j}d_{k}$ have {\it not} been regenerated by the right-handed sneutrino VEV. In this subsection, we analyze the lepton violating interactions in \eqref{123}. The question of baryon violation will be discussed in the next subsection.

It is well-known \cite{Sakharov}-\cite{Dreiner} that the lepton number violating terms in \eqref{123} influence the baryon asymmetry at high temperature in the early universe. The requirement that the existing baryon asymmetry is not erased before the electroweak phase transition typically implies \cite{Buchmuller} that
\begin{equation}
\left(\frac{\epsilon_{3}}{10^{-6}}\right)\left(\frac{\tan\beta}{10}\right) \lesssim 1 \ .
\label{125}
\end{equation}
Parameter $\epsilon_{3}$ for a given $\tan\beta$ can be explicitly evaluated for any $B$-$L$ MSSM vacuum using \eqref{121}. For example, consider the vacuum specified by point (P) in Figure 1. This has the values $\tan\beta=18$ and $c_{\mu}(0)=1.0$. RG running $c_{\mu}$ down to the electroweak scale, 
we find that $c_{\mu}(t_{EW})=0.855$ and, hence, that $\mu=0.855~{\cal{M}}$. 
The VEV of $\nu_{3}$ can be obtained  using \eqref{107}.
For the parameters of this vacuum, $\langle \nu_{3} \rangle=4.433~{\cal{M}}$. 
Finally, unless otherwise stated we will take the third family neutrino Yukawa coupling to be $\lambda_{\nu_{3}} \simeq10^{-10}$. This choice is motivated by the constraints on proton decay and will be discussed in the following subsection.
Putting these values into \eqref{121} gives $\epsilon_{3} 
\simeq5.185 \times 10^{-10}$ and, hence,
\begin{equation}
\left(\frac{\epsilon_{3}}{10^{-6}}\right)\left(\frac{\tan\beta}{10}\right) \simeq 0.933 \times 10^{-3}\ ,
\label{126}
\end{equation}
well below the necessary bound of unity. If we sample over all five vacua (P),(Q),(R),(S),(T) specified above, we find that 
\begin{equation}
0.688 \times 10^{-3}\lesssim \left(\frac{\epsilon_{3}}{10^{-6}}\right)\left(\frac{\tan\beta}{10}\right) \lesssim 1.04 \times 10^{-3} \ ,
\label{127}
\end{equation}
in each case below the bound in \eqref{125}. 
We conclude that our $B$-$L$ MSSM theory satisfies the conditions for baryon asymmetry.

As discussed in \cite{Buchmuller, Takayama}, theories with lepton number violating interactions of the form in \eqref{123} naturally solve many fundamental cosmological problems if the gravitino is the lightest supersymmetric partner (LSP). The lifetime of the gravitino is then 
found to be \cite{Buchmuller}
\begin{equation}
\tau_{3/2}
\simeq 10^{28}s \left(\frac{\epsilon_{3}}{10^{-7}}\right)^{-2}\left(\frac{\tan\beta}{10}\right)^{-2} \left(\frac{m_{3/2}}{10~GeV} \right)^{-3} \ . 
\label{128}
\end{equation}
Assuming that the lightest neutralino is the next-to-lightest superparticle (NLSP), one finds that
\begin{equation}
\tau_{NLSP}
\simeq 10^{-9}s \left(\frac{\epsilon_{3}}{10^{-7}}\right)^{-2}\left(\frac{\tan\beta}{10}\right)^{-2} \left(\frac{m_{\tilde{N}}}{200~GeV} \right)^{-3}  \ .
\label{129}
\end{equation}

These results are relevant to the $B$-$L$ MSSM theory discussed in this paper. First, it is possible to 
choose parameters so that the gravitino is, indeed, the LSP. Second, as can be seen from the spectra presented in the previous section at five different points, the lightest standard model sparticle is always the neutralino $\tilde{N}_{1}^{0}$. As an example, let us compute the lifetimes of the gravitino and the lightest neutralino at the point (P) in Figure 1. From Table 2, we see that $m_{\tilde{N}_{1}^{0}}=
147~GeV$. Hence, adjusting the gravitino mass to be, say, $m_{3/2}=80~GeV$, makes it the LSP while $\tilde{N}_{1}^{0}$ is the NLSP. Using this value for $m_{3/2}$ and \eqref{126}, it then follows from \eqref{128} that
\begin{equation}
\tau_{3/2} \simeq 3.45 \times 10^{28}s \ .
\label{130}
\end{equation}
Noting that the age of the universe is typically estimated to be 13.7 billion years, that is, $4.32 \times 
10^{17}$ seconds, we see that the gravitino lifetime greatly exceed this. Hence, the gravitino is the primary candidate for dark matter. On the other hand, using $m_{\tilde{N}_{1}^{0}}=100~GeV$ and \eqref{126}, we find from \eqref{129} that
\begin{equation}
\tau_{NLSP} \simeq 1.77 \times 10^{-6}s \ ,
\label{131}
\end{equation}
much to short-lived to form dark matter. Let us extend these results by evaluating the LSP and NLSP lifetimes at the five points (P),(Q),(R),(S),(T) specified above. Choosing $m_{3/2}$ to be $20~GeV$ lighter than the corresponding $\tilde{N}_{1}^{0}$ mass, we find using \eqref{127} that
\begin{equation}
1.65 \times 10^{28}s \lesssim \tau_{3/2} \lesssim 2.47 \times 10^{29}s
\label{132}
\end{equation}
and
\begin{equation}
1.45 \times 10^{-6}s \lesssim \tau_{NLSP} \lesssim 5.52 \times 10^{-6}s \ .
\label{133}
\end{equation}
We conclude that for a gravitino LSP, our $B$-$L$ MSSM theory  has a long-lived gravitino consistent with it being dark matter, as well as an NLSP which decays very rapidly.

\subsection{Baryon Number Violation}

Recall that since $U(1)_{B-L}$ contains matter parity, the dangerous lepton and baryon number violating  interactions in (7) are disallowed in the high energy superpotential. At much lower scales, the $B$-$L$ violating VEV $\langle \nu_{3} \rangle$ can potentially re-introduce these terms. As discussed above, however, this VEV induces from the {\it dimension four} superpotential
only the lepton number violating interactions in \eqref{123}. The baryon number violating $u_{i}d_{j}d_{k}$ terms are {\it not} regenerated. Therefore, to this order, baryon number is conserved and the proton is completely stable. 
However, the superpotential can contain $B$-$L$ invariant {\it higher dimensional} terms proportional to $\nu_{3}u_{i}d_{j}d_{k}$. When the sneutrino develops a non-zero VEV, this generates effective dimension four operators of the form
\begin{equation}
\lambda_{ijk}'' u_{i}d_{j}d_{k} \ ,
\label{134}
\end{equation}
where 
\begin{equation}
\lambda_{ijk}''=\gamma_{ijk} {\frac{\langle \nu_{3} \rangle}{M_{c}}} \ ,
\label{134a}
\end{equation}
$\gamma_{ijk}=-\gamma_{ikj}$ are dimensionless parameters, and
$M_{c}$ is the compactification scale which we loosely identify with $M_{u}$ in \eqref{16}. For proton decay, the relevant operators are
\begin{equation}
\lambda_{11k}'' u_{1}d_{1}d_{k} 
\label{134b}
\end{equation}
with $k$ restricted to $k=2,3$.

\begin{figure}
 \centering
 \includegraphics[scale=0.7]{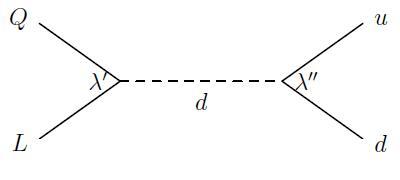}
 \caption{ Effective operators generated by the dimension 4 interactions $\lambda'_{ijk}L_{i}Q_{j}d_{k}$ and $\lambda_{ijk}'' u_{i}d_{j}d_{k}$.  When the external fields are light families, these graphs induce nucleon decay.  The solid lines represent fermions while the dashed line represents scalar propagators. }
 \label{fig:ProtonDecay}
\end{figure}

Lepton number violating terms of the form $\lambda'_{ijk}L_{i}Q_{j}d_{k}$ can combine with the baryon number violating interactions $\lambda_{ijk}'' u_{i}d_{j}d_{k}$ to produce the effective operators in Figure \ref{fig:ProtonDecay}. Generically, these operators can induce proton decay via several channels. For the specific $B$-$L$ MSSM theory in this paper, however, it follows from \eqref{123} that the relevant lepton number violating terms are restricted to
\begin{equation}
\epsilon_{3} \lambda_{d,k}L_{3}Q_{k}d_{k} 
\label{134c}
\end{equation}
with $k=2,3$. Since the $\tau^{+}$ and $B^{+}$-meson masses exceed that of the proton, in our specific theory the only potential decay channel is $p \rightarrow K^{+}+ {\bar{\nu}}_{3}$.
We find from \eqref{134a}, \eqref{134b} and \eqref{134c} that the product of the dimensionless couplings inducing this decay is
\begin{equation}
\lambda' \lambda''=\epsilon_{3} \lambda_{d,2} \gamma_{112} \frac{\langle \nu_{3} \rangle}{M_{c}} \ .
\label{138}
\end{equation}

As discussed in \cite{Barbier, Ub}, this channel will be suppressed below the experimental bound if 
\begin{equation}
\lambda' \lambda'' < {\cal{O}}(10^{-25}) \ .
\label{136}
\end{equation}
In estimating this bound, we have taken the mass of the intermediate squark in Figure 13 to be of ${\cal{O}}(1~TeV)$, corresponding to its derived values in Section 4. 
Under what conditions is \eqref{136} satisfied?
To be concrete, let us compute product \eqref{138} at the point (P) in Figure 1. As discussed above, here 
$\tan\beta=18$, $\mu=0.855{\cal{M}}$ and $\langle \nu_{3} \rangle=4.433 {\cal{M}}$. Leaving, for a moment, $\lambda_{\nu_{3}}$ arbitrary, one obtains $\epsilon_{3} \simeq 5.185 \lambda_{\nu_{3}}$. Using this and \eqref{16}, we find that for the 
$p \rightarrow K^{+}+{\bar{\nu}}_{3}$  channel 
\begin{equation}
\lambda' \lambda''=\epsilon_{3} \lambda_{d,2} \gamma_{112} \frac{\langle \nu_{3} \rangle}{M_{c}} = 6.89 \times 10^{-16} \lambda_{\nu_{3}} \gamma_{112} \ .
\label{140}
\end{equation}
Assuming, for simplicity, that $\gamma_{112}$ is of ${\cal{O}}(1)$, it follows that bound \eqref{136} will be satisfied by taking 
\begin{equation}
\lambda_{\nu_{3}} \lesssim 10^{-10} \ .
\label{140a}
\end{equation}
We arrive at a similar conclusion for each of the remaining four points (Q),(R),(S) and (T).
This explains our choice of the upper bound $\lambda_{\nu_{3}} \simeq 10^{-10}$ in the previous subsection.  Of course, choosing $\lambda_{\nu_{3}}<10^{-10}$ and/or $\gamma_{112} < 1$ will  suppresses proton decay even further below the experimental bound. 

Finally, although we relegate a detailed discussion of $B$-$L$ MSSM neutrino masses to a future publication, it follows from \eqref{lhn2} in Appendix A that the choice $\lambda_{\nu_{3}}\simeq 10^{-10}$ will generate a third family left-handed neutrino mass of order
\begin{equation}
m_{N_{3}} \simeq \frac{(\lambda_{\nu_{3}} \langle \nu_{3} \rangle)^{2}}{m_{{\tilde{N}}^{0}}}  \simeq 10^{-6} eV \ .
\label{140b}
\end{equation}
In evaluating $m_{N_{3}}$, we have taken the MSSM neutralino mass $m_{{\tilde{N}}^{0}}$ to be of 
${\cal{O}}(100 \ GeV)$ corresponding to the derived values in Section 4. The mass \eqref{140b} is consistent with an ``inverted'' hierarchy of neutrino masses, where the third family neutrino is the lightest. We emphasize that our entire analysis has been for the specific $B$-$L$ MSSM theory with diagonal Yukawa interactions and spontaneous $B$-$L$ breaking exclusively by  an expectation value of $\nu_{3}$. For generic off-diagonal couplings and $B$-$L$ breaking involving the remaining two sneutrinos, the analysis of baryon asymmetry, proton decay and the neutrino mass hierarchy can be considerably generalized. 

\section*{Appendix A - Mass Diagonalization}

The $B$-$L$ MSSM theory considered in this paper is a minimal extension of the standard supersymmetric model. Hence, portions of the analysis of the mass spectrum is similar to that of the MSSM; see for example \cite{Zc}. However, the addition of right-handed neutrino multiplets, the extra $U(1)_{B-L}$ gauge factor, the associated new soft SUSY breaking terms and the masses/field mixings induced by the $\langle \nu_{3} \rangle$ vacuum expectation value substantially complicate these calculations. In this Appendix, we briefly present the analysis of the mass spectrum used in this paper, emphasizing the differences from the MSSM.

\subsection*{Neutralinos:}

We begin by discussing the mass eigenstates of the neutralinos.  Restricting ourselves, for the time-being, to the MSSM portion of our theory,
the neutralino mass matrix will be of the form
\begin{equation}
     \begin{pmatrix}  
       M_{1} & 0  & -\frac{g_{Y} \langle \bar{H} \rangle}{\sqrt{2}}  &  \frac{g_{Y} \langle H \rangle}{\sqrt{2}}  \\
       0 & M_{2}  & \frac{g_{2} \langle \bar{H} \rangle}{\sqrt{2}}  &  -\frac{g_{2} \langle H \rangle}{\sqrt{2}}  \\
      -\frac{g_{Y} \langle \bar{H} \rangle}{\sqrt{2}} & \frac{g_{2} \langle \bar{H} \rangle}{\sqrt{2}} &  0  &  -\mu  \\
       \frac{g_{Y} \langle H \rangle}{\sqrt{2}} & -\frac{g_{2} \langle H \rangle}{\sqrt{2}}  &   -\mu & 0 \\
       \end{pmatrix} 
       \label{matrix1}
\end{equation}
in the basis $(\lambda_{Y}, \lambda_{{W}^{0}}, {\tilde{\bar{H}}}^{0}, {\tilde{H}}^{0})$.
In this matrix, the $\mu$ entries arise from the Higgsino portion of the $\mu$ term,  the $M_{i}$ entries are from the associated soft breaking terms and the off-diagonal entries are derived from the gauge superfield couplings in the Higgs kinetic energies.  The physical neutralinos  are the eigenstates of this matrix, which we label as ${{\tilde{N}}^{0}_{i}}$ for $i=1,..,4$. The associated masses are 
denoted $m_{{\tilde{N}}^{0}_{i}}$. Note that when restricted to the MSSM portion of our theory, the masses of both the left- and right-chiral neutrino fermions, $\psi_{N_{i}}$ and $\psi_{\nu_{i}}$ for $i=1,2,3$ respectively,  arise from pure Dirac  interactions and do not couple with the neutralino mass matrix.
 
The full $B$-$L$ MSSM theory discussed in this paper, however, is more complicated.
There is a fifth neutral gaugino, that is, the superpartner $\lambda_{B-L}$ of the $A_{B-L}$ boson. Although this field does not mix with the above MSSM neutralinos, it does mix with the third family right-handed neutrino through the $\nu_{3}$ kinetic energy.  It follows that a complete description requires extending the above basis to  $(\psi_{N_{3}}, \psi_{\nu_{3}}, \lambda_{B-L},\lambda_{Y}, \lambda_{{W}^{0}}, {\tilde{\bar{H}}}^{0}, {\tilde{H}}^{0})$. For this basis, the mass matrix takes the form
\begin{equation}
     \begin{pmatrix}
       0 &   \lambda_{\nu_{3}} \langle H \rangle & 0 & 0 & 0 & 0 &  \lambda_{\nu_{3}} \langle \nu_{3} \rangle \\ 
        \lambda_{\nu_{3}} \langle H \rangle & 0 & \sqrt{2} g_{B-L} \langle \nu_{3} \rangle & 0 & 0 & 0 & 0 \\
        0 & \sqrt{2} g_{B-L} \langle \nu_{3} \rangle & M_{4} & 0 & 0 & 0 & 0 \\
       0 & 0 & 0& M_{1} & 0  & -\frac{g_{Y} \langle \bar{H} \rangle}{\sqrt{2}}  &  \frac{g_{Y} \langle H \rangle}{\sqrt{2}}  \\
       0 & 0 & 0 & 0 & M_{2}  & \frac{g_{2} \langle \bar{H} \rangle}{\sqrt{2}}  &  -\frac{g_{2} \langle H \rangle}{\sqrt{2}}  \\
      0 & 0 & 0 & -\frac{g_{Y} \langle \bar{H} \rangle}{\sqrt{2}} & \frac{g_{2} \langle \bar{H} \rangle}{\sqrt{2}} &  0  &  -\mu  \\
      \lambda_{\nu_{3}} \langle \nu_{3} \rangle & 0 & 0 & \frac{g_{Y} \langle H \rangle}{\sqrt{2}} & -\frac{g_{2} \langle H \rangle}{\sqrt{2}}  &   -\mu & 0 \\
       \end{pmatrix} \ .
       \label{matrix2}
\end{equation}
We note that we have dropped all terms in this matrix proportional to the third family left-handed sneutrino expectation value. This has been calculated in \cite{tamaz,Spina} and found to be
\begin{equation}
\langle N_{3} \rangle =\left(\frac{\mu \langle {\bar{H}} \rangle-{\tilde{A}}_{\nu_{3}}\langle H \rangle}{m_{N_{3}}^{2}-g_{B-L}^{2} \langle \nu_{3} \rangle ^{2}}\right) \lambda_{\nu_{3}} \langle \nu_{3} \rangle \ .
\label{lVEV1}
\end{equation}
For the vacua in this paper, we find 
\begin{equation}
\langle N_{3} \rangle \ll \lambda_{\nu_{3}} \langle \nu_{3} \rangle
\label{lVEV2}
\end{equation}
and, hence, such terms can be safely ignored.

In the limit of vanishingly small $\lambda_{\nu_{3}}$, matrix \eqref{matrix2} splits into the MSSM neutralino matrix in \eqref{matrix1}, the $2 \times 2$ mass matrix
\begin{equation}
     \begin{pmatrix}
        0 & \sqrt{2} g_{B-L} \langle \nu_{3} \rangle \\
        \sqrt{2} g_{B-L} \langle \nu_{3} \rangle & M_{4}  \\
       \end{pmatrix} 
       \label{matrix3}
\end{equation}
mixing the right-handed neutrino $\psi_{\nu_{3}}$ and the $B$-$L$ gaugino $\lambda_{B-L}$ and a vanishing left-handed neutrino mass.
Diagonalizing \eqref{matrix3} assuming $M_{4} \ll 2\sqrt{2}g_{B-L} \langle \nu_{3} \rangle$, which is the case for the vacua described in this paper, gives
\begin{equation}
m_{\psi_{\nu_{3}'}}= M_{A_{B-L}}(1-\frac{M_{4}}{2 \sqrt{2}g_{B-L} \langle \nu_{3} \rangle}) \ , 
 m_{\lambda_{B-L}'}=M_{A_{B-L}}(1+\frac{M_{4}}{2 \sqrt{2}g_{B-L} \langle \nu_{3} \rangle}) 
\label{matrixeigs}
\end{equation}
where we have used \eqref{67}.
Note that for $M_{4} \rightarrow 0$, that is, the supersymmetric limit in this sector, 
\begin{equation}
m_{\psi_{\nu_{3}'}}=  m_{\lambda_{B-L}'}= M_{A_{B-L}} \ .
\label{equals}
\end{equation}
Hence, in addition to the $B$-$L$ vector boson and gaugino masses becoming identical, as they must, there is also a predicted degeneracy between the right-handed neutrino mass eigenstate and the $A_{B-L}$ mass. These degeneracies split as $M_{4}$ is turned on. 
Now consider $\lambda_{\nu_{3}}$ small, but non-zero. This gives small corrections to the $B$-$L$ gaugino and right-handed neutrino masses. In addition, it generates a non-vanishing left-handed neutrino mass \cite{Spina,Spinb}. For the vacua described in this paper, this is well-approximated by
\begin{equation}
m_{\psi_{N_{3}'}} \simeq 
\frac{ (\lambda_{\nu_{3}} \langle \nu_{3} \rangle)^{2}} {m_{{\tilde{N}}^{0}}} \ ,
\label{lhn2}
\end{equation}
where $m_{{\tilde{N}}^{0}}$ is a typical MSSM neutralino mass.

For the actual calculations in this paper, however, we do not use any of these approximations. Rather, we consider the complete mass matrix \eqref{matrix2} and diagonalize it numerically.

\subsection*{Charginos:}

Now consider the mass matrix for the charginos.  Note that the non-vanishing sneutrinos VEVs in the $B$-$L$ MSSM theory lead to off-diagonal chargino mixing terms which are not present in the MSSM.  However, these are proportional to either $\langle N_{3} \rangle$ or $\lambda_{\nu_{3}} \langle \nu_{3} \rangle$ and, thus,  are very small, far below the order of approximation used in this paper. It follows that the chargino mass matrix is very nearly that of the MSSM given, for example, in \cite{Zc}. In this paper, the chargino masses are calculated by diagonalizing their MSSM mass matrix.

\subsection*{Squarks and Sleptons:}

Next, let us analyze the squark and slepton mass matrices. These have supersymmetric contributions from both the D- and F-terms in the scalar potential energy. First consider the D-terms.
In the MSSM, left-handed squarks and sleptons get well-known D-term masses proportional to the square of the Higgs VEV's. However, as discussed in \cite{BLPaper}, in the $B$-$L$ MSSM each squark and slepton gets an additional--and significant--D-term contribution to its squared mass proportional to $g_{B-L}^{2} {\langle \nu_{3} \rangle}^{2}$. For example, these contributions to the left-handed squark and slepton squared masses appear in (\ref{66}). In evaluating scalar masses in this paper, we include these additional $B$-$L$ contributions. Note that there will also be D-term contributions to some scalar masses arising from the left-handed sneutrino VEV $\langle N_{3} \rangle$. However, as discussed above, this is very small and will be ignored in our calculations.

Now consider the F-term contributions. Recall that we have assumed in 
(\ref{8}) that the squark/slepton Yukawa couplings are diagonal and from (\ref{140a}) that $\lambda_{\nu_{3}}$ must be very small. It follows that the non-zero $\langle \nu_{3} \rangle$ and $\langle N_{3} \rangle$ VEV's in the $B$-$L$ MSSM do not induce significant contributions and, hence, the F-term scalar masses can be well-approximated using the pure MSSM.  Note that since the Yukawa couplings have been chosen  to be diagonal, the only possible mixing is between the left and 
right-handed states of each scalar field.  For the ``up'' scalars $\phi_{u}$, this mixing arises from 
\begin{equation}
  \vert \frac{\partial W}{\partial H } \vert^{2} = \vert \mu \bar{H} + \lambda_{\phi} \phi_{u,L}^{*} \phi_{u,R}  \vert^{2} 
\end{equation} 
when the Higgs obtains a VEV.
Similarly for the ``down" type scalars.  
Since the Yukawa couplings for the third family are considerably larger than the others, it is reasonable to drop the first and second family contributions to the F-term scalar masses.  

Finally, note that there are also non-supersymmetric contributions from the quadratic and cubic soft terms. The quadratic masses are diagonal and straightforward. A contribution to mixing terms arises from the soft cubic couplings. Recall from (\ref{26}) that the A-parameters are assumed to be proportional to the corresponding Yukawa couplings.  Hence, we drop all such mixings except for the third family squarks and sleptons. 

 Putting everything together, in the basis $(\phi_{L}, \phi_{R})$ the squark and down-slepton mass squared matrices ${\bf m}_{t}^{2},{\bf m}_{b}^{2}$, and ${\bf m}_{\tau}^{2}$ are given by
\begin{equation}
  {\bf m}_{t}^{2} = 
  \begin{pmatrix}  
    m_{Q_{3}}^{2} + m_{t}^{2}  &  \langle H \rangle A_{t}^{*} - \mu \lambda_{t} \langle \bar{H} \rangle  \\
    \langle H \rangle A_{t} - \mu^{*} \lambda_{t} \langle \bar{H} \rangle  &   m_{\tilde{u}_{3}}^{2} + m_{t}^{2} \\
  \end{pmatrix}
\end{equation} 
\begin{equation}
  {\bf m}_{b}^{2} = 
  \begin{pmatrix}  
    m_{Q_{3}}^{2} + m_{b}^{2}  &  \langle \bar{H} \rangle A_{b}^{*} - \mu \lambda_{b} \langle H \rangle  \\
    \langle \bar{H} \rangle A_{b} - \mu^{*} \lambda_{b} \langle H \rangle  &   m_{\tilde{d}_{3}}^{2} + m_{b}^{2} \\
  \end{pmatrix}
\end{equation}
\begin{equation}
  {\bf m}_{\tau}^{2} = 
  \begin{pmatrix}  
    m_{L_{3}}^{2} + m_{\tau}^{2}  &  \langle \bar{H} \rangle A_{\tau}^{*} - \mu \lambda_{\tau} \langle H \rangle\\
    \langle \bar{H} \rangle A_{\tau} - \mu^{*} \lambda_{\tau} \langle H \rangle  &   m_{\tilde{e}_{3}}^{2} + m_{\tau}^{2}  \\
  \end{pmatrix}
  .
\end{equation}
These can be diagonalized by a unitary matrix.  For example, in the case $\phi = t_{3}$ this unitary matrix can be written as
\begin{equation}
 \begin{pmatrix} \tilde{t}_{3}^{(1)} \\ \tilde{t}_{3}^{(2)} \\ \end{pmatrix}  =  \begin{pmatrix} c_{\tilde{t}}  & -s_{\tilde{t}}^{*}  \\   s_{\tilde{t}}  &   c_{\tilde{t}}  \\   \end{pmatrix}  \begin{pmatrix}  \tilde{t}_{3}^{L} \\ \tilde{t}_{3}^{R} \\ \end{pmatrix} .
\end{equation}
To obtain the masses $m_{\tilde{t}_{3}^{(1)}}^{2} m_{\tilde{t}_{3}^{(2)}}^{2}$ for the eigenstates $\tilde{t}_{3}^{(1)},\tilde{t}_{3}^{(2)}$, matrix ${\bf m}_{t}^{2 \dagger} {\bf m}_{t}^{2}$ is diagonalized and one
takes the square root of the eigenvalues.  
However, if any of these are degenerate, the Takachi diagonalization process \cite{Choi} must be used. The sneutrino mass squares were analyzed in detail in \cite{BLPaper}. Hence, we do not discuss the up-slepton mass matrix in this paper.

\subsection*{Higgs Bosons:}
Next, consider the masses of the five Higgs bosons.  We note that scalar potential is perturbed from the MSSM potential by the VEVs of both the right handed and left handed neutrinos.  However, similar to previous discussions, we note that these perturbations are small and choose to approximate it with the MSSM scalar potential. 

  As is well known, in the MSSM the tree level mass of the lightest neutral Higgs is bounded above by the mass of the Z boson \cite{Zc}.  However, a neutral Higgs boson with mass below $M_{Z}$ in the MSSM has been ruled out experimentally. Hence, one must explore the radiative corrections to Higgs masses.  It turns out these corrections to the lightest neutral Higgs mass are quite sizable and allow the MSSM to remain a viable theory, albeit with a restricted range of parameters.  

We note that there exists an extensive literature on the radiative corrections to the MSSM Higgs masses,  with leading corrections calculated up to the three-loop level \cite{threeloops}.
For the purposes of this paper, we consider the lightest neutral Higgs mass to the first few leading terms in the one-loop correction, as given in \cite{Dabelstein}.  That is, our approximation for the lightest Higgs mass is  
\begin{eqnarray}
m_{h^{0}}^{2} & = &\frac{1}{2} ( m_{A^{0}}^{2} + M_{Z}^{2} + \omega_{t}) \nonumber \\ 
  &-& \sqrt{\frac{( m_{A^{0}}^{2} + M_{Z}^{2} )^{2} + \omega_{t}^{2} }{4}  - m_{A^{0}}^{2} M_{Z}^{2} \cos^{2} 2\beta + \frac{\omega_{t} \cos 2\beta}{2} (m_{A^{0}}^{2} - M_{Z}^{2})} \nonumber   \\
   & & \label{cat1}
\end{eqnarray}
where
\begin{eqnarray}
\omega_{t} & =& \frac{3}{4 \pi^{2}}  \lambda^{2}_{t} m_{t}^{2} \Big\lbrace \ln \big(\frac{m_{\tilde{t}_{3}^{(1)}} m_{\tilde{t}_{3}^{(2)}}}{m_{t}^{2}} \big)  +  c_{\tilde{t}_{1}}^{2} s_{\tilde{t}_{2}}^{2} \frac{\big( m_{\tilde{t}_{3}^{(2)}}^{2} - m_{\tilde{t}_{3}^{(1)}}^{2} \big)}{m_{t}^{2}} \ln \big( \frac{m_{\tilde{t}_{3}^{(2)}}^{2}}{m_{\tilde{t}_{3}^{(1)}}^{2}} \big) \nonumber \\  &+&  c_{\tilde{t}_{1}}^{4} s_{\tilde{t}_{2}}^{4} \frac{\big( m_{\tilde{t}_{3}^{(2)}}^{2} - m_{\tilde{t}_{3}^{(1)}}^{2} \big)^{2}}{m_{t}^{4}} \Big(1 - \frac{1}{2}( \frac{ m_{\tilde{t}_{3}^{(2)}}^{2} + m_{\tilde{t}_{3}^{(1)}}^{2}}{ m_{\tilde{t}_{3}^{(2)}}^{2} - m_{\tilde{t}_{3}^{(1)}}^{2}}) \ln \big(\frac{m_{\tilde{t}_{3}^{(2)}}^{2}}{m_{\tilde{t}_{3}^{(1)}}^{2}}\big) \Big)   \Big\rbrace  \ .  \label{cat2} 
\end{eqnarray}   

\noindent  The one-loop corrections to the remaining neutral and charged Higgs masses are also presented in \cite{Dabelstein}. It is useful, however, to note two things; first, 
the one-loop corrections to the $A^{0}$ mass are sub-leading when inserted into \eqref{cat1}
and, second, the radiative corrections to the $A^{0}$,$H^{(0)}$ and $H^{\pm}$ masses do not substantially change conclusions about their lower bounds derived from their tree-level expressions. That is,
the masses for the remaining neutral and charged Higgs can be well-approximated by their tree level expressions. These are given by \cite{Dabelstein}
\begin{eqnarray}
m_{A^{0}}^{2} &=& 2 \vert \mu \vert ^{2} + m_{H}^{2} + m_{\bar{H}}^{2} \label{wed1} \\
m_{H^{0}}^{2} &=& \frac{1}{2} (m_{A^{0}}^{2} + M_{Z}^{2}) + \sqrt{\frac{(m_{A^{0}}^{2} + M_{Z}^{2})^{2}}{4} - m_{A^{0}}^{2} M_{Z}^{2} \cos^{2} (2\beta) } \Big) \label{wed2} \\
m_{H^{\pm}}^{2} &=& m_{A^{0}}^{2} + M_{W}^{2}  \label{wed3} \ .
\end{eqnarray} 

We have followed the standard notation of labeling the Higgs mass eigenstates as $h^{0}$($H^{0}$) for the lightest(heaviest) neutral Higgs, $A^{0}$ for the CP odd neutral Higgs and $H^{\pm}$ for the respective $\pm$-charged Higgs.  It is also useful to recall that
\begin{equation}
  M_{W}^{2} = \frac{1}{2} g_{2}^{2} ( \langle H \rangle^{2} +  \langle \bar{H} \rangle^{2} ) \ .
\end{equation}
The definition for $M_{Z}$ was given in (\ref{28}) and (\ref{59}).
Finally, note from \eqref{cat1} and \eqref{wed2} that
\begin{equation}
m_{H^{0}} > m_{h^{0}} 
\label{thurs1}
\end{equation}
at tree-level and using  \cite{Dabelstein} that this inequality persists when the one-loop corrections are included.

\section*{Appendix B - Experimental Bounds on Higgs Masses}

In this Appendix, we review the experimental bounds on the Higgs masses.  We point out that these bounds are highly model dependent and, even within the MSSM, depend on assumptions imposed on the soft SUSY breaking parameters, the structure of mass matrices and so on.  To illustrate the non-trivial nature of these bounds, we briefly review one scenario of the MSSM and the experimental results for the Higgs masses.  Since Higgs fields carry no $B$-$L$ charge, to the one-loop level the following discussion is directly relevant to our $U(1)_{B-L}$ MSSM theory.

\begin{figure}
 \centering
 \includegraphics[scale=0.7]{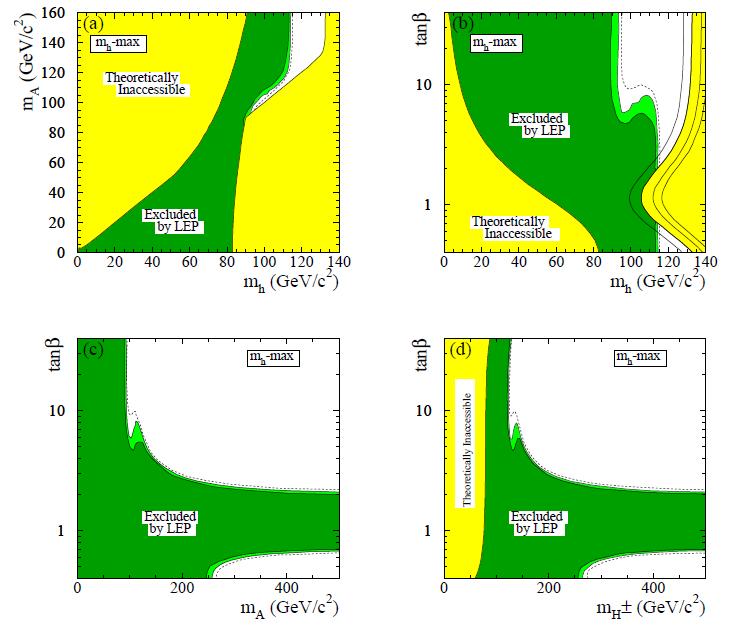}
 \caption{ {\scriptsize Exclusions at the 95\% confidence level (medium-grey or light-green) and at the 99.7\% confidence level (dark-grey or dark-green) for the CP conserving, $m_{h^{0}}-max$ scenario and for $m_{t}=174.3$ GeV/$c^{2}$.  The figure shows the theoretically inaccessible domains (light-grey or yellow) and the regions excluded in this search, in four projections of the MSSM parameters: (a): ($m_{h^{0}}$, $m_{A^{0}})$; (b): ($m_{h^{0}}$, $\tan\beta$); (c): ($m_{A^{0}}$, $\tan\beta$); (d): ($m_{H^{\pm}}$, tan$\beta$).
The dashed lines indicate the boundaries of the region which are expected to be excluded at 95\% confidence level based on Monte Carlo simulations with no signal. In the ($m_{h^{0}}$, $\tan\beta$) projection (plot (b)), the upper boundary of the parameter space is indicated for four values of the top quark mass; from left to right: $m_{t}=$169.3, 174.3, 179.3, and 183.0 GeV/$c^{2}$.  }} 
 \label{fig:LEPHiggs}
\end{figure}

We will discuss what is commonly referred to as the $m_{h^{0}}-max$ scenario, in which one assumes CP conservation and that the off-diagonal terms in the stop mixing matrix are large.  For this scenario, the Higgs mass bounds are reported in  \cite{Zb,LEP} and reproduced here in Figure \ref{fig:LEPHiggs}.  
We begin our analysis with plot (c) of Figure 14, which presents the experimentally excluded zone in the $m_{A^{0}}$-tan$\beta$ plane. Recall from \eqref{31} that to guarantee that the Yukawa parameters are  perturbative, we have restricted the range of tan$\beta$ to be 
\begin{equation}
4 \lesssim \tan\beta \lesssim 50 \ .
\label{ovrut1}
\end{equation}
Note from (c) that the lower bound on $m_{A_{0}}$ can be as small as $\sim 93$ GeV, but only for large values of tan$\beta$. Furthermore, for $93$ GeV $\lesssim m_{A_{0}} \lesssim 120$ GeV the lower bound on tan$\beta$ must exceed 4 in a complicated way. Therefore, to simplify the  calculations  in this paper we will always take
\begin{equation}
m_{A^{0}} \gtrsim 120 \ GeV  \ .
\label{ovrut2}
\end{equation}
With this assumption, there is no restriction on the the value of tan$\beta$ beyond those of \eqref{ovrut1}.

We see from plots (a) and (b) respectively that the lower bounds on the lightest Higgs mass $m_{h^{0}}$, and, hence, using \eqref{thurs1}, on $m_{H^{0}}$, are highly dependent on the values of $m_{A^{0}}$ and $\tan\beta$.  First, consider plot (a). For
\begin{equation}
m_{A^{0}} \gtrsim 120 \ GeV  \ \ \Longrightarrow \ \ m_{h^{0}} \gtrsim 114.4 \ GeV \ ,
\label{ovrut3}
\end{equation}
the oft-quoted lower bound on the lightest neutral Higgs mass. Note that for $93$ GeV $\lesssim m_{A^{0}} \lesssim 120$ GeV there is a small, almost linear, allowed region where $94$ GeV $\lesssim m_{h^{0}} \lesssim 114.4$ GeV. Theoretically, finding parameters in this region would requiring an unnaturally  high degree of fine-tuning. This is another justification for choosing the bound in \eqref{ovrut2}.
Second, let us analyze the bounds in plot (b). It follows from  \eqref{ovrut1} and \eqref{ovrut3} that plot (b) puts no further restrictions on the parameters.

Now consider the remaining plot (d) in Figure 14. This presents the experimental exclusion zone for $m_{H^{\pm}}$ as a function of tan$\beta$. Note from \eqref{wed3} that when $m_{A^{0}}$ exceeds its lower bound in \eqref{ovrut2}, then $m_{H^{\pm}}$ must be larger than $\sim144$ GeV. That is,
\begin{equation}
m_{A^{0}} \gtrsim 120 GeV \ \ \Longrightarrow \ \ m_{H^{\pm}} \gtrsim 144 GeV \ .
\label{ovrut4}
\end{equation}
With the exception of the  small region $144$ GeV $\lesssim m_{H^{\pm}} \lesssim  160$ GeV, where, roughly, tan$\beta$ must exceed 6, the bounds \eqref{ovrut1} and \eqref{ovrut4} guarantee that plot (d) puts no further constraints on the parameters.

To conclude: in the numerical analysis in this paper we use the $m_{h^{0}} - max$ scenario to put lower bounds on all five Higgs masses. This is accomplished by first calculating $m_{A^{0}}$
using \eqref{wed1}, accepting the result if $m_{A^{0}} \gtrsim 120$ GeV and rejecting it if $m_{A^{0}}$ has a smaller value. Assuming $m_{A^{0}}$ satisfies this inequality, we then calculate $m_{h^{0}}$ from \eqref{cat1} and \eqref{cat2}. Note that to evaluate $\omega_{t}$ in \eqref{cat2}, one must specify a value for the top quark mass. In order to compare our results to Figure 14, we take
\begin{equation}
m_{t}=174.3 \ GeV \ .
\label{ovrut5}
\end{equation}
The result is accepted if $m_{h^{0}} \gtrsim 114.4$ GeV and is rejected otherwise. If $m_{h^{0}}$ exceeds this bound we continue by computing $m_{H^{\pm}}$ using \eqref{wed3}. If it lies in the range $144$ GeV $ \lesssim m_{H^{\pm}} \lesssim 160$ GeV, then tan$\beta$ is restricted to exceed $6$. If it does we accept the result and reject it if tan$\beta$ is below this value. For $m_{H^{\pm}}  \gtrsim 160$ GeV, there are no further constraints and the result is accepted. In this way, for a given point in the $c_{q}(0)$-$c_{\nu_{3}}(0)$ projection we can map out the allowed Higgs scalar region in the $c_{\mu}(0)$-tan$\beta$ plane. Outside of this region, at least one of the five Higgs masses is less than the experimental lower bound.

For a more thorough discussion of this and many other scenarios in the MSSM, see \cite{LEP}.

\section*{Appendix C - Kinetic Mixing}

As is well-known, see for example \cite{Yd,km1}, in theories with at least two $U(1)$ gauge factors mixing can occur between the kinetic terms of the associated vector supermultiplets. More specifically, in the $B$-$L$ MSSM theory discussed in this paper there can be mixing between the $U(1)_{Y}$ and $U(1)_{B-L}$ gauge kinetic terms. To analyze this, consider the hypercharge and $B$-$L$ generators $\bf Y/2$ and $\bf Y_{B-L}$ defined in Section 2. Then
\begin{equation}
Tr( \frac{\bf Y}{2})^{2}=11, \quad  Tr {\bf Y_{B-L}}^{2}=16,  \quad Tr \frac{\bf Y}{2} {\bf Y_{B-L}}=8 \ .
\label{c1}
\end{equation}
The fact that $Tr \frac{\bf Y}{2} {\bf Y_{B-L}} \neq 0$ implies that kinetic mixing does indeed occur, introducing a new dimensionless coupling for the mixed term which we denote by $g_{M}$. Following \cite{Yd}, we construct the $2 \times 2$ matrix $g_{\alpha \beta}$ with components $g_{11}=g_{Y}$, $g_{22}=g_{B-L}$, $g_{12}=g_{21}=g_{M}$ and define the new generators
\begin{equation}
{\bar{\bf Y}}_{\alpha}= \sum_{\beta=1}^{2} {\bf Y}'_{\beta}g_{\beta \alpha}, \quad \alpha=1,2  
\label{c2}
\end{equation}
where ${\bf Y}'_{1}={\bf Y}/2$ and ${\bf Y}'_{2}={\bf Y}_{B-L}$. It then follows that at the one-loop level the RGEs for the gauge parameters generalize to
\begin{equation}
\frac{dg_{\alpha \beta}}{dt}=\frac{1}{16\pi^{2}} \beta_{\alpha \beta} , \quad \beta_{\alpha \beta}=g_{\alpha \gamma}Tr {\bar{\bf Y}}_{\gamma} {\bar{\bf Y}}_{\beta} \ .
\label{c3}
\end{equation}
The expressions for $ \beta_{\alpha \beta}$ are cubic polynomials in the gauge couplings and easily worked out using $\bf Y/2$ and $\bf Y_{B-L}$. If one assumes that $g_{M} \ll g_{Y},g_{B-L}$, then
\begin{equation}
\beta_{11}=\beta_{Y}=11 g_{Y}^{3}, \quad \beta_{22}=\beta_{B-L}=16 g_{B-L}^{3}, \quad \beta_{12}=\beta_{M}= 8 g_{Y}^{2}g_{B-L} \ .
\label{c4} 
\end{equation}
Note that the first two $\beta$-functions are exactly the ones used to scale $g_{Y}$ and $g_{B-L}$ in this paper.  However, it is important to be cognizant of the existence and scaling of $g_{M}$ and its physical implications.

Using the scaling solutions for $g_{Y}$ and $g_{B-L}$, one can solve the RGE for $g_{M}$. The result is
\begin{equation}
g_{M}(t)=g_{M}(0)-\frac{4 g(0)}{\sqrt{33}} \left( {\rm arctan}( \frac{1}{3} \sqrt{11+\frac{33g(0)^{2}|t|}{2 \pi^{2}}}) - {\rm arctan} (\frac{\sqrt{11}}{3})\right)
\label{c5}
\end{equation}
with $g(0)$ given in (\ref{15}). Note that $g_{M}(t)$ decreases monotonically as one scales down from the unification scale $t=0$ to the electroweak scale $t_{EW} \simeq -33.3$. We find that
\begin{equation}
g_{M}(t_{EW})=g_{M}(0) - .1489 \ .
\label{c6}
\end{equation}
The initial value $g_{M}(0)$ is a model and threshold dependent quantity. As with the soft SUSY breaking parameters, its calculation requires specifying an explicit ultraviolet vacuum--which is beyond the scope of this paper. Hence, as with the soft parameters, we treat it as an arbitrary parameter. 
To make the following discussion explicit, we find it convenient to choose $g_{M}(0)=.1489/2$. It follows that $g_{M}$ has the positive value $.0745$ at the unification scale, passes through zero at lower energy and becomes $g_{M}(t_{EW})=-.0745$ at the  electroweak scale. Over this entire range the assumption that $g_{M} \ll g_{Y},g_{B-L}$ remains valid.

What is the physical consequence of $g_{M}(t_{EW}) \neq 0$? It follows from (\ref{c2}) that, in addition to the hypercharge generator $g_{Y}(\frac{\bf Y}{2})$, the mixing induces a new generator 
\begin{equation}
g_{X}( a\frac{\bf Y}{2} + b{\bf Y}_{B-L}) \ ,
\label{c7}
\end{equation}
where
\begin{equation}
g_{X}a= g_{M} , \quad g_{X}b=g_{M}+g_{B-L} \ .
\label{c8}
\end{equation}
It was shown in \cite{Spinb} that the parameter controlling the physical $Z$-$Z'$ mixing is given to leading order by
\begin{equation}
2\xi=\frac{2a \sqrt{g_{2}^{2}+g_{Y}^{2}}}{b^{2}g_{X}} \left(\frac{{\langle H \rangle}^{2}+{\langle {\bar{H}}\rangle}^{2}}{{\langle \nu_{3} \rangle}^{2}} \right) \ .
\label{c9}
\end{equation}
Using the data in this paper and the choice $g_{M}(0)=.0745$, this becomes
\begin{equation}
2\xi \simeq .5292 \left(\frac{M_{Z}}{M_{A_{B-L}}}  \right)^{2}  \ .
\label{c10}
\end{equation}
The hierarchy $M_{Z}/M_{A_{B-L}}$
is sufficiently large everywhere in the allowed regions discussed in Section 4 that 
\begin{equation}
2\xi \lesssim {\cal{O}}(10^{-3}) \ ,
\label{c11}
\end{equation}
which is phenomenologically acceptable \cite{Spinb}. For example, at the points (P) and (S), where  $M_{A_{B-L}}=1252$ GeV and $M_{A_{B-L}}=3005$ GeV, (\ref{c10}) becomes $2\xi=2.796 \times 10^{-3}$ and  $2\xi=4.853 \times 10^{-4}$ respectively. Finally, we note that taking larger $g_{M}(0)$ will reduce the mixing at the electroweak scale, with $2\xi$ vanishing for $g_{M}(0)=.1489$. On the other hand, for smaller values of 
$g_{M}(0)$, such as $g_{M}(0)=0$, the $Z$-$Z'$ mixing continues to satisfy (\ref{c11}) over most of each allowed region in Section 4.

We conclude that there is a reasonable range of $g_{M}(0)$ for which 1) the scaling equations for the gauge parameters are well-approximated by dropping the $g_{M}$ contributions and 2) for which the induced $Z$-$Z'$ mixing is naturally small and phenomenologically acceptable. The analysis in this paper is meant to establish the existence of  $B$-$L$/electroweak symmetry breaking with a reasonable spectrum over a subspace of a large initial parameter space, and not to give precision values of all quantities. It is consistent with this goal, and {\it greatly} simplifies the calculations, to ignore the $U(1)_{Y}$-$U(1)_{B-L}$ kinetic mixing.


\section*{Acknowledgments}
B.A.O. would like to thank W. Buchm\"uller for helpful discussions at the KITP Workshop: Strings at the LHC and in the Early Universe.  The work of M.A. and B.A.O. is supported in part by the DOE under contract No. DE-AC02-76-ER-03071.  B.A.O. acknowledges partial support from the NSF RTG grant DMS-0636606.


\begin{thebibliography}{99}



\bibitem{a} The Heterotic String, David J. Gross, Jeffrey A. Harvey, Emil J. Martinec, Ryan Rohm, Phys.Rev.Lett.54:502-505 (1985).

\bibitem{b}Heterotic and Type I String Dynamics from Eleven-Dimensions,
Petr Horava, Edward Witten, Nucl.Phys.B460:506-524 (1996), 
[hep-th/9510209].

\bibitem{c} Strong Coupling Expansion of Calabi-Yau Compactification,
Edward Witten, Nucl.Phys.B471:135-158 (1996),
[hep-th/9602070].

\bibitem{d} Eleven-Dimensional Supergravity on a Manifold with Boundary.
Petr Horava, Edward Witten, Nucl.Phys.B475:94-114 (1996), 
[hep-th/9603142].

\bibitem{e} On the Four-Dimensional Effective Action of Strongly Coupled Heterotic String Theory,
A.~Lukas, B.~A.~Ovrut and D.~Waldram,
Nucl.\ Phys.\  B {\bf 532}, 43 (1998),
[arXiv:hep-th/9710208].

\bibitem{f} The Universe as a Domain Wall,
Andre Lukas, Burt A. Ovrut, K.S. Stelle, Daniel Waldram, Phys.Rev.D59:086001(1999),
hep-th/9803235.

\bibitem{g} Heterotic M-theory in Five Dimensions,
A.~Lukas, B.~A.~Ovrut, K.~S.~Stelle and D.~Waldram,
Nucl.\ Phys.\  B {\bf 552}, 246 (1999),
[arXiv:hep-th/9806051].


\bibitem{Ca}  Aspects of (2,0) String Compactifications, J. Distler and B.R. Greene,
Nucl.Phys.B {\bf 304}, 1 (1988).

\bibitem{Cb}  Some Three Generation (0,2) Calabi-Yau Models, S. Kachru,
Phys.Lett.B {\bf 349}, 76 (1995),
hep-th/9501131.

\bibitem{Cc}  Heterotic GUT and Standard Model Vacua from Simply Connected Calabi-Yau Manifolds,
R. Blumenhagen, S. Moster and T. Weigand,
Nucl.Phys.B {\bf 751}, 186 (2006),
hep-th/0603015.

\bibitem{Cd} Heterotic compactification, an algorithmic approach,
L.B. Anderson, Y.H. He and A. Lukas,
JHEP {\bf 0707}, 049 (2007),
hep-th/0702210.

\bibitem{Ce} The Edge Of Supersymmetry: Stability Walls in Heterotic Theory,
L.B. Anderson, J. Gray, A. Lukas and B. Ovrut,
Phys.Lett.B {\bf 677}, 190 (2009),
arXiv:0903.5088 [hep-th].

\bibitem{Gaaa} Vector Bundles and F Theory,
Robert Friedman, John Morgan, Edward Witten, Commun.Math.Phys.187:679-743 (1997), 
hep-th/9701162.

\bibitem{Gaa} Principal Bundles on Elliptic Fibrations, R. Donagi, Asian J. Math, Vol. 1, 214-223 (1997),
alg-geom/9702002.

\bibitem{Ga} Nonperturbative Vacua and Particle Physics in M Theory,
Ron Donagi, Andre Lukas, Burt A. Ovrut, Daniel Waldram, JHEP 9905:018 (1999), 
hep-th/9811168.

\bibitem{Gc} Standard Models from Heterotic M Theory,
Ron Donagi, Burt A. Ovrut, Tony Pantev, Daniel Waldram,  Adv.Theor.Math.Phys.5:93-137 (2002), 
hep-th/9912208.

\bibitem{Hb} Standard Model Bundles, Ron Donagi, Burt A. Ovrut, Tony Pantev, Dan Waldram, Adv.Theor.Math.Phys.5:563-615 (2002), math/0008010.

\bibitem{Hd}  SU(4) Instantons on Calabi-Yau Threefolds with Z(2) x Z(2) Fundamental Group.
Ron Donagi, Burt A. Ovrut, Tony Pantev, Rene Reinbacher, JHEP 0401:022 (2004), 
hep-th/0307273.


\bibitem{Ia}  Massless Spectra of Three Generation U(N) Heterotic String Vacua,
R. Blumenhagen, S. Moster, R. Reinbacher and T Weigand,
JHEP {\bf 0705}, 041 (2007),
hep-th/0612039.

\bibitem{Ib} Monad Bundles in Heterotic String Compactifications,
L.B. Anderson, Y.H. He and A. Lukas,
JHEP {\bf 0807}, 104 (2008),
arXiv:0805.2875 [hep-th].


\bibitem{Ja} Moduli Dependent Spectra of Heterotic Compactifications,
Ron Donagi, Yang-Hui He, Burt A. Ovrut, Rene Reinbacher, Phys.Lett.B598:279-284 (2004), 
hep-th/0403291.

\bibitem{Jb} The Particle Spectrum of Heterotic Compactifications,
Ron Donagi, Yang-Hui He, Burt A. Ovrut, Rene Reinbacher, JHEP 0412:054 (2004), 
hep-th/0405014.

\bibitem{Kb} The Spectra of Heterotic Standard Model Vacua,
Ron Donagi, Yang-Hui He, Burt A. Ovrut, Rene Reinbacher, JHEP 0506:070 (2005), 
hep-th/0411156.

\bibitem{Kc} Heterotic Standard Model Moduli,
Volker Braun, Yang-Hui He, Burt A. Ovrut, Tony Pantev, JHEP 0601:025 (2006), 
hep-th/0509051.

\bibitem{Lb} Yukawa Couplings in Heterotic Standard Models,
Volker Braun, Yang-Hui He, Burt A. Ovrut,  JHEP 0604:019 (2006), 
hep-th/0601204.

\bibitem{Lc} Two Higgs Pair Heterotic Vacua and Flavor-Changing Neutral Currents, 
Michael Ambroso, Volker Braun, Burt A. Ovrut,  JHEP 0810:046 (2008), 
arXiv:0807.3319 [hep-th].

 \bibitem{Ld} Yukawa Couplings in Heterotic Compactification,
L.B. Anderson, J. Gray, D. Grayson, Y.H. He and A. Lukas, 
arXiv:0904.2186 [hep-th].

\bibitem{Maaa}  Superpotentials and Membrane Instantons,
Jeffrey A. Harvey, Gregory W. Moore, hep-th/9907026.

\bibitem{Maa} Instabilities in Heterotic M Theory Induced by Open Membrane Instantons,
Gregory W. Moore, Grigor Peradze, Natalia Saulina, Nucl.Phys.B607:117-154 (2001), 
hep-th/0012104.

\bibitem{Ma} Nonperturbative Superpotential from Membrane Instantons in Heterotic M Theory,
Eduardo Lima, Burt A. Ovrut, Jaemo Park, Rene Reinbacher, Nucl.Phys.B614:117-170 (2001), 
hep-th/0101049.

\bibitem{Mc} Superpotentials for Vector Bundle Moduli,
Evgeny I. Buchbinder, Ron Donagi, Burt A. Ovrut, Nucl.Phys.B653:400-420 (2003), 
hep-th/0205190.

\bibitem{Nb} Stabilizing Moduli with a Positive Cosmological Constant in Heterotic M-Theory,
Volker Braun, Burt A. Ovrut, JHEP 0607:035 (2006), 
hep-th/0603088.

\bibitem{Ob} Visible Branes with Negative Tension in Heterotic M Theory,
Ron Y. Donagi, Justin Khoury, Burt A. Ovrut, Paul Steinhardt, Neil Turok, JHEP 0111:041 (2001), 
hep-th/0105199.

\bibitem{Qa} Elliptic Calabi-Yau Threefolds with Z(3) x Z(3) Wilson Lines,
Volker Braun, Burt A. Ovrut, Tony Pantev, Rene Reinbacher, UPR-1089-T, JHEP 0412:062 (2004), 
hep-th/0410055.

\bibitem{Ra} Vector Bundle Extensions, Sheaf Cohomology, and the Heterotic Standard Model,
Volker Braun, Yang-Hui He, Burt A. Ovrut, Tony Pantev, Adv.Theor.Math.Phys.10:4(2006), 
hep-th/0505041.

\bibitem{Sa} A Heterotic Standard Model,
Volker Braun, Yang-Hui He, Burt A. Ovrut, Tony Pantev, Phys.Lett.B618:252-258 (2005), 
hep-th/0501070.

\bibitem{donagi}  An SU(5) heterotic standard model,  V.~Bouchard and R.~Donagi,  Phys.\ Lett.\  B {\bf 633}, 783 (2006),  [arXiv:hep-th/0512149].

\bibitem{exact}  The Exact MSSM Spectrum from String Theory, V.~Braun, Y.~H.~He, B.~A.~Ovrut, and T.~Pantev,
JHEP0605:043,2006, [arXiv:hep-th/0512177].

\bibitem{lukas}  Exploring Positive Monad Bundles And A New Heterotic Standard Model,  L.~B.~Anderson, J.~Gray, Y.~H.~He and A.~Lukas,  arXiv:0911.1569 [hep-th].


\bibitem{Sakharov}  Violation of CP Invariance, c Asymmetry, and Baryon Asymmetry of theUniverse,  A.~D.~Sakharov,  Pisma Zh.\ Eksp.\ Teor.\ Fiz.\  {\bf 5}, 32 (1967)  [JETP Lett.\  {\bf 5}, 24 (1967\ SOPUA,34,392-393.1991\ UFNAA,161,61-64.1991)].

\bibitem{Lepto}  Leptogenesis as the origin of matter, W.~Buchmuller, R.~D.~Peccei and T.~Yanagida,  Ann.\ Rev.\ Nucl.\ Part.\ Sci.\  {\bf 55}, 311 (2005)  [arXiv:hep-ph/0502169].

\bibitem{Campbell}  Cosmological baryon asymmetry constraints on extensions of the standard model,  B.~A.~Campbell, S.~Davidson, J.~R.~Ellis and K.~A.~Olive,   Phys.\ Lett.\  B {\bf 256}, 484 (1991).
  
\bibitem{Dreiner}  Sphaleron Erasure Of Primordial Baryogenesis,  H.~K.~Dreiner and G.~G.~Ross,  Nucl.\ Phys.\  B {\bf 410}, 188 (1993)  [arXiv:hep-ph/9207221].

\bibitem{Barbier}  R-parity violating supersymmetry,  R.~Barbier {\it et al.},  Phys.\ Rept.\  {\bf 420}, 1 (2005)  [arXiv:hep-ph/0406039].  

\bibitem{Ub} Discrete Gauge Symmetries And The Origin Of Baryon And Lepton Number Conservation In Supersymmetric Versions Of The Standard Model,  L. E. Ibanez, G. G. Ross, Nucl.Phys.B368, 3 (1992).

\bibitem{Salam}  Supersymmetry, Parity And Fermion-Number Conservation,  A.~Salam and J.~A.~Strathdee,  Nucl.\ Phys.\  B {\bf 97}, 293 (1975).
  
\bibitem{Fayet}
  Supergauge Invariant Extension Of The Higgs Mechanism And A Model For The Electron And Its Neutrino,  P.~Fayet,  Nucl.\ Phys.\  B {\bf 90}, 104 (1975).
  
\bibitem{Wein}  Supersymmetry At Ordinary Energies. 2. R Invariance, Goldstone Bosons, And  Gauge Fermion Masses,
  G.~R.~Farrar and S.~Weinberg,  Phys.\ Rev.\  D {\bf 27}, 2732 (1983).
  
\bibitem{Ud} Phenomenology of the Production, Decay, and Detection of New Hadronic States Associated with Supersymmetry, G. R. Farrar, P. Fayet, Phys.Lett.B76, 5 (1978). 

\bibitem{Ua}  R Symmetry in MSSM and Beyond with Several Consequences, G. Lazarides, Q. Shafi,
Phys.Rev.D58, 071702 (1998), arXiv:hep-ph/9803397.

\bibitem{Uc}  Softly Broken Supersymmetry And SU(5),  S.~Dimopoulos and H.~Georgi, Nucl.\ Phys.\  B  193, 150 (1981).

\bibitem{Va} New Contributions to Neutrinoless Double-Beta Decay in Supersymmetric     
Theories,  R. N. Mohapatra, Phys. Rev. D34, 3457 (1986).                                       
 
\bibitem{Vb} SM Extensions with Gauged B-L, F. Zwirner, review presented at the NO-VE International Workshop on Neutrino Oscillations, Venice, Italy, April 17, 2008.

\bibitem{Vc}  Radiative B-L Symmetry Breaking in Supersymmetric Models, S. Khalil, A. Masiero,
Phys.Lett.B665:374 (2008), arXiv:0710.3525.

\bibitem{Vd}  B-L Mediated SUSY Breaking with Radiative B-L Symmetry Breaking, T. Kikuchi, T. Kubo,
AIP Conf.Proc.1078:402-404 (2009)  arXiv:0809.2011.

\bibitem{Ve}  Implications of Supersymmetric Models with Natural R-parity Conservation,  
Stephen P. Martin, Phys.Rev. D54 2340-2348 (1996),  arXiv:hep-ph/9602349.

\bibitem{Vf}  Some Simple Criteria for Gauged R-Parity,  S.P. Martin, Phys.Rev.D46, 2769 (1992),
hep-ph/9207218.


\bibitem{BLLetter}  The B-L/Electroweak Hierarchy in Heterotic String and M-Theory,  M.~Ambroso and B.~Ovrut,  JHEP {\bf 0910}, 011 (2009)  [arXiv:0904.4509 [hep-th]].
    
\bibitem{BLPaper}  The B-L/Electroweak Hierarchy in Smooth Heterotic Compactifications,  M.~Ambroso and B.~A.~Ovrut, accepted in Intl Journal of Modern Physics A (2010),
 arXiv:0910.1129 [hep-th].


\bibitem{tamaz}  B-L Cosmic Strings in Heterotic Standard Models,
T.~Brelidze and B.~A.~Ovrut, 
arXiv:1003.0234 [hep-th/ph].
  
\bibitem{Buchmuller}  Gravitino dark matter in R-parity breaking vacua,  W.~Buchmuller, L.~Covi, K.~Hamaguchi, A.~Ibarra and T.~Yanagida,  JHEP {\bf 0703}, 037 (2007)  [arXiv:hep-ph/0702184].  

\bibitem{Takayama}  Gravitino dark matter without R-parity, F.~Takayama and M.~Yamaguchi,  Phys.\ Lett.\  B {\bf 485}, 388 (2000)  [arXiv:hep-ph/0005214].


\bibitem{Spina}  Minimal gauged $U(1)_{B-L}$ model with spontaneous R-parity violation,  V.~Barger, P.~Fileviez Perez and S.~Spinner,  Phys.\ Rev.\ Lett.\  {\bf 102}, 181802 (2009)  [arXiv:0812.3661 [hep-ph]].
  
\bibitem{Spinb}  Spontaneous R-Parity Breaking in SUSY Models,  P.~Fileviez Perez and S.~Spinner,
  Phys.\ Rev.\  D {\bf 80}, 015004 (2009)  [arXiv:0904.2213 [hep-ph]].  
  
\bibitem{Spinc}  The Right Side of Tev Scale Spontaneous R-Parity Violation,  L.~L.~Everett, P.~Fileviez Perez and S.~Spinner,  Phys.\ Rev.\  D {\bf 80}, 055007 (2009)  [arXiv:0906.4095 [hep-ph]].
  
\bibitem{Spind}  Spontaneous R-Parity Breaking and Left-Right Symmetry,  P.~Fileviez Perez and S.~Spinner,
  Phys.\ Lett.\  B {\bf 673}, 251 (2009)  [arXiv:0811.3424 [hep-ph]].
  
\bibitem{Lara}  The Edge Of Supersymmetry: Stability Walls in Heterotic Theory, Lara B. Anderson, James Gray, Andre Lukas, Burt Ovrut, Phys.Lett.B677:190-194,2009, arXiv:0903.5088 [hep-th].

\bibitem{Yd} Renormalisation of the Fayet-Iliopoulos D-term, I. Jack, D. R.~T. Jones, 
Phys.Lett.B 473, 102 (2000), arXiv:hep-ph/9911491.


\bibitem{W}  Soft Breaking of Supersymmetry, L. Girardello, M. T. Grisaru,
Nuc. Phys. B194, 65 (1982).

\bibitem{Xa}  Supersymmetric Higgs and Radiative Electroweak Breaking, L.E.Ibanez, G.G.Ross,
ComptesRendusPhysique8:1013-1028 (2007),  arXiv:hep-ph/0702046.

\bibitem{Xb}  Five--Branes and Supersymmetry Breaking in M--Theory, 
A. Lukas, B. A. Ovrut, D. Waldram,  JHEP 9904,009 (1999), arXiv:hep-th/9901017.

\bibitem{Xd}  Supersymmetry Breaking and Soft Terms in M-Theory, 
H.P. Nilles, M. Olechowski, M. Yamaguchi, Phys.Lett. B415,24-30 (1997),  arXiv:hep-th/9707143.

\bibitem{Zc} A Supersymmetry Primer,  S.~P.~Martin,  [arXiv:hep-ph/9709356]. 

\bibitem{Zd}  Supersymmetric unification without low energy supersymmetry and  signatures  for fine-tuning at the LHC,  N.~Arkani-Hamed and S.~Dimopoulos,  JHEP {\bf 0506}, 073 (2005)  [arXiv:hep-th/0405159].

\bibitem{Zb}  Review of Particle Physics, C. Amsler {\it et al.}  [Particle Data Group],  Phys.Lett.B667, 1 (2008) and 2009 partial update for the 2010 edition

\bibitem{Ze}  Precision prediction of gauge couplings and the profile of a string theory,  D.~M.~Ghilencea and G.~G.~Ross,  Nucl.\ Phys.\  B {\bf 606}, 101 (2001)  [arXiv:hep-ph/0102306].

\bibitem{Yg} The Soft Supersymmetry-Breaking Lagrangian: Theory and Applications, 
D.J.H. Chung, L.L. Everett, G.L. Kane, S.F. King, J. Lykken, Lian-Tao Wang, Phys.Rept. 407,1-203 (2005),  arXiv:hep-ph/0312378.

\bibitem{Ya}  Regularization Dependence of Running Couplings in Softly Broken Supersymmetry, 
S. P. Martin,  M. T. Vaughn, Phys.Lett.B 318, 331 (1993), arXiv:hep-ph/9308222.

\bibitem{Yb} Two Loop Renormalization Group Equations For Soft Supersymmetry Breaking Couplings, S. P. Martin, M. T. Vaughn, Phys.Rev.D50, 2282 (1994), arXiv:hep-ph/9311340.

\bibitem{Yc}Two-Loop Renormalization Group Equations for Soft Supersymmetry-Breaking Scalar Interactions: Supergraph Method,  Y. Yamada, Phys. Rev. D50, 3537 (1994),  arXiv:hep-ph/9401241.

\bibitem{Ye} The Fayet-Iliopoulos D-term and its Renormalisation in Softly-Broken Supersymmetric Theories,  I. Jack, D. R. T. Jones, S. Parsons, Phys.Rev.D62, 125022 (2000), arXiv:hep-ph/0007291.
 
\bibitem{Yf}The Fayet-Iliopoulos D-term and its Renormalisation in the MSSM,
 I. Jack, D. R. T. Jones, Phys.\ Rev.\  D63, 075010 (2001), arXiv:hep-ph/0010301.

\bibitem{Xc}Four-Dimensional Effective Supergravity and Soft Terms in M-Theory,  
K. Choi, H. B. Kim, C. Munoz, Phys.Rev. D57,7521-7528 (1998), arXiv:hep-th/9711158.

\bibitem{Xe} Towards a Theory of Soft Terms for the Supersymmetric Standard Model, 
A. Brignole, L.E. Ibanez, C. Mu–oz, Nucl.Phys.B422,125 (1994), arXiv:hep-ph/9308271.
  
\bibitem{Sola}  Quantum SUSY signatures in low-energy and high-energy processes,  J.~Sola, Pramana {\bf 51}, 239 (1998) [arXiv:hep-ph/9807244].


\bibitem{Choi}  The neutralino sector in the U(1)-extended supersymmetric standard model,  S.~Y.~Choi, H.~E.~Haber, J.~Kalinowski and P.~M.~Zerwas,  Nucl.\ Phys.\  B {\bf 778}, 85 (2007)  [arXiv:hep-ph/0612218].


\bibitem{threeloops}  Higgs boson mass in supersymmetry to three loops,  R.~V.~Harlander, P.~Kant, L.~Mihaila and M.~Steinhauser,  Phys.\ Rev.\ Lett.\  {\bf 100}, 191602 (2008)  [arXiv:0803.0672 [hep-ph]].

\bibitem{Dabelstein}  The One loop renormalization of the MSSM Higgs sector and its application to the neutral scalar Higgs masses,  A.~Dabelstein,  Z.\ Phys.\  C {\bf 67}, 495 (1995)  [arXiv:hep-ph/9409375].

\bibitem{LEP} Search for neutral MSSM Higgs bosons at LEP,  S.~Schael {\it et al.}  [ALEPH, DELPHI, L3 and OPAL Collaboration],  Eur.\ Phys.\ J.\  C {\bf 47}, 547 (2006)  [arXiv:hep-ex/0602042].


\bibitem{km1} Kinetic Mixing and the Supersymmetric Gauge Hierarchy,
Keith R. Dienes, Christopher F. Kolda, John March-Russell,
Nucl.Phys.B492:104-118 (1997),
hep-ph/9610479.


\end{thebibliography}
\end{document}